\documentclass[%
 reprint,
 superscriptaddress,
 showkeys,
 amsmath,amssymb,
 aps,
]{revtex4-1}

\usepackage{graphicx}
\usepackage{dcolumn}
\usepackage{bm}
\usepackage{float}
\usepackage{xcolor}
\usepackage{hyperref}
\usepackage{tabularx}
\usepackage{multirow}

\begin{document}

\preprint{APS/123-QED}

\title{Fine Structure of the Isovector Giant Dipole Resonance in $^{142-150}$Nd and $^{152}$Sm}

\author{L.~M.~Donaldson}
\email{lmdonaldson@tlabs.ac.za}
\affiliation{iThemba LABS, Old Faure Road, Faure 7131, South Africa}
\affiliation{School of Physics, University of the Witwatersrand, Johannesburg 2050, South Africa}

\author{J.~Carter}
\affiliation{School of Physics, University of the Witwatersrand, Johannesburg 2050, South Africa}

\author{P.~von~Neumann-Cosel}
\affiliation{Institut {f\"u}r Kernphysik, Technische Universit{\"a}t Darmstadt, D-64289 Darmstadt, Germany}

\author{V.~O.~Nesterenko}
\affiliation{Bogoliubov Laboratory of Theoretical Physics, Joint Institute for Nuclear Research, Dubna 141980, Russia}

\author{R.~Neveling}
\affiliation{iThemba LABS, Old Faure Road, Faure 7131, South Africa}

\author{P.-G.~Reinhard}
\affiliation{Institut f{\"u}r Theoretische Physik II, Universit{\"a}t Erlangen, D-91058 Erlangen, Germany}

\author{I.~T.~Usman}
\affiliation{School of Physics, University of the Witwatersrand, Johannesburg 2050, South Africa}

\author{P.~Adsley}
\affiliation{iThemba LABS, Old Faure Road, Faure 7131, South Africa}
\affiliation{School of Physics, University of the Witwatersrand, Johannesburg 2050, South Africa}
\affiliation{Department of Physics, Stellenbosch University, Matieland 7602, South Africa}

\author{C.~A.~Bertulani}
\affiliation{Department of Physics and Astronomy, Texas A\&M University-Commerce, Commerce TK 75429, USA}

\author{J.~W.~Br{\"u}mmer}
\affiliation{Department of Physics, Stellenbosch University, Matieland 7602, South Africa}

\author{E.~Z.~Buthelezi}
\affiliation{iThemba LABS, Old Faure Road, Faure 7131, South Africa}

\author{G.~R.~J.~Cooper}
\affiliation{School of Geosciences, University of the Witwatersrand, Johannesburg 2050, South Africa}

\author{R.~W.~Fearick}
\affiliation{Department of Physics, University of Cape Town, Rondebosch 7700, South Africa}

\author{S.~V.~F{\"o}rtsch}
\affiliation{iThemba LABS, Old Faure Road, Faure 7131, South Africa}

\author{H.~Fujita}
\affiliation{Research Center for Nuclear Physics, Osaka University, Ibaraki, Osaka 567-0047, Japan}

\author{Y.~Fujita}
\affiliation{Research Center for Nuclear Physics, Osaka University, Ibaraki, Osaka 567-0047, Japan}

\author{M.~Jingo}
\affiliation{School of Physics, University of the Witwatersrand, Johannesburg 2050, South Africa}

\author{N.~Y.~Kheswa}
\affiliation{iThemba LABS, Old Faure Road, Faure 7131, South Africa}

\author{W.~Kleinig}
\affiliation{Bogoliubov Laboratory of Theoretical Physics, Joint Institute for Nuclear Research, Dubna 141980, Russia}

\author{C.~O.~Kureba}
\affiliation{School of Physics, University of the Witwatersrand, Johannesburg 2050, South Africa}
\affiliation{Department of Physics and Astronomy, Botswana International University of Science and Technology, P. Bag 16, Palapye, Botswana}

\author{J.~Kvasil}
\affiliation{Institute of Particle and Nuclear Physics, Charles University, CZ-18000, Prague 8, Czech Republic}

\author{M.~Latif}
\affiliation{School of Physics, University of the Witwatersrand, Johannesburg 2050, South Africa}

\author{K.~C.~W.~Li}
\affiliation{Department of Physics, Stellenbosch University, Matieland 7602, South Africa}

\author{J.~P.~Mira}
\affiliation{iThemba LABS, Old Faure Road, Faure 7131, South Africa}

\author{F.~Nemulodi}
\affiliation{iThemba LABS, Old Faure Road, Faure 7131, South Africa}

\author{P.~Papka}
\affiliation{iThemba LABS, Old Faure Road, Faure 7131, South Africa}
\affiliation{Department of Physics, Stellenbosch University, Matieland 7602, South Africa}

\author{L.~Pellegri}
\affiliation{iThemba LABS, Old Faure Road, Faure 7131, South Africa}
\affiliation{School of Physics, University of the Witwatersrand, Johannesburg 2050, South Africa}

\author{N.~Pietralla}
\affiliation{Institut {f\"u}r Kernphysik, Technische Universit{\"a}t Darmstadt, D-64289 Darmstadt, Germany}

\author{V.~Yu.~Ponomarev}
\affiliation{Institut {f\"u}r Kernphysik, Technische Universit{\"a}t Darmstadt, D-64289 Darmstadt, Germany}

\author{B.~Rebeiro}
\affiliation{Department of Physics and Astronomy, University of the Western Cape, Bellville, Cape Town 7535, South Africa}

\author{A.~Richter}
\affiliation{Institut {f\"u}r Kernphysik, Technische Universit{\"a}t Darmstadt, D-64289 Darmstadt, Germany}

\author{N.~Yu.~Shirikova}
\affiliation{Bogoliubov Laboratory of Theoretical Physics, Joint Institute for Nuclear Research, Dubna 141980, Russia} 

\author{E.~Sideras-Haddad}
\affiliation{School of Physics, University of the Witwatersrand, Johannesburg 2050, South Africa}

\author{A.~V.~Sushkov}
\affiliation{Bogoliubov Laboratory of Theoretical Physics, Joint Institute for Nuclear Research, Dubna 141980, Russia}

\author{F.~D.~Smit}
\affiliation{iThemba LABS, Old Faure Road, Faure 7131, South Africa}

\author{G.~F.~Steyn}
\affiliation{iThemba LABS, Old Faure Road, Faure 7131, South Africa}

\author{J.~A.~Swartz}
\affiliation{iThemba LABS, Old Faure Road, Faure 7131, South Africa}
\affiliation{Department of Physics, Stellenbosch University, Matieland 7602, South Africa}

\author{A.~Tamii}
\affiliation{Research Center for Nuclear Physics, Osaka University, Ibaraki, Osaka 567-0047, Japan}

\date{November 20, 2020; Accepted: December 7, 2020}

\begin{abstract}

\begin{description}

\item[Background]
Inelastic proton scattering at energies of a few hundred MeV and very-forward scattering angles including $0^\circ$ has been established as a tool for the study of electric-dipole strength distributions in nuclei. The present work reports a systematic investigation of the chain of stable even-mass Nd isotopes representing a transition from spherical to quadrupole-deformed nuclei.

\item[Purpose]
Extraction of the equivalent photo-absorption cross sections and analysis of their fine structure in the energy region of the IsoVector Giant Dipole Resonance (IVGDR).

\item[Method]
Proton inelastic scattering reactions of 200 MeV protons were measured at the iThemba Laboratory for Accelerator Based Sciences in Cape Town, South Africa. The scattering products were momentum-analysed by the K600 magnetic spectrometer positioned at $\theta_{\mathrm{Lab}}=0^\circ$. Using dispersion-matching techniques, energy resolutions of $\Delta E \approx 40 - 50$ keV (Full Width at Half Maximum) were obtained. After subtraction of background and contributions from other multipoles, the spectra were converted to photo-absorption cross sections using the equivalent virtual-photon method. Wavelet-analysis techniques are used to extract characteristic energy scales of the fine structure of the IVGDR from the experimental data.

\item[Results]
Comparisons between the extracted experimental energy scales and those energy scales obtained from the Quasiparticle-Phonon Model (QPM) and Skyrme Separable Random Phase Approximation (SSRPA) predictions provide insight into the role of different giant-resonance damping mechanisms.

\item[Conclusions]
Fine structure of the IVGDR is observed even for the most deformed nuclei studied. Fragmentation of the one particle-one hole (1p1h) strength seems to be the main source in both spherical and deformed nuclei. Some impact of the spreading due to coupling of the two particle-two hole (2p2h) states to the 1p1h doorway states is seen in the spherical/transitional nuclei, where calculations beyond the 1p1h level are available.

\end{description}
\end{abstract}


\maketitle

\section{\label{sec:Intro}Introduction}

Giant resonances are essentially a collective motion of many, if not all, particles in the nucleus~\cite{Harakeh}. In quantum-mechanical terms, they correspond to a transition between the ground state and the collective state, with strengths described by transition amplitudes. It is for this reason that giant resonances serve as a prime example of collective modes in the nucleus. A smooth mass-number dependence of the resonance parameters is characteristic of all giant resonances  and, as such, a study into their properties yields information about the non-equilibrium dynamics and the bulk properties of the nucleus~\cite{Drozdz}. The first indication of the existence of giant resonances was the observation of the dominant giant resonance structure in photo-absorption spectra now known as the IsoVector Giant Dipole Resonance (IVGDR)~\cite{Harakeh}.

Although photo-absorption has been used most extensively to study the IVGDR, direct nuclear reactions such as inelastic scattering are generally just as effective in the study of giant resonances provided that the appropriate kinematics are selected ~\cite{Spe81}. In recent years, methods for the extraction of electric-dipole strength distributions in nuclei via relativistic Coulomb excitation have been developed~\cite{vonNeumann-Cosel2019a}. These make use of proton inelastic scattering with energies of a few hundred MeV at scattering angles close to $\theta_{\mathrm{Lab}}=0^\circ$, which requires special experimental techniques~\cite{Neveling2011b,Tam09}. Under these kinematic conditions, the background from other nuclear processes has been found to be small in heavy nuclei and its contribution can easily be subtracted~\cite{Tamii2011,Pol2012,Krumbholz2015,Hashimoto2015,Don18,Bassauer2020,Bassauer2020a}.

An important property of any giant resonance is its width, since it provides valuable information on the excitation and decay of the giant resonance. It is often expressed by~\cite{Harakeh}
\begin{equation}
\Gamma = \Delta\Gamma + \Gamma^\uparrow + \Gamma^\downarrow 
\label{eq:width}
\end{equation}
with contributions from the following three processes: 
$(i)$ Landau damping ($\Delta\Gamma$), in which the initial dipole excitation fragments into 1p1h states,  $(ii)$ direct particle emission from the 1p1h excitations, which give rise to an escape width $\Gamma^\uparrow$, and 
$(iii)$ coupling to more complex 2p2h states through the residual two-body interaction~\cite{Drozdz} and finally to $n$p$n$h states. 
This coupling to more and more complex states results in a spreading width $\Gamma^\downarrow$ and ultimately terminates in the compound nucleus states. 
This particular coupling mechanism is referred to as the doorway states mechanism. 
Fine structure in the response of the nucleus may be induced by the chaos of nuclear states mentioned above~\cite{Lacroix1999}. 
Insight into the dominant damping mechanisms of nuclear giant resonances is provided in the properties of the fine structure~\cite{vonNeumann-Cosel2019b}, which, in the case of the IVGDR, is understood to be the result of characteristic energy scales or energies of the coupling steps.

The contributions in Eq.~(\ref{eq:width}) cannot be easily separated but coincidence decay experiments have shown \cite{Harakeh} that the spreading width increases with mass number and makes the largest contribution in heavy nuclei, while the escape width can be large (several MeV) in light nuclei but is of the order of a hundred keV only in heavy nuclei.
The distribution of IVGDR strength due to Landau damping in the spherical nuclei studied here is of the order of 1 MeV as exemplified in the theoretical results presented below.

An additional contribution to the width comes from resonance splitting owing to nuclear deformation. There exists a clear correlation between the IVGDR width and the nuclear deformation parameter in a certain mass region~\cite{Ish11}. 
For heavy quasi-spherical nuclei with $N = 82$, the IVGDR occurs in the form of a narrow, single peak. Slight deformations with increasing $N$ result merely in an increase in the IVGDR width, whereas strong deformations result in the splitting of the IVGDR into two distinct components corresponding to different $K$ quantum numbers~\cite{Berm75}.

\begin{figure}
	\includegraphics[width=\columnwidth]{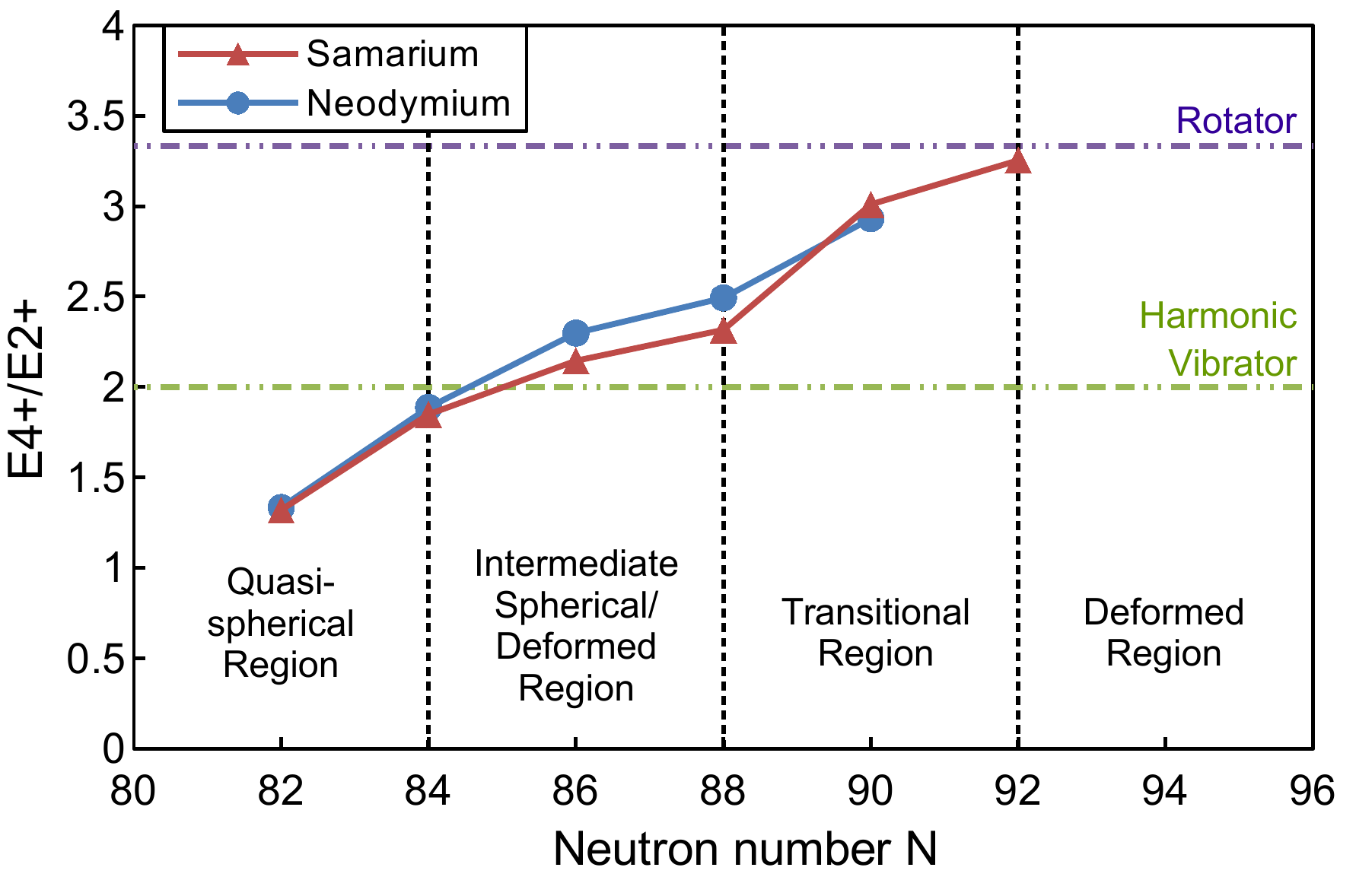}
	\caption{Variation of the ratio of the energies of the first $4^{+}$ and $2^{+}$ states, $E(4^+_1)/E(2^+_1)$,  with respect to the neutron number $N$ for the neodymium (blue) and samarium (red) isotope chains.}
	\label{fig:ratio}
\end{figure}

In the early 1950s, the discovery was made that some nuclei are deformed in their ground states. Research in this area has focused primarily on the rare-earth (for which $82 < N < 120$) and actinide regions since a large number of deformed nuclei are located near to or on the line of beta-stability in these regions. These regions are, as a result, easily accessible experimentally~\cite{Beng84}. The increased permanent deformation of the ground state with increasing neutron number is typified by the even-even neodymium and samarium isotope chains. 
This is shown in Fig.~\ref{fig:ratio} by the ratio $E(4^+_1)/E(2^+_1)$, which is a useful method for determining the extent of nuclear deformation in a nucleus~\cite{Casten}. Figure~\ref{fig:ratio} separates both the neodymium and samarium isotope chains into four regions, namely, the quasi-spherical region, the intermediate spherical/deformed region, the transitional region and the deformed region. The transitional region is particularly interesting since this region is comprised of nuclei that alter the properties of nuclear surfaces dramatically. In the range where the neutron number varies from 88 to 92, there is essentially a phase transition from a spherical vibrator to an axial rotor.
The IVGDR in the even-even Nd and Sm isotope chains has been studied using photo-absorption experiments at Saclay~\cite{Carlos1971,Carlos1974} and Coulomb excitation at the iThemba Laboratory for Accelerator Based Sciences (iThemba LABS)~\cite{Don18}. It is important to note, however, that the energy resolution of the Nd and Sm photo-absorption data is very poor; Refs.~\cite{Carlos1971,Carlos1974} report that the width of the quasi-monochromatic $\gamma$-ray beam obtained from the annihilation in flight of monochromatic positrons was approximately 300 keV. The Coulomb-excitation study of the Nd and Sm isotope chains at iThemba LABS was a two-part study. The first part, published in Ref.~\cite{Don18}, investigated the shape evolution of the IVGDR from spherical to deformed nuclei in the rare-earth region as well as the reproducibility of the photo-absorption results of Refs.~\cite{Carlos1971,Carlos1974}. The non-trivial disagreements observed between the two sets of data with respect to the distribution of the IVGDR strength are discussed in detail Ref.~\cite{Don18}. The second - current - part makes use of the same data as in Ref.~\cite{Don18} but focuses on investigating the fine structure of the IVGDR observed using the high energy-resolution capabilities at iThemba LABS. The present paper provides additional experimental details not included in Ref.~\cite{Don18} as well as the results of the fine-structure investigation.  

\section{\label{sec:Exp}Experimental Details}

The experiments were performed using a dispersion-matched 200 MeV proton beam produced by the Separated Sector Cyclotron (SSC) at iThemba LABS. Protons were inelastically scattered off self-supporting $^{142,144,146,148,150}$Nd and $^{152}$Sm targets with areal densities ranging from 1.8 mg/cm$^{2}$ to 2.6 mg/cm$^{2}$. All of the targets were isotopically enriched to values greater than 96\% except for the $^{148}$Nd target, which was enriched to 90\%. The calibration targets used were $^{24}$Mg and $^{26}$Mg. The reaction products were momentum-analysed by the K600 magnetic spectrometer in $0^\circ$ mode \cite{Neveling2011b} with the acceptance defined by a circular collimator with an opening angle of $\theta_{\mathrm{Lab}}={\pm}1.91^{\circ}$. Under these kinematic conditions, the dominant reaction mechanism is relativistic Coulomb excitation. Two Multi-Wire Drift Chambers (MWDCs) followed by two rectangular plastic scintillators comprised the focal-plane detection system, which made it possible to do particle tracking in the focal plane in order to determine the horizontal and vertical focal-plane coordinates ($x_{\mathrm{fp}},y_{\mathrm{fp}}$) as well as the focal-plane angle ($\theta_{\mathrm{fp}}$).

Faint-beam and dispersion-matching techniques were implemented in order to exploit the high energy-resolution capabilities of the K600 magnetic spectrometer. As a result, energy resolutions $\Delta E = 42 - 50$ keV Full Width at Half Maximum (FWHM) were achieved. Further details regarding the experimental setup and data extraction (beyond the descriptions provided in Section~\ref{sec:Extract} below) can be found in Refs.~\cite{Neveling2011b,Don16}.  

\section{\label{sec:Extract}Data Extraction and Optimisation}

\subsection{\label{sec:PID}Particle Identification}

Particle identification was based on the combination of information on the Time-Of-Flight (TOF) selection and the energy loss of the particles in the scintillation detectors. 

\subsubsection{Time-Of-Flight Selection}

The TOF is determined from a fast coincidence between the radiofrequency signal of the cyclotron and the signals from the plastic scintillators. The magnetic rigidity $R$ of the spectrometer determined by the radius of curvature $r$ and the magnetic field $B$
\begin{equation}
    R=rB=\frac{p}{q}
\end{equation}
selects particles with the same $p/q$ ratio, where $p$ denotes the momentum and $q$ the charge. The combination of the $q$-dependent energy loss $\Delta E$ on the TOF thus allows distinction of different particle types.  

\begin{figure}
	\includegraphics[width=\columnwidth]{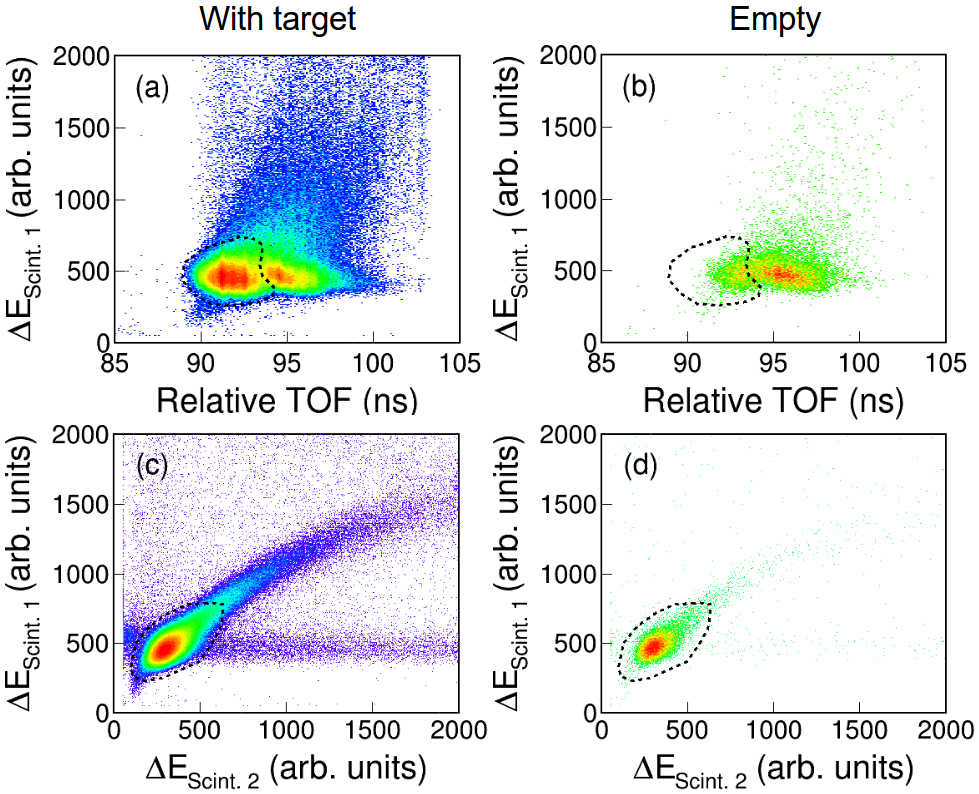}
	\caption{Particle and background identification for the $^{144}$Nd($p,p^{\prime}$) reaction with $E_{\mathrm{p}}=200$ MeV and $\theta_{\mathrm{Lab}}=0^{\circ}{\pm}1.91^{\circ}$. (a) Two-dimensional spectrum of the pulse height in the first scintillator detector versus the relative TOF.  The software gate indicated with dashed contour lines was used to select the events of interest. 
	(b) Same as (a) but for an empty target. 
	The background component caused by beam halo is identified. 
	(c) Two-dimensional correlation of the energy losses through the first and second scintillator. 
	(d) Same as (c) but for an empty target.
	The software gates used to select the events of interest are indicated by dashed contour lines.}
	\label{fig:PID}
\end{figure}

In the case of $0^\circ$ proton inelastic-scattering experiments, the combined information is particularly useful to distinguish between protons scattered from the target and beam-halo events. To determine the TOF and energy-loss characteristics of the beam-halo events only, an empty-target measurement can be done. Figures \ref{fig:PID} (a) and (b) display the two-dimensional spectra of the pulse height in the first scintillator detector versus the TOF values for a $^{144}$Nd-target measurement and an empty-target measurement, respectively. As can be seen from Fig.~\ref{fig:PID}, the majority of target-related events can be distinguished from the beam-halo events, which can be removed using a software gate on the target-related events as indicated by the dashed line.  

\subsubsection{$\Delta E$ Correlation Technique}

The type of particle and its associated kinetic energy will determine its loss of energy in the scintillation detectors. For this experiment, which suffers from a significant background of low-energy protons caused by small-angle elastic scattering off the target foil followed by re-scattering off any exposed part inside the spectrometer, a $\Delta E_{\mathrm{Scint. 1}}$ versus $\Delta E_{\mathrm{Scint. 2}}$ spectrum is valuable to further isolate the protons of interest. 
The dependence of pulse height on the position of the detected particle along the length of the scintillation detector was removed by taking the geometrical average of the signals from the two photomultiplier tubes that are mounted on opposite ends. Figures \ref{fig:PID} (c) and (d) show the $\Delta E_{\mathrm{Scint. 1}}$ versus $\Delta E_{\mathrm{Scint. 2}}$ spectra for a $^{144}$Nd-target measurement and an empty-target measurement, respectively. The region containing the high-energy protons of interest is bordered with a dashed line in Fig.~\ref{fig:PID} (c). This dashed region is reproduced in Fig.~\ref{fig:PID} (d) to illustrate the high-energy nature of the beam-halo events, which can mostly be removed by the TOF selection. However, there are still instrumental-background events due to beam halo as well as the previously mentioned target-related background in our identified region of interest. Because of this, in addition to the particle-identification techniques applied, there is a need for further background subtraction to remove these background events.   

\subsection{\label{sec:BackSub}Background Subtraction}

The instrumental background can be readily characterised if the spectrometer is operated in vertical focus mode, i.e., the inelastically scattered protons from the target are focused around the vertical focal-plane position $y_{\mathrm{fp}}=0$, while the instrumental background is evenly distributed in the vertical direction~\cite{Neveling2011b,Jingo2018}. In this particular experiment, however, centering the spectrum around $y_{\mathrm{fp}}=0$ resulted in an undesirable amount of beam-halo events. The spectrum was, therefore, shifted slightly by changing the vertical beam position on the target resulting in an unequal distribution of background events above and below the area of interest. The method for background subtraction described in Refs.~\cite{Neveling2011b,Jingo2018} relies on the $x_{\mathrm{fp}}$ versus $y_{\mathrm{fp}}$ spectrum having two equal background regions. Consequently, the usual method had to be adjusted slightly in this case.

\begin{figure*}
\includegraphics[width=0.55\textwidth]{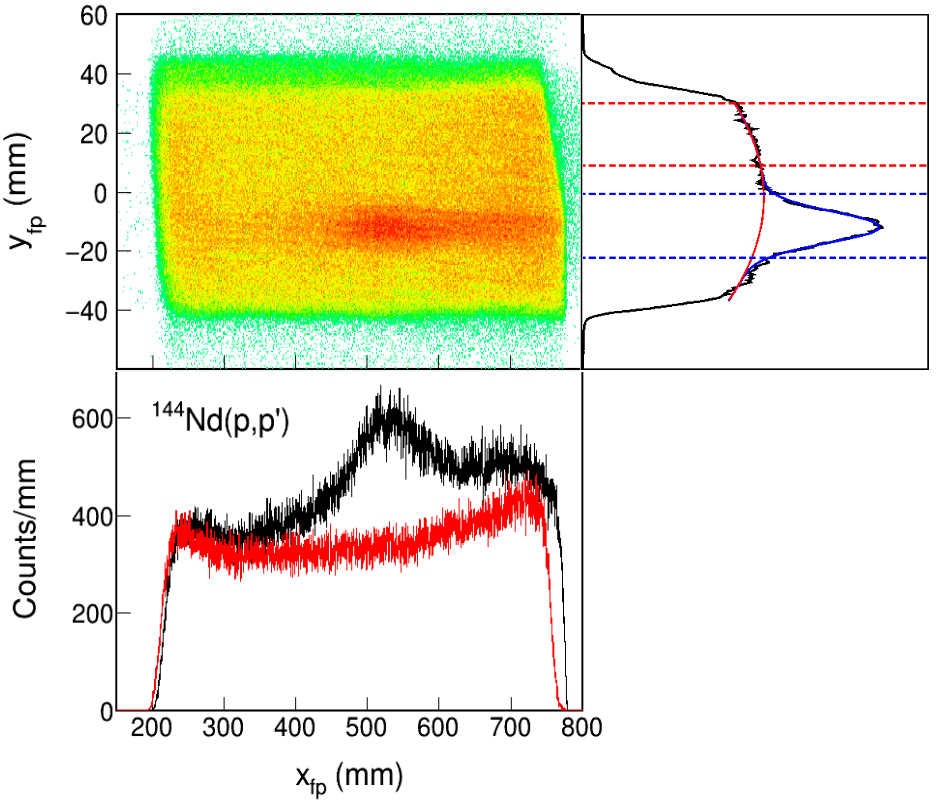}
\caption{\label{fig:BackGSub}
Top left: Two-dimensional scatterplot of the horizontal and vertical focal-plane coordinates ($x_{\mathrm{fp}}$,$y_{\mathrm{fp}}$) for the $^{144}$Nd(p,p$^{\prime}$) reaction at $E_{\mathrm{p}}=200$ MeV. 
Top right: Projection of the $y_{\mathrm{fp}}$ events showing a total fit (in blue), where the events of interest sit on top of the background fitted with a quadratic function (in red). The area between the dashed red lines corresponds to the region used to approximate the background under the central region of interest indicated with dashed blue lines. 
Bottom left: Focal-plane position spectrum showing the components described in the background subtraction procedure, i.e., the raw data (in black) and the background component corresponding to the area between the dashed red lines identified in the top right panel (in red).}
\end{figure*}

Figure \ref{fig:BackGSub} displays a two-dimensional spectrum of the horizontal ($x_{\mathrm{fp}}$) and vertical ($y_{\mathrm{fp}}$) focal-plane coordinates for the $^{144}$Nd(p,p$^{\prime}$) reaction with $E_{\mathrm{p}}=200$ MeV in the top left panel as well as the $y_{\mathrm{fp}}$ projection in the top right panel. The total area of the projection is comprised of the events of interest and the background events, with the assumption that the profile of each remains uniform throughout the spectrum. These components were fitted with a Gaussian peak and a quadratic background (in red). The total fit, which is the addition of these components, is shown in blue. Fitting the spectrum in this way allowed for the total number of background events directly underneath the area of interest to be determined. An area from the background section, which contained the same number of background events, was then identified and subtracted. This background component is indicated by the dashed red lines and the red component in the top-right and bottom-left panels of Fig.~\ref{fig:BackGSub}, respectively. This ensured that the correct amount of background was subtracted without inducing structure in the region of interest. 

\subsection{\label{sec:Lineshape}Lineshape Correction}

The lineshape-correction procedure was done using the $^{24}$Mg and $^{26}$Mg calibration targets, which show sharp peaks in the region of interest. During the experiment, a kinematic correction procedure \cite{Fujita2002,Don16} was performed as the first step towards obtaining good energy resolution. In order to achieve this, the magnetic spectrometer field was adjusted such that the nuclear states appeared approximately upright in the two-dimensional $\theta_{\mathrm{fp}}$ versus $x_{\mathrm{fp}}$ spectrum, thus ensuring that particles from the target emerging at the same excitation energy of the residual nucleus but at different $\theta_{\mathrm{fp}}$ converge at the same $x_{\mathrm{fp}}$. To do this, changes were made to the K-coil and H-coil of the K600 spectrometer \cite{Neveling2011b} for first-order focusing and corrections of second-order aberrations, respectively. 

\begin{figure}[!ht]
	\includegraphics[width=\columnwidth]{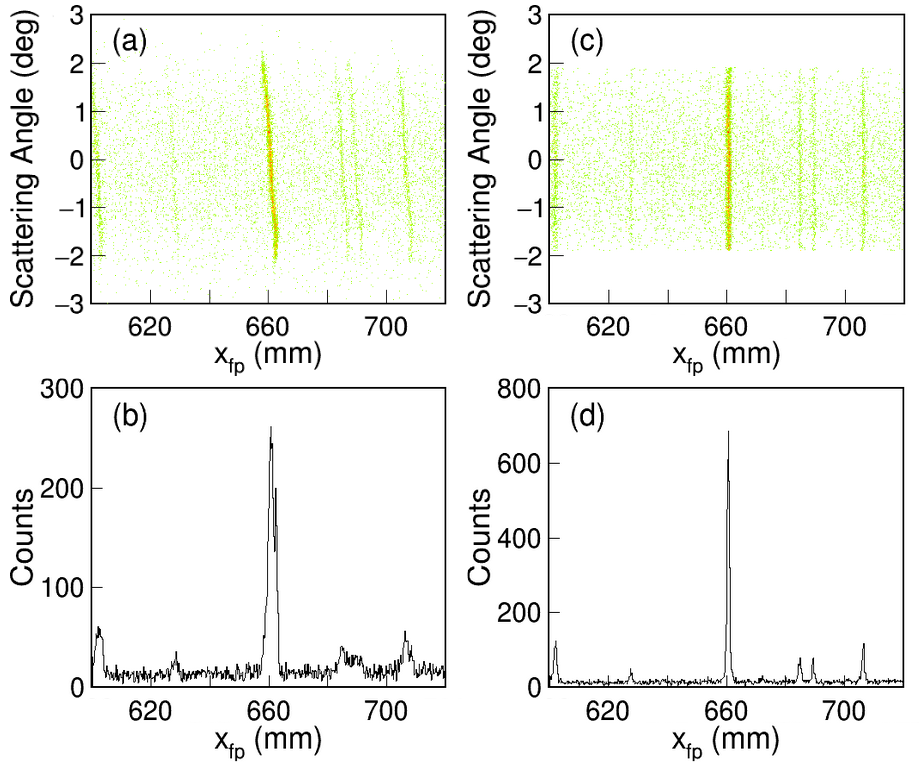}
	\caption{An illustration of the effect of the lineshape correction. (a) Two-dimensional scatterplot of $\theta_{\mathrm{scat}}$ versus $x_{\mathrm{fp}}$ in the vicinity of the prominent 10.711 MeV peak in $^{24}$Mg before lineshape correction. (b) Corresponding focal-plane position spectrum. The plots (c) and (d) are the same as (a) and (b), respectively, but after the lineshape correction.}
	\label{fig:LCEffect}
\end{figure}

Although the use of the K- and H-coils during the experiment provides a reasonable starting point, additional offline adjustments to ensure upright and well-resolved states are required to obtain the best-possible energy resolution of the spectra. Figure \ref{fig:LCEffect}(a) shows the scatterplot of $\theta_{\mathrm{scat}}$ versus $x_{\mathrm{fp}}$ for the $^{24}$Mg target before the lineshape-correction procedure, where $\theta_{\mathrm{scat}}$ is the horizontal component of the scattering angle of the reaction reconstructed using $\theta_{\mathrm{fp}}$ information. As can be seen, a dependence of the focal-plane position on the scattering angle is still visible, which leads to broadening in the focal-plane position spectrum as seen in Fig. \ref{fig:LCEffect}(b). 

To correct for the remaining angular dependence of the focal-plane position, a function dependent on the focal-plane coordinates was subtracted from the original $x_{\mathrm{fp}}$ values as follows: 
\begin{eqnarray}
    x_\mathrm{corr} =  x_\mathrm{fp} && - \Big[\sum_{n=1}^{5} a_n (\theta_{\mathrm{scat}})^n \nonumber  + \sum_{n=1}^2 b_n(y_{\mathrm{fp}} + y_{\mathrm{offset}})^n \nonumber \\
    && +  c(x_{\mathrm{fp}}-x_0){{\theta_{\mathrm{scat}}}^2}\Big].
\end{eqnarray}
The dependence on $y_\mathrm{fp}$ was included to compensate for the unequal distribution of background events below and above the area of interest described in Sec.~\ref{sec:BackSub}. 
By including this term in the correction (and thus centering the events around zero), any dependence of $x_{\mathrm{fp}}$ on $y_\mathrm{fp}$ does not change the $x_{\mathrm{fp}}$ position but simply rotates it around the vertical centre of the lineshape. This correction was then applied to the Nd and Sm data.

\subsection{\label{sec:DDCS}Double-Differential Cross Sections}

\begin{figure}
	\includegraphics[width=0.8\linewidth]{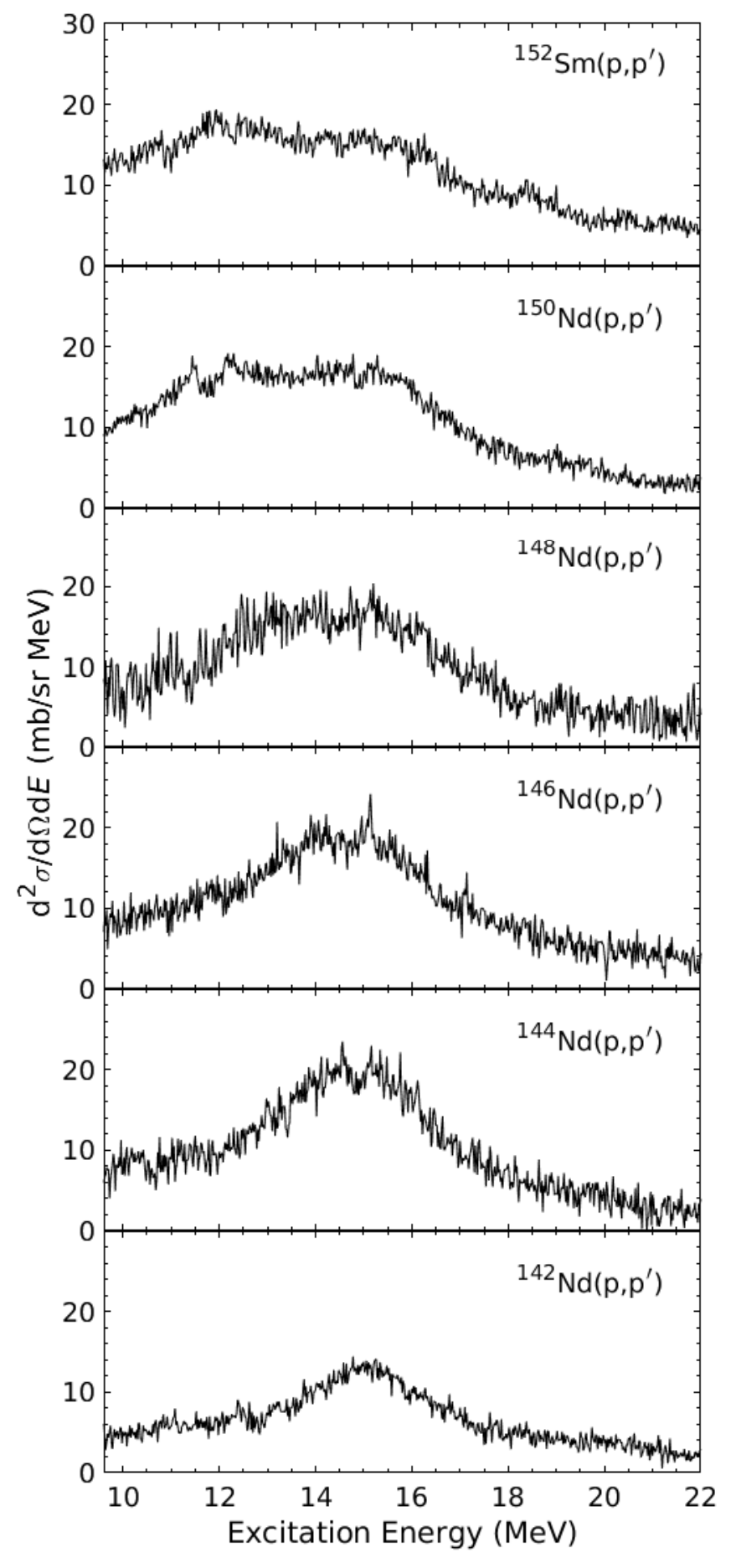}
	\caption{Experimental double-differential cross sections for the $^{142, 144, 146, 148, 150}$Nd(p,p$^{\prime}$) and $^{152}$Sm(p,p$^{\prime}$) reactions at $E_{\mathrm{p}} = 200$ MeV and $\theta_{\mathrm{Lab}}=0^{\circ} \pm 1.91^{\circ}$.}
	\label{fig:DDCS}
\end{figure}

The double-differential cross sections (with a systematic uncertainty of $\pm 7$\%) obtained following the above-mentioned procedures and those in Ref.~\cite{Don16} are displayed in Fig.~\ref{fig:DDCS} binned to 20 keV. The experimental energy resolutions (FWHM) achieved were 50 keV, 42 keV, 46 keV, 43 keV, 45 keV and 42 keV for $^{142, 144, 146, 148, 150}$Nd and $^{152}$Sm, respectively. The broad structure visible for all isotopes between approximately $E_{\mathrm{x}}=12$ MeV and $E_{\mathrm{x}}=18$ MeV corresponds to the excitation of the IVGDR. Statistical errors in this region are of the order of 2-4\%. Pronounced fine structure is visible over the excitation-energy region of the IVGDR for all isotopes, even in the most deformed nuclei.  

\section{\label{sec:Conversion}Conversion from ($\mathrm{p,p}^{\prime}$) to Equivalent Photo-Absorption Cross Sections}

Obtaining equivalent photo-absorption cross sections that are comparable to the $E1$ response spectra provides a better representation of the IVGDR in terms of position and width than that provided by the spectra shown in Fig.~\ref{fig:DDCS}. This is because the Coulomb-excitation probability has a strong energy dependence. The fine-structure analysis of the IVGDR will thus be performed on the converted spectra. 

\begin{figure}
	\includegraphics[width=0.74\linewidth]{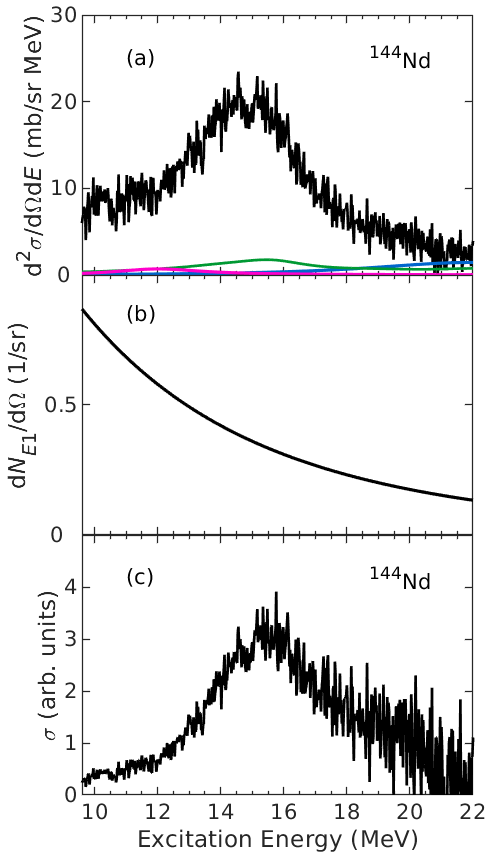}
	\caption{Conversion process from $^{144}$Nd(p,p$^{\prime}$) to photo-absorption cross sections: (a) Double-differential (p,p$^{\prime}$) cross section and background components. 
	The green and pink lines describe the contribution from the ISGMR and ISGQR, respectively. 
	The blue line describes the phenomenological component explained in the text. 
	(b) Virtual-photon spectrum. 
	(c) Equivalent photo-absorption spectrum resulting from Eq.~\ref{eqn:virtphoton}.}
	\label{fig:ConvOverview}
\end{figure}

The conversion of the measured (p,p$^{\prime}$) spectra to equivalent photo-absorption cross sections is a three-stage process: subtraction of the nuclear background, calculation of the virtual-photon spectrum and implementation of the equivalent virtual-photon method. 
By way of example, Fig.~\ref{fig:ConvOverview} provides an overview of the conversion process for $^{144}$Nd, which is outlined below.

\subsection{Subtraction of the Nuclear Background}

Although proton inelastic scattering at incident energies of several hundred MeV at very-forward scattering angles predominantly excites the IVGDR, other resonances of different multipolarities also contribute to the spectra. In similar measurements with heavy nuclei, these contributions were found to be small~\cite{Tamii2011, Pol2012, Krumbholz2015, Hashimoto2015} but, in order to isolate the IVGDR strength reliably, they must be taken into consideration and subtracted before a conversion to equivalent photo-absorption cross sections can be performed.

\begin{figure}
	\includegraphics[width=0.88\linewidth]{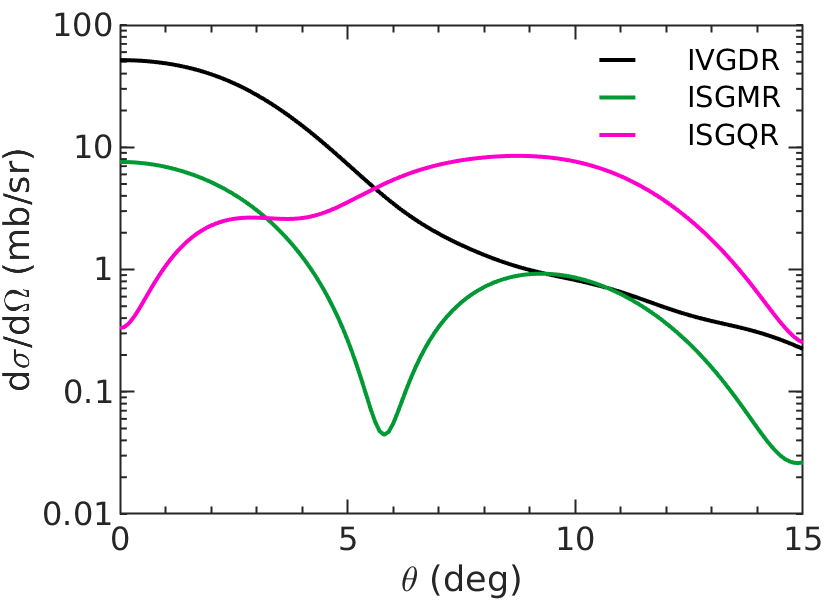}
	\caption{Example of the DWBA07 calculations of the IVGDR (black), ISGMR (green) and ISGQR (pink) differential cross sections in (p,p$^{\prime}$) scattering at $E_{\mathrm{p}} = 200$ MeV off $^{142}$Nd.}
	\label{fig:DWBA}
\end{figure}

The most important among these contributions are the IsoScalar Giant Quadrupole Resonance (ISGQR), the IsoScalar Giant Monopole Resonance (ISGMR), and a phenomenological background displayed in Fig.~\ref{fig:ConvOverview}(a) by the pink, green and blue lines, respectively. The way in which the ISGQR and ISGMR contributions to the spectrum in Fig.~\ref{fig:ConvOverview}(a) were estimated is described in a previous paper \cite{Don18} but will be summarised here for completeness. Distorted wave Born approximation calculations were performed with the code DWBA07 \cite{Raynal07} using Quasi-particle Phonon Model (QPM) transition amplitudes and the Love-Franey effective interaction \cite{Love81, Franey85} as input (analogous to Ref.~\cite{Pol2012}) in order to determine the computed angular distributions of the ISGQR and ISGMR cross sections. 
A representative example of the DWBA07 calculations for the $^{142}$Nd isotope is shown in Fig.~\ref{fig:DWBA}. 
Since the ISGMR and ISGQR are collective excitations exhausting a large fraction of the respective sum rules, the angular distributions of the other studied nuclei are the same except for a small correction due to the target mass dependence of the recoil term.

Taking the experimental angular acceptance into account, these calculations provide a relationship between the theoretical cross sections and transition strengths under the assumption of a dominant one-step reaction mechanism, which is well fulfilled for $E_{\mathrm{p}} = 200$ MeV. 
The ISGQR and ISGMR strength distributions were then converted to (p,p$^{\prime}$) cross sections using this proportionality. The positions and widths of the ISGQR component were taken from a recent study of the ISGQR across the even-even Nd isotope chain \cite{Kureba2014, Kureba2018} with methods analogous to Refs.~\cite{Shevchenko2004, Shev2009, Usman2011191}. For the ISGMR contribution, the isoscalar giant-resonance strength distributions for the Sm isotope chain reported by Itoh et al. \cite{Itoh2003} could be directly applied in the case of $^{152}$Sm. The results from Ref.~\cite{Itoh2003} were also applied to the corresponding Nd isotones, which have very similar deformation parameters, incorporating a correction for the global mass dependence of the ISGMR \cite{Harakeh}. The $B(E1)$ transition strengths (and, by implication, the photo-absorption cross sections) cannot, however, be extracted using this approach since the Coulomb-nuclear interference term breaks the proportionality \cite{Don18}.

The phenomenological background incorporates all unknown multipolarity contributions as well as quasi-free scattering and describes the behaviour of the cross section at higher excitation energies where the Coulomb excitation contribution is negligible. It was approximated by finding the maximum of the cross section between 20 MeV and 23 MeV and using a width that best described that region of the spectrum. In a study of $^{208}$Pb \cite{Pol14}, where an experimental extraction of the angular distribution of the background was possible, a similar description for the shape of this component was found.

\subsection{Virtual-Photon Production Function}

The equivalent virtual-photon method describes the excitation of a target nucleus as the absorption of equivalent photons whose spectrum is determined by the Fourier transform of the projectile's time-dependent electromagnetic field~\cite{Bert88}. More in-depth descriptions of the equivalent virtual-photon method can be found in Refs.~\cite{Bert88,Bertulani2009}. In order to make use of this method, the virtual $E1$ photon spectrum was calculated for each isotope using the eikonal approximation \cite{Bert93} and averaged over the angular acceptance of the detector. The calculated virtual $E1$ photon spectrum for $^{144}$Nd is shown in Fig.~\ref{fig:ConvOverview}(b) as an example.

The impact of a deformed projectile or target nucleus on the virtual-photon spectrum has been investigated in Ref.~\cite{Bertulani1993}.
A calculation along these lines for the most deformed case $^{150}$Nd finds a very small increase ($< 2$\% for the $K = 0$ and $< 3$\% for the $K = 1$ component) for the present reaction. 
Thus, deformation effects on the virtual-photon spectrum are neglected for the conversion to photo-absorption cross sections discussed in the next section.

\subsection{Application of the Equivalent Virtual-Photon Method}

\begin{figure}
	\includegraphics[width=0.74\linewidth]{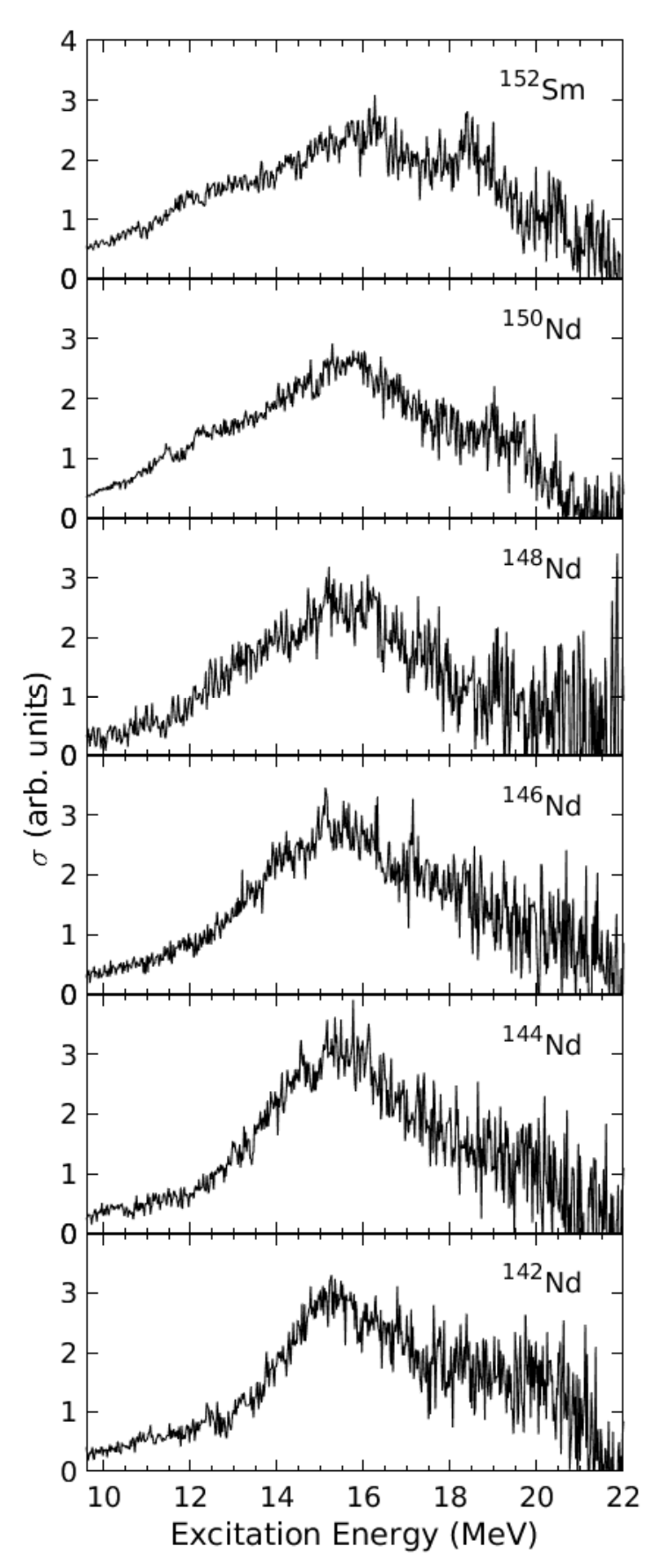}
	\caption{Equivalent photo-absorption cross sections obtained using (p,p$^{\prime}$) scattering at $E_{\mathrm{p}} = 200$ MeV off Nd and Sm isotopes.}
	\label{fig:AbsCs}
\end{figure}

As a final step, the equivalent photo-absorption cross section for each isotope was obtained (see Fig.~\ref{fig:ConvOverview}(c)) using the following equation
\begin{equation}
\label{eqn:virtphoton}
    \frac{d^2\sigma}{d\Omega dE_{\gamma}} = \frac{1}{E_{\gamma}} \frac{dN_{E1}}{d\Omega} \sigma_{\gamma}^{\pi\lambda}(E_\gamma) \;\, ,
\end{equation}
where Coulomb-nuclear interference is assumed to be negligible as a result of the kinematics of the experiment~\cite{vonNeumann-Cosel2019a}. Figure \ref{fig:AbsCs} shows the resulting equivalent photo-absorption cross sections for all isotopes binned to 20 keV. The present setup at $\theta_{\mathrm{Lab}}=0^{\circ}$ at iThemba LABS does not allow for the accurate determination of the vertical component of the scattering angle, which limits the angular resolution of the measurement \cite{Neveling2011}. As stated in Ref.~\cite{Don18}, we, therefore, refrain from extracting absolute photo-absorption cross sections. The excitation-energy dependence of the conversion is, however, unaffected. 

\section{\label{sec:FineStructure}Fine-Structure Investigation}

\subsection{\label{sec:WaveletTech}Wavelet-Analysis Technique}

Wavelets are functions that satisfy a set of predetermined mathematical requirements and are useful in the representation of data or other functions. This concept is not unique to wavelet analysis since the same idea is used in Fourier analysis, where functions are represented by a superposition of sine and cosine functions. Wavelet algorithms differ in that they are capable of processing data at different scales (or resolutions). Thus, they serve as a powerful tool to analyse the role that these scales play in the interpretation of the data~\cite{Gra95}. Observing the data through a large window would mean that broader features of that data are extracted. Similarly, observing through a small window would mean that the finer features would be seen. Using an analogy, the wavelet-analysis technique, therefore, allows for the forest as well as its individual trees to be seen~\cite{Gra95}.

In the case of Fourier analysis, the sine and cosine functions used to represent the data are non-local and extend to infinity. The disadvantage is that any sharp discontinuities in the data are approximated badly. The functions that form the basis of wavelet analysis, however, are approximating functions contained in finite domains, which makes them ideal for the analysis of data with sharp discontinuities. Wavelet analysis is, therefore, well suited to the fine-structure analysis of giant resonances in nuclei \cite{vonNeumann-Cosel2019b}.  

The wavelet-analysis procedure involves the selection of a wavelet prototype, which is referred to as a {\textit{mother wavelet}}. The time-based analysis is conducted with a contracted, high-frequency version of this mother wavelet. In contrast, the frequency analysis is performed using a dilated, low-frequency version. The selection of a particular wavelet as the mother wavelet depends on its adherence to a set of predefined mathematical criteria outlined in the formalism below. 

\subsubsection{Wavelet-Analysis Formalism}

A real or complex function, $\Psi(x)$, may only be used as a mother wavelet if
\begin{equation}
{\int^{\infty}_{-\infty}} {\Psi}(x) dx = 0 
\end{equation}
and
\begin{equation}
K_{\Psi} = {\int^{\infty}_{-\infty}} \vert {\Psi}^{2}(x) \vert dx < \infty \, ,
\end{equation}
where $K_{\Psi}$ is the wavelet norm or rather, the admissibility constant. 
This varies depending on the wavelet under consideration. 

The first condition imposes the requirement of an oscillating function for which the mean value should be zero. The second condition restricts the function to a finite domain. Since wavelets are localised both in time and in frequency, they are unaffected by the properties of the data far away from the region of interest and as such, are ideal for the description of the local behaviour. 

A full discussion of the most frequently used functions for wavelet analysis can be found in Ref.~\cite{Shevchenko2008}. For the analysis of the fine structure of nuclear giant resonances, the Morlet wavelet is most suitable. This is because it contains a Gaussian envelope on top of a periodic structure, and the detector response is well approximated by a Gaussian lineshape. In the present fine-structure analysis, the complex Morlet wavelet (see Fig. 1 in Ref.~\cite{Coo08}) was used and is given by
\begin{equation}
\label{eq:complexMor}
\Psi(x) = \frac{1}{\sqrt{{\pi}f_{\mathrm{b}}}} \exp(2{\pi}i{f_{\mathrm{c}}}x) \exp \left( - \frac{x^2}{f_{\mathrm{b}}} \right) \, ,
\end{equation}
using $f_{\mathrm{b}}=2$, which controls the wavelet bandwidth and $f_{\mathrm{c}}=1$, which is the centre frequency of the wavelet~\cite{Teolis98}.

A wavelet transform results in a set of coefficients that are representative of the data, which rely on the two parameters of Eq.~(\ref{eq:complexMor}). There are two classes of wavelet transforms available: the Continuous Wavelet Transform (CWT) and the Discrete Wavelet Transform (DWT). For the purposes of the present analysis, only the CWT will be detailed. The DWT as well as a comparison between the two transforms can be found in Ref.~\cite{Shevchenko2008}.   

\subsubsection{Continuous Wavelet Transform}
\label{subsec:cwt}

The coefficients of the wavelet transform for a spectrum, $\sigma(E)$, following its convolution with the (generally complex-conjugated) wavelet function are given by:
\begin{equation}
C({\delta}E,E_\mathrm{x}) = {\frac{1}{\sqrt{{\delta}E}}}{\int}{\sigma(E)}{\Psi^{*}} \left( \frac{E_\mathrm{x}-E}{{\delta}E} \right) dE \;\, ,
\label{eq:cwt}
\end{equation} 
where ${\delta}E$ is the bin size and is responsible for the scaling of the function. The parameter $E_\mathrm{x}$ shifts the position of the wavelet in excitation energy and thus provides access to the scale-localisation information. In the CWT, the scale and location parameters ${\delta}E$ and $E_\mathrm{x}$ are varied continuously. When the form of a scaled and shifted wavelet, $\Psi(x)$, is similar to the original spectrum, $\sigma(E)$, the values of $C({\delta}E,E_\mathrm{x})$ will be large. Similarly, if the form differs greatly, the coefficients obtained will be small. 

\begin{figure}
	\includegraphics[width=\columnwidth]{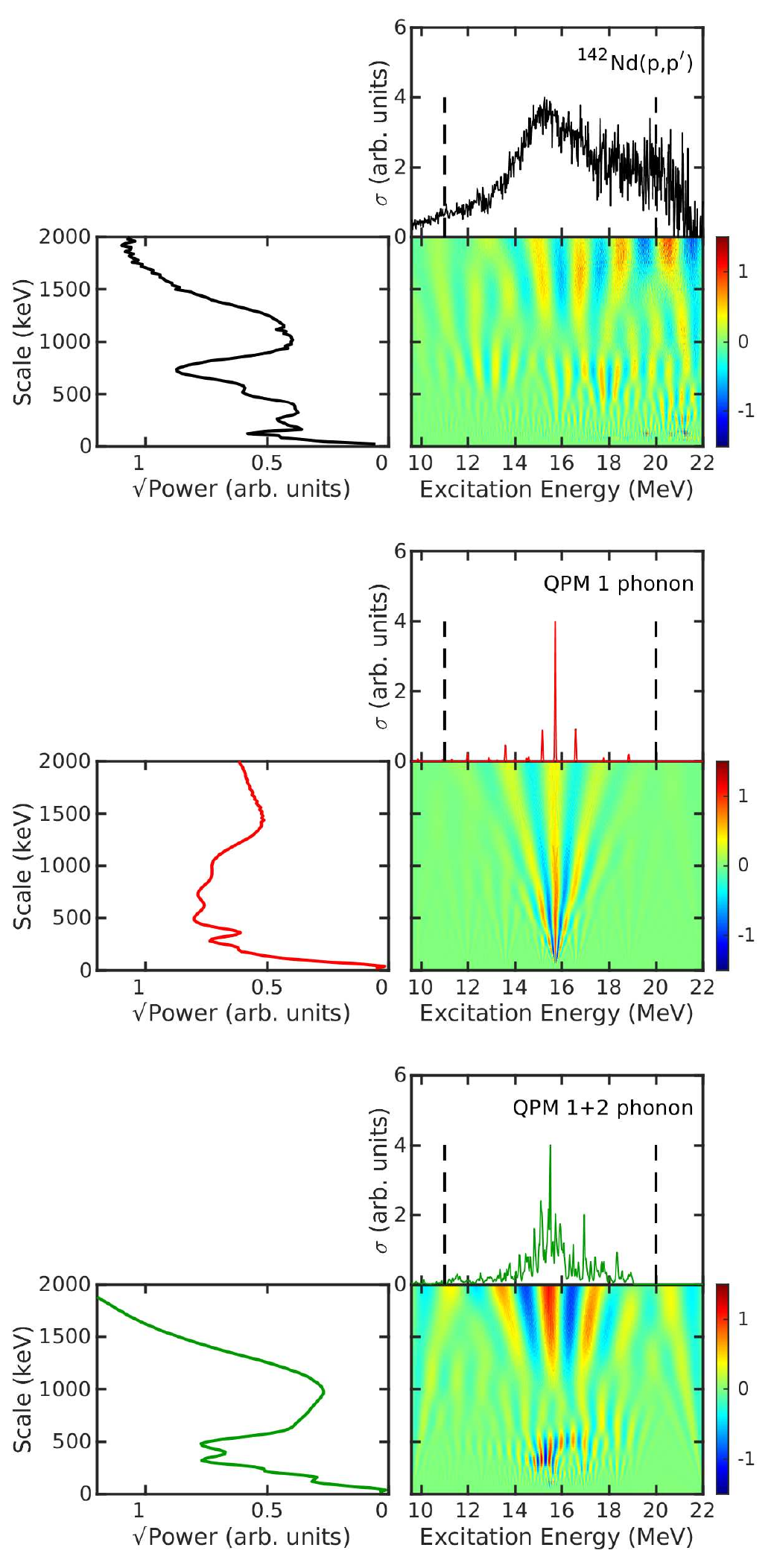}
	\caption{Top set (right column): CWT analysis of the equivalent photo-absorption spectrum for $^{142}$Nd measured at $E_\mathrm{p}=200$ MeV and $\theta_{\mathrm{Lab}}=0^{\circ} \pm 1.91^{\circ}$. 
	Top set (lower right): Density plot of the real part of the CWT coefficients of the data. 
	This CWT plot displays a wavelet scale that is equivalent to the Fourier scale. 
	Top set (left column): The corresponding power spectrum for the excitation-energy region indicated by the vertical dashed lines (11 MeV $\leq \mathrm{E_x} \leq$ 20 MeV). 
	Middle and bottom set: Same as the top set but for the QPM 1 phonon and QPM 1+2 phonon calculations, respectively.}
	\label{fig:142Nd-WA-Full}
\end{figure}

The application of Eq.~(\ref{eq:cwt}) is illustrated in Fig.~\ref{fig:142Nd-WA-Full} for the experimental photo-absorption spectrum of the IVGDR in $^{142}$Nd (top) and two examples of the corresponding model calculations (middle and bottom). The two-dimensional distributions below the spectra display the wavelet coefficient as a function of the parameters $E_{\rm x}$ and $\delta E$.
One observes large amplitudes localised at certain ranges of energy and scale values. The sign change from positive (red) to negative (blue) amplitudes is due to the oscillatory structure of the mother wavelet (see Eq.~(\ref{eq:complexMor})).

As a natural consequence of wavelet analysis, wavelet energy scales can be extracted from the wavelet coefficient plot as peaks in the corresponding power spectrum obtained by squaring the complex CWT coefficients and summing these as follows 
\begin{equation}
P({\delta}E) = {\frac{1}{N}}\sum_i \vert C_i(\delta E) C_i^*(\delta E) \vert   \;\, ,
\label{eq:power}
\end{equation} 
where $P({\delta}E)$ is the power as a function of scale ${\delta}E$ summed at each scale value over the index $i = N$ with $N$ the number of energy bins on the excitation-energy axis.

We note that the choice of the complex-Morlet mother wavelet yields the equivalent Fourier scale, while other mother wavelets require a scaling factor different from unity between their own intrinsic scale and the equivalent Fourier scale. A particular value of scale in a CWT plot corresponds to the excitation-energy difference between consecutive minima (or maxima) in the coefficient plot (referred to as a ``length-like" scale).

By way of example, the two-dimensional coefficient plot shown at the bottom of Fig.~\ref{fig:142Nd-WA-Full} for the $^{142}$Nd QPM 1+2 phonon calculation displays a repetitive oscillatory structure from negative coefficients (blue) to positive coefficients (red) at a wavelet energy scale of 2000 keV with a distinct peak in the coefficient plot at $E_{\rm x}$ = 15.5 MeV corresponding to the main $E{\rm 1}$ strength in the panel above it. Half of the 2000 keV wavelet energy scale is the width 1000 keV of this peak (FWHM) and is referred to as a ``width-like" scale. 

Peaks and points of inflection in the power spectrum (here and throughout, shown as the square root of power for better clarity) indicate the presence of particular scales. 
The presence of the same scale in both the experimental and a theoretical model prediction is taken to be that both scales agree within error bars of one standard deviation. 
Assuming a normal distribution, the standard deviation of the peak values is obtained by dividing the FWHM by a factor of 2.355 in the usual way. 
Descriptions of the applications of the CWT to other high energy-resolution nuclear giant-resonance spectra can be found in Refs.~\cite{Shevchenko2008,Shevchenko2004,Shev2009,Pol14,Usman2011191,Petermann2010,Usman2016,Fearick2018,Kureba2018,Jingo2018,vonNeumann-Cosel2019b,Kalmykov2006} and applications to $\alpha$ clustering in Refs.~\cite{Wakasa2007,Bailey2019}. We note that in some of the previous applications, the scale values quoted correspond to width-like scales. 

\subsection{\label{sec:Modelcomparison}Model Predictions}
\label{subsec:models}

\subsubsection{Skyrme Separable Random Phase Approximation}

Calculations for the IVGDR are performed within the Skyrme Separable Random Phase Approximation (SSRPA) approach \cite{Nes06}. The method is fully self-consistent since both the mean field and residual interaction are derived from the same initial Skyrme functional. The residual interaction includes all the functional contributions as well as the Coulomb direct and exchange terms. The self-consistent factorisation of the residual interaction significantly reduces the computational effort while maintaining high accuracy of the calculations \cite{Nes06, Nes08, SSRPA08}. We use the Skyrme parameterisation SLy6 \cite{Chabanat98}, which provides a good description of the IVGDR in medium- and heavy-mass deformed nuclei \cite{SSRPA08}. The code used exploits the 2D grid in cylindrical coordinates. Pairing with volume delta forces is treated at the BCS level \cite{Bender00}. A large two-quasiparticle basis up to 100 MeV is taken into account. 
The Thomas-Reiche-Kuhn sum rule (including the enhancement factor $\kappa$) for isovector  $E1$ strength \cite{Ring1980} is exhausted by 102\% ($^{146}$Nd), 100\% ($^{148}$Nd), 97\% ($^{150}$Nd) and 98\% ($^{152}$Sm).

The axial quadrupole deformation characterised by the parameter $\beta_2$ is generally determined by minimisation of the total energy. However, since $^{144,146,148,150}$Nd and  $^{152}$Sm are rather soft, their energy curves $E(\beta_2)$  are  flat and give a large uncertainty in determination of the theoretical ground-state equilibrium deformation. Thus, we adopt the experimental values of $\beta_2$ given in Ref.~\cite{Don18}, which are $\beta_2$ = 0.13, 0.15, 0.20, 0.28, and 0.31 for $^{144,146,148,150}$Nd and $^{152}$Sm, respectively. For spherical $^{142}$Nd, a negligible deformation, $\beta_2$ = 0.001, is used. For IVGDR branches with $K=0$ and 1, the photo-absorption (in ${\rm fm}^2$) is computed as
\begin{eqnarray}
\label{13}
&& \sigma (E1K; E_{\rm x})
= \frac{16}{9}\pi^3 \alpha \sum_{K=0,1}(2-\delta_{K,0})
 \\ \nonumber
&\times & \sum_{\nu} E_{\nu} \: \big| \langle \nu|\:
\hat{M}(1EK) \: |{0} \rangle \big|^2 \: \xi_{\Delta}(E_{\rm x} -E_{\nu}) \;\, ,
\end{eqnarray}
where $\alpha$ is the fine-structure constant. Further, $|\nu \rangle$ and $E_{\nu}$ are the wave function  and energy of $\nu$-th SSRPA state, and
\begin{equation}
\hat{M}(E1K)= 
\frac{N}{A} \sum_{i=1}^Z \: r_i  Y_{1\mu}(\Omega_i)-
\frac{Z}{A} \sum_{i=1}^N \: r_i  Y_{1\mu}(\Omega_i)
\end{equation}
is the isovector dipole transition operator, which includes the centre of mass recoil correction. The smoothing Lorentz weighting reads
\begin{equation}
\xi_{\Delta}(E_{\rm x}-E_{\nu}) = \frac{1}{2 \pi}
\frac{\Delta}{(E_{\rm x}-E_{\nu})^2 + (\Delta/2)^2} \;\, .
\label{17}
\end{equation}
For accurate comparison between SSRPA and experimental results, a smoothing parameter equivalent to the experimental energy resolution is used. 
The strength is then summed over the appropriate number of bins.

\subsubsection{Quasi-particle Phonon Model}

The Quasi-particle Phonon Model (QPM) \cite{Sol} calculations were performed assuming either a  spherical ($^{142,144}$Nd) or a deformed ($^{148,150}$Nd,$^{152}$Sm) nature of the ground state of the corresponding group of nuclei. The transitional nucleus, $^{146}$Nd, was considered under both assumptions, spherical and deformed. The Woods-Saxon potential with parameters from global parameterisations was used as a mean field. The same sets were used for all spherical and all deformed nuclei under consideration and can be found in Refs.~\cite{WSs} and \cite{WSd}, respectively. The same values of the quadrupole deformation $\beta_2$ as in the SSRPA calculations were used for the mean field of the deformed nuclei; a weak hexadecapole with $\beta_4 = 0.08$ was also added in all cases. The strength of the pairing interaction, treated within the BCS with a constant matrix element, was adjusted to experimental values of the pairing energies.
 
The QPM employs a residual interaction in a separable form (the dipole-dipole one for the 1$^-$ states). The strength of the isoscalar residual interaction is adjusted to exclude the spurious state, i.e., to obtain the zero energy for the lowest QRPA solution. However, the isovector strength parameter is obtained from the correct  description of the IVGDR peak energy and in deformed nuclei, it is done for the $K=1$ branch, and the same value is used in the $K=0$ calculation.

The results of the QPM calculations in deformed nuclei are presented in the QRPA (or one-phonon) approximation for the $K=0$ and $K=1$ components. In spherical nuclei, the QRPA results are completed by the calculation in which the doorway one-phonon $1^-$ states interact with the background two-phonon configurations. The latter calculations are referred to as the ``1+2 phonon" calculations. The corresponding spectra are obtained by diagonalisation of the model Hamiltonian on the set of states which are described by a wave function which contains both one- and two-phonon configurations. The two-phonon 1$^-$ configurations were made up from the phonons with multipolarities from $1^{\pm}$ to $9^{\pm}$.

\subsection{\label{sec:WaveletAnalysis}Wavelet Analysis of Data and Model Predictions}

In this section, the results of the wavelet analysis of the experimental equivalent photo-absorption cross-sections and the photo-absorption predictions are presented. The discussion is subdivided into the quasi-spherical $^{142,144}$Nd, the intermediate spherical/deformed $^{146}$Nd, and the transitional-deformed $^{148,150}$Nd, $^{152}$Sm as classified in Fig.~\ref{fig:ratio}.

\subsubsection{Quasi-spherical $^{142,144}$Nd}
\label{subsubsec:142Nd}

\begin{figure}[b]
	\includegraphics[width=\linewidth]{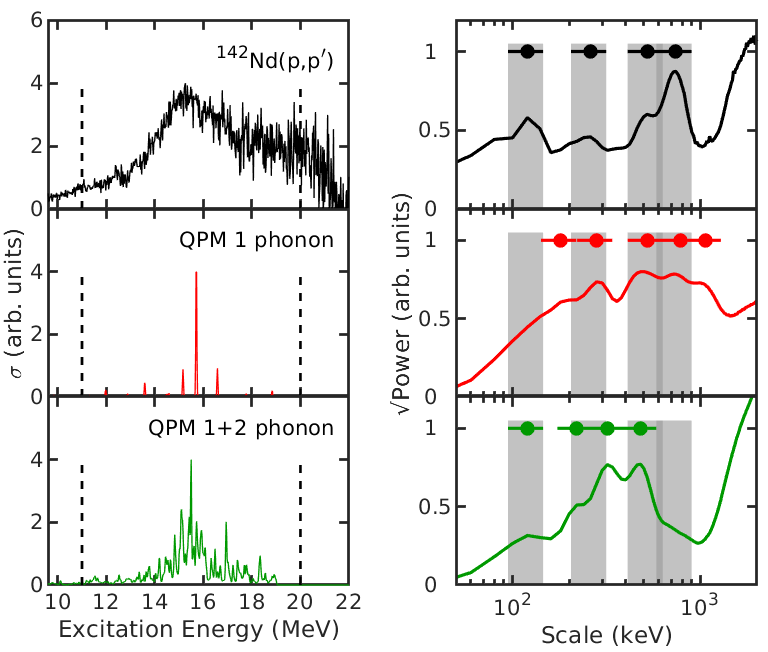}
	\caption{Left column: Equivalent photo-absorption spectrum for $^{142}$Nd (top) in comparison with the QPM 1 phonon (middle) and QPM 1+2 phonon (bottom) model predictions. The vertical dashed lines indicate the excitation-energy region from 11 to 20 MeV over which the wavelet coefficients were summed in order to determine the corresponding power spectra. Right column: Corresponding power spectra where the positions of the scales are indicated by filled circles together with error bars for the width. The experimental scales are also indicated by vertical grey bars in each panel to allow direct visual comparison with the theoretical predictions.}
	\label{fig:142Nd-WA}
\end{figure}

Referring to the wavelet analysis example for $^{142}$Nd shown in Fig.~\ref{fig:142Nd-WA-Full}, the top left-hand panel of Fig.~\ref{fig:142Nd-WA} again shows the $^{142}$Nd(p,p$^\prime$) equivalent photo-absorption cross section together with its associated power spectrum in the top right-hand panel, now rotated so that the scale axis is displayed horizontally. The vertical dashed lines appearing in the panels on the left side of Fig.~\ref{fig:142Nd-WA} indicate the excitation-energy region from 11 to 20 MeV over which the wavelet coefficients were summed in order to determine the corresponding power spectra.

\begin{figure}
	\includegraphics[width=\linewidth]{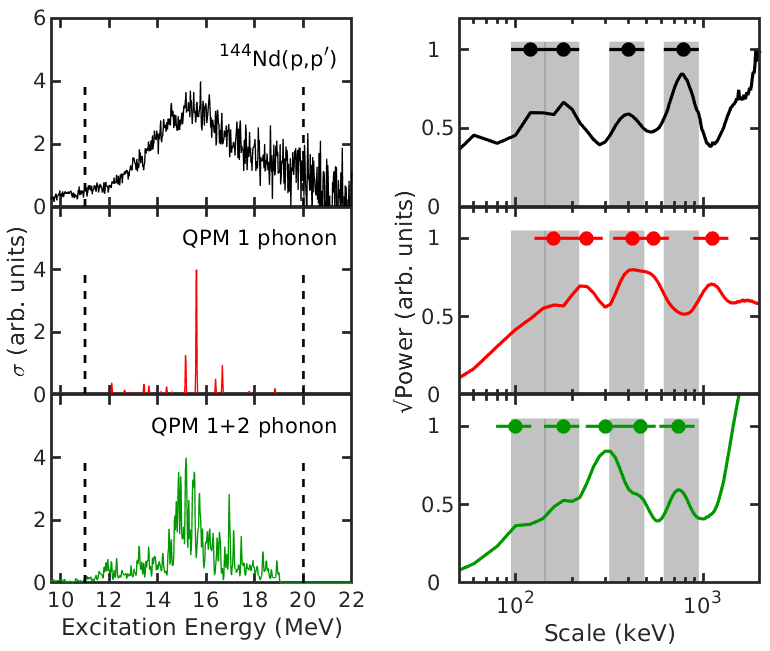}
	\caption{As for Fig.~\ref{fig:142Nd-WA}, but for $^{144}$Nd.}
	\label{fig:144Nd-WA}
\end{figure}

Following the procedure outlined in Section~\ref{subsec:cwt}, wavelet length-like energy scales are identified from the peaks and points of inflection in the power spectrum and are displayed as filled circles, with error bars representing one standard deviation of the corresponding width-like scale. 
The experimental results are also represented by vertical grey bars repeated in all right-side panels in order to facilitate the determination of similar energy scales in the corresponding power spectra for the theoretical predictions of the QPM 1-phonon calculation (red) and 1+2 phonon calculation (green) in the middle and bottom panels, respectively. The corresponding energy scales are indicated by red and green filled circles and error bars. Table~\ref{tab:142Nd} lists the extracted experimental and theoretical energy scales. When they agree within error bars, they are placed in the same column. The application of the above procedures to $^{144}$Nd is displayed in Fig.~\ref{fig:144Nd-WA} with the extracted energy scales listed in Table~\ref{tab:144Nd}.

\begin{table}[t]
\centering
\caption{Energy scales extracted for $^{142}$Nd.}
\begin{tabularx}{\linewidth}{lXXXXl} 
\hline\hline
Dataset        & \multicolumn{5}{l}{ Scales (keV)}  \\ 
\hline
Exp.           & \hspace{0.2cm} 120 & \hspace{0.2cm} 260 & \hspace{0.2cm} 520 & \hspace{0.2cm} 740 &                 \\
QPM 1 phonon   & \hspace{0.2cm}     & 180 280            & \hspace{0.2cm} 520 & \hspace{0.2cm} 780 & \hspace{0.2cm} 1060           \\
QPM 1+2 phonon & \hspace{0.2cm} 120 & 220 320            & \hspace{0.2cm} 480 &     &                 \\
\hline\hline
\end{tabularx}
\label{tab:142Nd}
\end{table}

\begin{table}[b]
\centering
\caption{Energy scales extracted for $^{144}$Nd.}
\begin{tabularx}{\linewidth}{lXXXXXXl} 
\hline\hline
Dataset        & \multicolumn{7}{l}{ Scales (keV)}  \\ 
\hline
Exp.           & 120 & 180 &     & 400 &     & 780 &                   \\
QPM 1 phonon   &     & 160 & 240 & 420 & 540 &     & 1120              \\
QPM 1+2 phonon & 100 & 180 & 300 & 460 &     & 740 &                   \\
\hline\hline
\end{tabularx}
\label{tab:144Nd}
\end{table}

The similarity of the experimental equivalent photo-absorption spectra in $^{142,144}$Nd allows for the expectation that the respective power spectra, top right-side panels of Figs.~\ref{fig:142Nd-WA} and \ref{fig:144Nd-WA}, would also display similar wavelet energy scales. This is indeed the case, where four scales are identified between 100 and 1000 keV, with the middle two scales moving to smaller values for $^{144}$Nd. In both nuclei, the theoretical calculations are lacking the lowest scale in the data at 120 keV, and a scale at values slightly above 1000 keV is predicted without an experimental counterpart. The QPM 1 phonon predictions each have a single dominant doorway state close to the experimental maximum surrounded by well-spaced weaker doorway states. As such, the corresponding power spectra display scales that correspond well to the experimental scales in $^{142}$Nd. In $^{144}$Nd, there are two theoretical scales seen to correspond within error with the experimental scales at 180 and 400~keV. 

For both $^{142}$Nd and $^{144}$Nd, the basic shape of the IVGDR is predicted by the QPM 1+2 phonon calculations where the maxima correlate well with the corresponding maxima in the experimental spectra. Similar scales to the 1-phonon results are found in the range of several hundred keV with only small shifts in energy. The experimental scale at 780 keV in $^{144}$Nd is well reproduced in contrast to the 1-phonon results, but the corresponding scale in $^{142}$Nd at 740 keV is missed. 
However, for both nuclei the low-energy scales missing in the 1-phonon calculations are found indicating that these indeed result from 2p2h coupling.

The present findings conform with similar studies of the fine structure of the IVGDR in lighter spherical nuclei \cite{Jingo2018} and in $^{208}$Pb \cite{Pol14}. The scales in the calculations are generated mainly by the fragmentation of the 1p1h strength, i.e., Landau damping. Different to the present cases, no scales related to the spreading width could be identified in $^{208}$Pb, but this may be due to the low density of 2p2h states in a doubly magic nucleus.  

\subsubsection{Intermediate Spherical/Deformed $^{146}$Nd}

\begin{figure}
	\includegraphics[width=\linewidth]{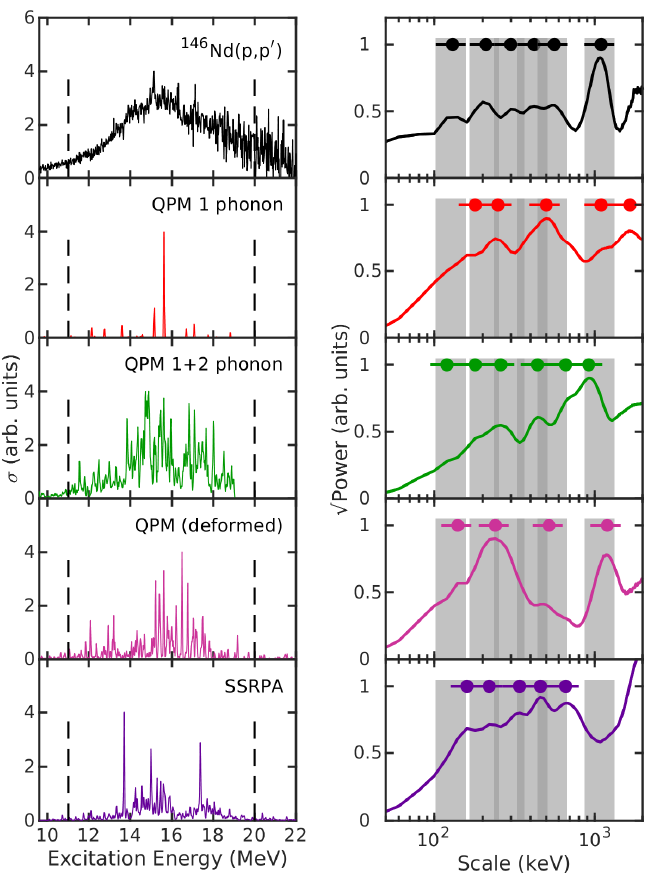}
	\caption{Left column: Equivalent photo-absorption spectrum for $^{146}$Nd (top) in comparison with the QPM 1 phonon, QPM 1+2 phonon, QPM (deformed) and SSRPA model predictions. Right column: The corresponding power spectra.}
	\label{fig:146Nd-WA}
\end{figure}

The results for the intermediate spherical/deformed nucleus $^{146}$Nd are shown in Fig.~\ref{fig:146Nd-WA}. Now, in addition to the spherical QPM results, the theoretical calculations have been extended to include the QPM (deformed) and SSRPA predictions (see Section~\ref{subsec:models}) because of the onset of deformation in $^{146}$Nd. A comparison between experimental scales and the various model predictions is given in Table~\ref{tab:146Nd}.

\begin{table}[b]
\centering
\caption{Energy scales extracted for $^{146}$Nd.}
\begin{tabularx}{\linewidth}{lXXXXXXl}
\hline\hline
Dataset          & \multicolumn{7}{l}{Scales (keV)}        \\ 
\hline
Exp.             & 130 & 210 & 300 & 420 & 560 &                 1100 &       \\
QPM 1 phonon     &     & 180 & 250 &     & 500 &                 1100 & 1660  \\
QPM 1+2 phonon   & 120 & 180 & 260 & 440 & 660 & \hspace{0.01cm} 920  &       \\
QPM (deformed)   & 140 & 240 &     &     & 520 &                 1200 &       \\
SSRPA            & 160 & 220 & 340 & 460 & 660 &                      &       \\ 
\hline\hline
\end{tabularx}
\label{tab:146Nd}
\end{table}

The experimental equivalent photo-absorption cross section in the top left-side panel of Fig.~\ref{fig:146Nd-WA} shows an overall peak structure similar to the quasi-spherical $^{142,144}$Nd but the resonance peak has a larger width. The corresponding power spectrum in the top right-side panel displays a pattern very much like the adjacent less-deformed $^{144}$Nd (see Fig.~\ref{fig:144Nd-WA}) but exhibits more structure with six (instead of four in $^{144}$Nd) scales identified. The QPM 1 phonon prediction shows a single dominant doorway state close to the experimental peak maximum surrounded by well-spaced weaker doorway states as was the case for the lower-mass quasi-spherical nuclei, and produces similar scales as can be seen when comparing Tables \ref{tab:144Nd} and \ref{tab:146Nd}.

The QPM 1+2 phonon prediction, in contrast, shows strong fragmentation and a width that already approaches that of the experimental case. The corresponding scales extracted from the power spectrum  exhibit larger differences to the 1-phonon result than in the spherical cases highlighting the relevance of spreading width contributions to the fine structure. In particular, as in spherical nuclei, the lowest experimental scale at 130 keV and the scale at 420 keV can now be reproduced. 
Indeed, the same number of scales is seen as in the data and fair agreement of absolute values is achieved except for the largest scale (1100 keV experimentally versus 920 keV theoretically).
 
As deformation increases, it is now possible to consider approaches starting from a deformed single-particle basis. The QPM (deformed) prediction produces a more compact strength function but yields four out of the six experimental scales with absolute values close to experiment. The SSRPA results in the lower panel of the left side of Fig.~\ref{fig:146Nd-WA} show somewhat less fine structure than seen in the QPM predictions. The experimental scales are well reproduced except for the largest one. We note that both calculations find a scale at about 150 keV close to the lowest experimental scale. It is, therefore, not possible in the present case to assign this scale uniquely to spreading due to coupling with 2p2h states.

\subsubsection{Transitional-deformed $^{148}$Nd, $^{150}$Nd and $^{152}$Sm}

\begin{figure}[t]
	\includegraphics[width=\linewidth]{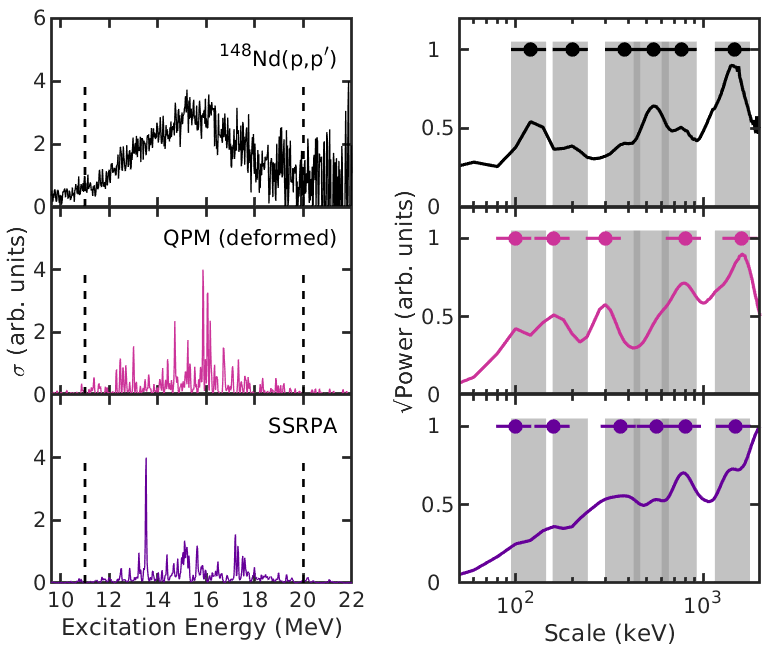}
	\caption{Left column: Equivalent photo-absorption spectrum for $^{148}$Nd (top) in comparison with the QPM (deformed) (middle) and SSRPA (bottom) model predictions. Right column: The corresponding power spectra.}
	\label{fig:148Nd-WA}
\end{figure}

\begin{figure}
	\includegraphics[width=\linewidth]{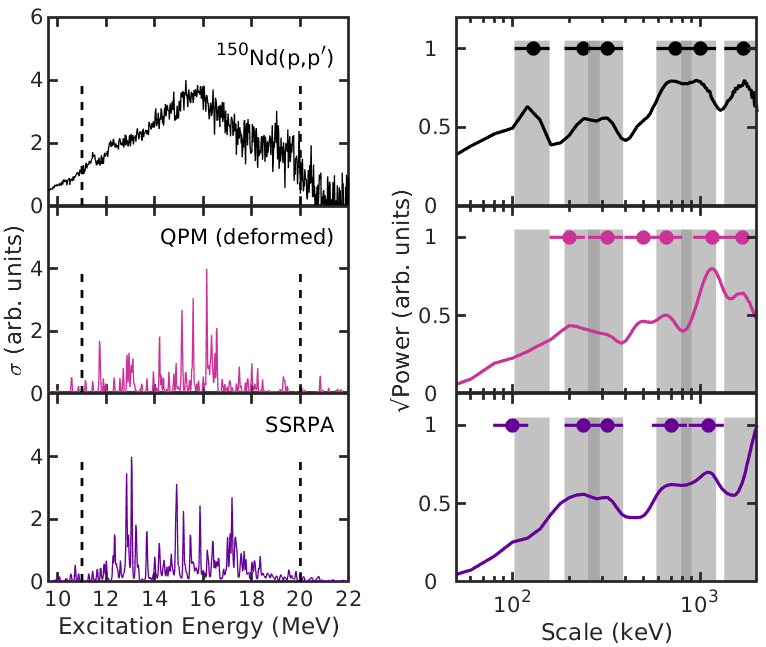}
	\caption{As for Fig.~\ref{fig:148Nd-WA} but for $^{150}$Nd.}
	\label{fig:150Nd-WA}
\end{figure}

\begin{figure}
	\includegraphics[width=\linewidth]{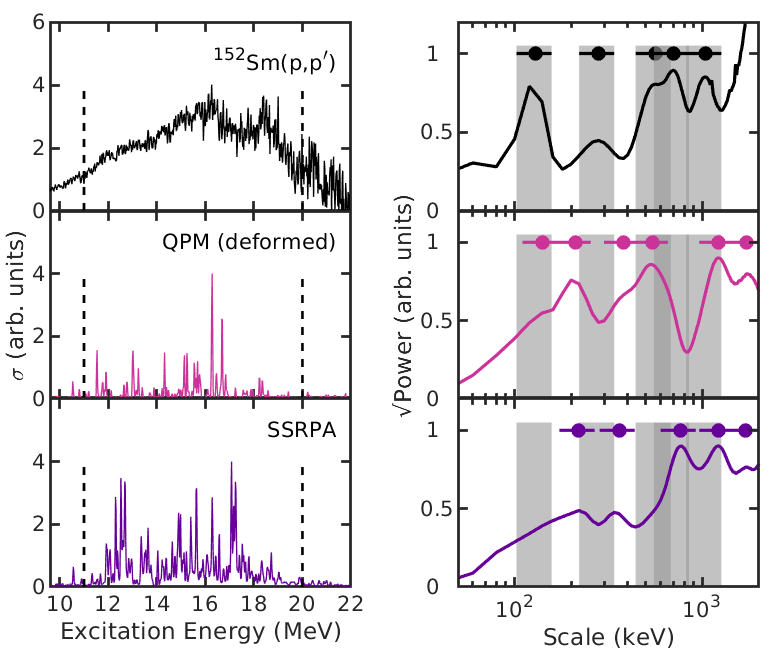}
	\caption{As for Fig.~\ref{fig:148Nd-WA} but for $^{152}$Sm.}
	\label{fig:152Sm-WA}
\end{figure}

Results for the transitional-deformed nuclei $^{148}$Nd, $^{150}$Nd and $^{152}$Sm are given in Figs.~\ref{fig:148Nd-WA}-\ref{fig:152Sm-WA}, respectively. The calculations used for comparison in this deformed region of interest are the QPM (deformed) and SSRPA model predictions (see Section~\ref{subsec:models}). The comparisons between the experimental scales and those for the model predictions are summarised in Tables~\ref{tab:148Nd}-\ref{tab:152Sm}.

\begin{table}[t]
\centering
\caption{Energy scales extracted for $^{148}$Nd.}
\begin{tabularx}{\linewidth}{lXXXXXl} 
\hline\hline
Dataset      & \multicolumn{6}{l}{ Scales (keV)}  \\ 
\hline
Exp.           & 120 & 200 & 380 & 540 & 760 & 1460             \\
QPM (deformed) & 100 & 160 & 300 &     & 800 & 1600             \\
SSRPA          & 100 & 160 & 360 & 560 & 800 & 1480             \\
\hline\hline
\end{tabularx}
\label{tab:148Nd}
\end{table}

\begin{table}
\centering
\caption{Energy scales extracted for $^{150}$Nd.}
\begin{tabularx}{\linewidth}{lXXXXXXl} 
\hline\hline
Dataset      & \multicolumn{7}{l}{ Scales (keV)}  \\ 
\hline
Exp.           & 130 & 240 & 320 &     & 740 & 1000 & 1700      \\
QPM (deformed) &     & 200 & 320 & 500 & 660 & 1160 & 1680       \\
SSRPA          & 100 & 240 & 320 &     & 700 & 1100             \\
\hline\hline
\end{tabularx}
\label{tab:150Nd}
\end{table}

\begin{table}
\centering
\caption{Energy scales extracted for $^{152}$Sm.}
\begin{tabularx}{\linewidth}{lXXXXXXl} 
\hline\hline
Dataset      & \multicolumn{7}{l}{ Scales (keV)}  \\ 
\hline
Exp.           & 130 & 280 &     & 560 & 700 & 1040 &            \\
QPM (deformed) & 140 & 210 & 380 & 540 &     & 1220 & 1720       \\
SSRPA          &     & 220 & 360 &     & 760 & 1220 & 1700        \\
\hline\hline
\end{tabularx}
\label{tab:152Sm}
\end{table}

The transitional-deformed $^{148}$Nd, $^{150}$Nd and $^{152}$Sm nuclei are considered together, noting that deformation increases when moving to $^{148}$Nd with a further increase moving to $^{150}$Nd and $^{152}$Sm, which are essentially equally deformed isotones. Experimental equivalent photo-absorption cross sections shown in the top left-side panels of Figs.~\ref{fig:148Nd-WA}-\ref{fig:152Sm-WA} show an increase in the width of the IVGDR with deformation, although no double-hump structure due to $K$ splitting is visible in $^{150}$Nd and $^{152}$Sm \cite{Don18} in contrast to the observations from previous $(\gamma,x$n) experiments \cite{Carlos1971,Carlos1974}. The corresponding power spectra display patterns similar to the less-deformed nuclei with a comparable number of scales (five to six). 

Photo-absorption spectra from the QPM (deformed) and SSRPA models applicable to the deformed cases in the middle and lower left-side panels of Figs.~\ref{fig:148Nd-WA}-\ref{fig:152Sm-WA} show a high degree of fragmentation. Differences (discussed further below) are observed in the low-energy region of the IVGDR ($11-15$ MeV), where the SSRPA calculations find significantly more strength than the QPM (deformed) calculations. The corresponding power shown respectively in the middle and lower right-side panels have varying degrees of overlap with the experimental scales but typically give a comparable number of scales, indicating that it is likely that the experimentally observed fine structure arises mainly from fragmentation of the 1p1h strength (Landau damping). Theoretical calculations that include 2p2h degrees of freedom would, however, be highly beneficial to verify this. 

\subsubsection{$K$-splitting in $^{152}$Sm}

\begin{figure}[b]
	\includegraphics[scale=0.35]{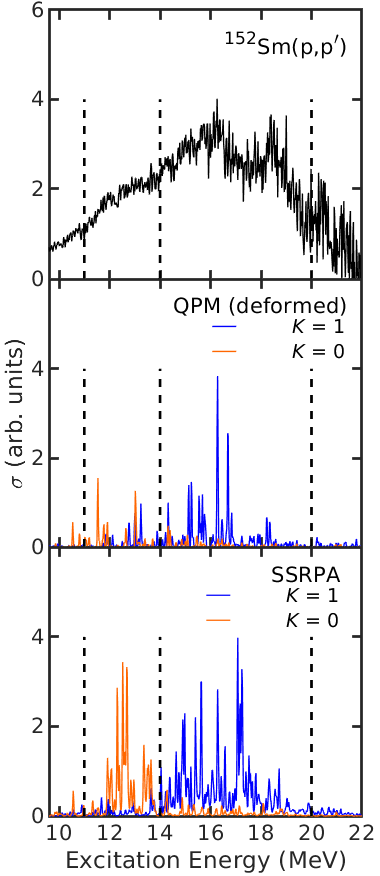}
	\caption{Excitation-energy spectra for the $^{152}$Sm equivalent photo-absorption cross section (top); the QPM (deformed) predictions for the $K=0$ (orange) and $K=1$ (blue) components (middle); and the SSRPA predictions for the $K=0$ (orange) and $K=1$ (blue) components (bottom).}
	\label{fig:152Sm-Ksplit-Ex}
\end{figure}
As pointed out above, the theoretical results show a clear difference in the lower-energy region of the IVGDR for the most deformed nuclei. This is illustrated in the middle and bottom panels of Fig.~\ref{fig:152Sm-Ksplit-Ex} for the example of $^{152}$Sm, where the $K = 0$ and 1 components are shown separately in orange and blue, respectively, and where $K$ is the projection of the nuclear momentum, $I$, onto the axis of symmetry. In both calculations, one finds a well-defined lower excitation-energy region from 11 to 14 MeV for the $K = 0$ component and a higher excitation-energy region from 14 to 20 MeV for the $K = 1$ component, but the $K =0$ contribution is much weaker in QPM (deformed). The experimental spectrum does not allow such a clear separation, since the typical widths of the $K$ components are comparable or larger than the spacing of the centroids. Nevertheless, one can expect dominance of either $K = 0$ or 1 in the chosen excitation-energy windows. 

\begin{figure}[t]
	\includegraphics[width=\linewidth]{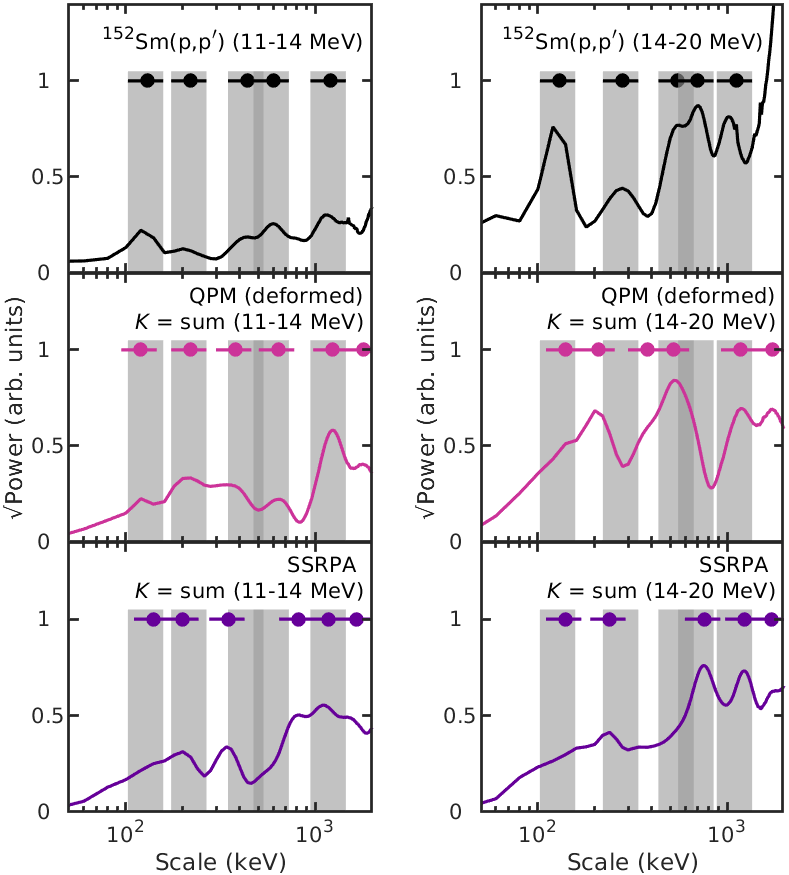}
	\caption{Wavelet power spectra for excitation energy regions of dominant $K = 0$ (left) and $K = 1$ (right) strength in $^{152}$Sm.}
	\label{fig:152Sm-Ksplit}
\end{figure}

The results of a wavelet analysis for these excitation-energy regions are shown in Fig.~\ref{fig:152Sm-Ksplit}. For consistency, the theoretical results are summed over both $K$ components in these intervals. A comparison between the extracted experimental and theoretical scales from the total spectra and from the dominant $K = 0$ and $K = 1$ regions is given in Table~\ref{tab:152Sm-Ksplit}.

\begin{table}[t]
\centering
\caption{Energy scales extracted from the total spectra and from the energy regions of dominant $K = 0$ ($11-14$ MeV) and $K = 1$ ($14-20$ MeV) strength in $^{152}$Sm.}
\begin{tabular}{p{0.07\linewidth}p{0.3\linewidth}p{0.072\linewidth}p{0.072\linewidth}p{0.072\linewidth}p{0.072\linewidth}p{0.072\linewidth}p{0.072\linewidth}p{0.072\linewidth}}
\hline\hline
\textit{K}           & Dataset        & \multicolumn{7}{l}{ Extracted Scales (keV)}      \\ 
\hline
\multirow{3}{*}{sum} & Exp.           & 130 & 280 &     & 560 & 700 & 1040 &           \\
                     & QPM (deformed) & 140 & 210 & 380 & 540 &     & 1220 & 1720       \\
                     & SSRPA          &     & 220 & 360 &     & 760 & 1220 & 1700       \\ 
\hline
\multirow{3}{*}{0}   & Exp.           & 130 & 220 & 440 & 600 &     & 1200 &          \\
                     & QPM (deformed) & 120 & 220 & 380 & 640 &     & 1240 & 1800      \\
                     & SSRPA          & 140 & 200 & 350 &     & 820 & 1180 & 1660      \\ 
\hline
\multirow{3}{*}{1}   & Exp.           & 130 & 280 &     & 550 & 700 & 1120 &           \\
                     & QPM (deformed) & 140 & 210 & 380 & 520 &     & 1180 & 1740       \\
                     & SSRPA          & 140 & 240 &     &     & 760 & 1240 & 1720          \\
\hline\hline
\end{tabular}
\label{tab:152Sm-Ksplit}
\end{table}

It can be seen in the top panels of Fig.~\ref{fig:152Sm-Ksplit} that the experimental power spectra for the $K = 0$ and $K = 1$ equivalent regions have noticeably different forms, resulting in two sets of different scales. This can also be said for the corresponding QPM (deformed) and SSRPA power spectra shown below. The power spectra from experiment and QPM (deformed) in the dominant $K = 1$ region are very similar to those deduced from the total spectra (Fig.~\ref{fig:152Sm-WA}). The deviations are somewhat larger for the SSRPA results, but still the wavelet power from the region of $K = 1$ dominance broadly follows that from the total spectrum. The scales deduced from the energy region  of $K = 0$ dominance differ, in particular in the region of scale energies between about 300 and 800 keV. 

One interesting aspect is the differences of the $K= 0$ and $K = 1$ wavelet power ratios in the experiment and models in Fig.~\ref{fig:152Sm-Ksplit}, where one can compare the relative ratios.
The data show significantly more power in the energy region of $K = 1$ dominance. This is qualitatively also seen in the calculations, but the effect is more pronounced in the QPM results reflecting the much smaller strength of the $K = 0$ component compared to the SSRPA results visible in Fig.~\ref{fig:152Sm-Ksplit-Ex}.

\section{\label{sec:conclusions}Conclusions}

We present new virtual-photon absorption data for the chain of stable Nd isotopes representing a transition from spherical to quadrupole-deformed nuclei and for the deformed nucleus $^{152}$Sm extracted from measurements of the (p,p$^\prime)$ reaction at 200 MeV and extreme forward scattering angles, $\theta_{\mathrm{Lab}}=0^\circ \pm 1.91^\circ$. Using dispersion-matching techniques, high energy-resolutions of the order $40 - 50$ keV (FWHM) were obtained. While the differences to previous photo-absorption data and the implications for $K$ splitting due to ground-state deformation have been discussed elsewhere \cite{Don18}, here we focus on the phenomenon of non-statistical cross-section fluctuations in the energy region of the IVGDR. The observation of this fine structure, even in the most deformed cases studied, is quite remarkable considering the extreme level densities (e.g.\ about $10^8$ $J^\pi = 1^-$ states per MeV at the IVGDR peak energy in $^{150}$Nd).

Wavelet-analysis techniques permit the quantification of features of the fine structure in terms of characteristic scales. Comparison is made with microscopic calculations of the photo-absorption strength functions based on RPA and in spherical nuclei extended to include 2p2h states. The agreement is mixed: neglecting trivial scales resulting from the experimental energy resolution and from the total width of the IVGDR, the number of scales can be approximately reproduced in most cases, but the agreement for absolute values varies. However, it is clear that the scales in the spherical and probably also in the deformed Nd nuclei mainly result from the fragmentation of the 1p1h strength into several dominant transitions serving as doorway states, but we note again that theoretical calculations including 2p2h degrees of freedom would be beneficial to clarify this for the deformed cases. This result is consistent with findings for the IVGDR in $^{208}$Pb \cite{Pol14}, $^{120}$Sn \cite{vonNeumann-Cosel2019b}, and in light nuclei \cite{Fearick2018}, although deformation is affected by alpha clustering for the latter. Thus, the origin of the fine structure of the IVGDR is fundamentally different from the case of the ISGQR, where coupling to low-lying phonons was identified as the driving mechanism \cite{Shevchenko2004,Shev2009} (except maybe for lighter nuclei \cite{Usman2011191,Usman2016}). In the spherical $^{142,144}$Nd nuclei, effects of the coupling to 2p2h states are seen in the QPM calculations including 2-phonon states. The overall agreement with the experimental scales is improved and the lowest scale at about 100 keV can be reproduced in contrast to calculations on the RPA level. The transitional nucleus $^{146}$Nd was studied with approaches starting from a spherical and deformed single-particle base. The spherical QPM calculation including 2-phonon states provides a superior description of the experimental strength function and the wavelet scales. 

To summarise, the wavelet analysis reveals information about the nature of the fine structure observed in the IVGDR. Landau damping seems to be the main source of the fine structure in both spherical and deformed nuclei. Some impact of the spreading due to coupling of the 2p2h states to the 1p1h doorway states is seen in the spherical/transitional nuclei, where such calculations beyond RPA are available. At present, it remains open whether information on the fine structure scales can be utilised to improve the development of global energy density functionals, where the isovector dipole response remains an open problem \cite{Bassauer2020}. 
This question will be addressed in future work, e.g.\ by systematic investigations of their dependence on bulk matter parameters \cite{Bassauer2020a}.
Further insight is expected from the comparison of experimental scales in selected nuclei with microscopic calculations incorporating all three mechanisms contributing to the resonance width simultaneously \cite{Tselyaev2016}.

\begin{acknowledgments}

We wish to thank J.L. Conradie, D.T. Fourie and the accelerator team at iThemba LABS for working tirelessly to provide a good-quality, low-halo proton beam. This work was supported by the National Research Foundation of South Africa under Grant Number 85509 and by the Deutsche Forschungsgemeinschaft (DFG, German Research Foundation) under Grant No.\ SFB 1245 (project ID 279384907). C.A.B.\ acknowledges support from the United States Department of Energy under Grant DE-FG02-08ER41533 and funding through the LANL Collaborative Research Program by the Texas A\&M System National Laboratory Office and Los Alamos National Laboratory. J.K.\ acknowledges support by the Czech Science Foundation under Grant Number P203-19-14048S. Furthermore, J.K.\ and V.O.N.\ acknowledge the support of the Votruba-Blokhintsev (Czech Republick - BLTP JINR) grant, and V.O.N.\ thanks the Hesenberg-Landau (Germany - BLTP JINR) grant. P.A. thanks the Claude Leon Foundation.

\end{acknowledgments}

\bibliography{Nd-Sm-Wavelets}

\begin{thebibliography}{59}%
\makeatletter
\providecommand \@ifxundefined [1]{%
 \@ifx{#1\undefined}
}%
\providecommand \@ifnum [1]{%
 \ifnum #1\expandafter \@firstoftwo
 \else \expandafter \@secondoftwo
 \fi
}%
\providecommand \@ifx [1]{%
 \ifx #1\expandafter \@firstoftwo
 \else \expandafter \@secondoftwo
 \fi
}%
\providecommand \natexlab [1]{#1}%
\providecommand \enquote  [1]{``#1''}%
\providecommand \bibnamefont  [1]{#1}%
\providecommand \bibfnamefont [1]{#1}%
\providecommand \citenamefont [1]{#1}%
\providecommand \href@noop [0]{\@secondoftwo}%
\providecommand \href [0]{\begingroup \@sanitize@url \@href}%
\providecommand \@href[1]{\@@startlink{#1}\@@href}%
\providecommand \@@href[1]{\endgroup#1\@@endlink}%
\providecommand \@sanitize@url [0]{\catcode `\\12\catcode `\$12\catcode
  `\&12\catcode `\#12\catcode `\^12\catcode `\_12\catcode `\%12\relax}%
\providecommand \@@startlink[1]{}%
\providecommand \@@endlink[0]{}%
\providecommand \url  [0]{\begingroup\@sanitize@url \@url }%
\providecommand \@url [1]{\endgroup\@href {#1}{\urlprefix }}%
\providecommand \urlprefix  [0]{URL }%
\providecommand \Eprint [0]{\href }%
\providecommand \doibase [0]{http://dx.doi.org/}%
\providecommand \selectlanguage [0]{\@gobble}%
\providecommand \bibinfo  [0]{\@secondoftwo}%
\providecommand \bibfield  [0]{\@secondoftwo}%
\providecommand \translation [1]{[#1]}%
\providecommand \BibitemOpen [0]{}%
\providecommand \bibitemStop [0]{}%
\providecommand \bibitemNoStop [0]{.\EOS\space}%
\providecommand \EOS [0]{\spacefactor3000\relax}%
\providecommand \BibitemShut  [1]{\csname bibitem#1\endcsname}%
\let\auto@bib@innerbib\@empty
\bibitem [{\citenamefont {Harakeh}\ and\ \citenamefont {Van~der
  Woude}(2001)}]{Harakeh}%
  \BibitemOpen
  \bibfield  {author} {\bibinfo {author} {\bibfnamefont {M.~N.}\ \bibnamefont
  {Harakeh}}\ and\ \bibinfo {author} {\bibfnamefont {A.}~\bibnamefont {Van~der
  Woude}},\ }\href@noop {} {\emph {\bibinfo {title} {Giant Resonances:
  Fundamental High-Frequency Modes of Nuclear Excitation}}}\ (\bibinfo
  {publisher} {Oxford University Press},\ \bibinfo {year} {2001})\BibitemShut
  {NoStop}%
\bibitem [{\citenamefont {Dro\ifmmode \dot{z}\else
  \.{z}\fi{}d\ifmmode~\dot{z}\else \.{z}\fi{}}\ \emph
  {et~al.}(1991)\citenamefont {Dro\ifmmode \dot{z}\else
  \.{z}\fi{}d\ifmmode~\dot{z}\else \.{z}\fi{}}, \citenamefont {Nishizaki},
  \citenamefont {Speth},\ and\ \citenamefont {Wambach}}]{Drozdz}%
  \BibitemOpen
  \bibfield  {author} {\bibinfo {author} {\bibfnamefont {S.}~\bibnamefont
  {Dro\ifmmode \dot{z}\else \.{z}\fi{}d\ifmmode~\dot{z}\else \.{z}\fi{}}},
  \bibinfo {author} {\bibfnamefont {S.}~\bibnamefont {Nishizaki}}, \bibinfo
  {author} {\bibfnamefont {J.}~\bibnamefont {Speth}}, \ and\ \bibinfo {author}
  {\bibfnamefont {J.}~\bibnamefont {Wambach}},\ }\href@noop {} {\bibfield
  {journal} {\bibinfo  {journal} {Phys. Rep.}\ }\textbf {\bibinfo {volume}
  {197}},\ \bibinfo {pages} {1} (\bibinfo {year} {1991})}\BibitemShut {NoStop}%
\bibitem [{\citenamefont {Speth}\ and\ \citenamefont {van~der
  Woude}(1981)}]{Spe81}%
  \BibitemOpen
  \bibfield  {author} {\bibinfo {author} {\bibfnamefont {J.}~\bibnamefont
  {Speth}}\ and\ \bibinfo {author} {\bibfnamefont {A.}~\bibnamefont {van~der
  Woude}},\ }\href@noop {} {\bibfield  {journal} {\bibinfo  {journal} {Rep.
  Prog. Phys.}\ }\textbf {\bibinfo {volume} {44}},\ \bibinfo {pages} {719}
  (\bibinfo {year} {1981})}\BibitemShut {NoStop}%
\bibitem [{\citenamefont {von Neumann-Cosel}\ and\ \citenamefont
  {Tamii}(2019)}]{vonNeumann-Cosel2019a}%
  \BibitemOpen
  \bibfield  {author} {\bibinfo {author} {\bibfnamefont {P.}~\bibnamefont {von
  Neumann-Cosel}}\ and\ \bibinfo {author} {\bibfnamefont {A.}~\bibnamefont
  {Tamii}},\ }\href@noop {} {\bibfield  {journal} {\bibinfo  {journal} {Eur.
  Phys. J. A}\ }\textbf {\bibinfo {volume} {55}},\ \bibinfo {pages} {110}
  (\bibinfo {year} {2019})}\BibitemShut {NoStop}%
\bibitem [{\citenamefont {Neveling}\ \emph
  {et~al.}(2011{\natexlab{a}})\citenamefont {Neveling}, \citenamefont {Fujita},
  \citenamefont {Smit}, \citenamefont {Adachi}, \citenamefont {Berg},
  \citenamefont {Buthelezi}, \citenamefont {Carter}, \citenamefont {Conradie},
  \citenamefont {Couder}, \citenamefont {Fearick}, \citenamefont {F\"{o}rtsch},
  \citenamefont {Fourie}, \citenamefont {Fujita}, \citenamefont {G\"{o}rres},
  \citenamefont {Hatanaka}, \citenamefont {Jingo}, \citenamefont {Krumbholz},
  \citenamefont {Kureba}, \citenamefont {Mira}, \citenamefont {Murray},
  \citenamefont {von Neumann-Cosel}, \citenamefont {O'Brien}, \citenamefont
  {Papka}, \citenamefont {Poltoratska}, \citenamefont {Richter}, \citenamefont
  {Sideras-Haddad}, \citenamefont {Swartz}, \citenamefont {Tamii},
  \citenamefont {Usman},\ and\ \citenamefont {van Zyl}}]{Neveling2011b}%
  \BibitemOpen
  \bibfield  {author} {\bibinfo {author} {\bibfnamefont {R.}~\bibnamefont
  {Neveling}}, \bibinfo {author} {\bibfnamefont {H.}~\bibnamefont {Fujita}},
  \bibinfo {author} {\bibfnamefont {F.~D.}\ \bibnamefont {Smit}}, \bibinfo
  {author} {\bibfnamefont {T.}~\bibnamefont {Adachi}}, \bibinfo {author}
  {\bibfnamefont {G.~P.~A.}\ \bibnamefont {Berg}}, \bibinfo {author}
  {\bibfnamefont {E.~Z.}\ \bibnamefont {Buthelezi}}, \bibinfo {author}
  {\bibfnamefont {J.}~\bibnamefont {Carter}}, \bibinfo {author} {\bibfnamefont
  {J.~L.}\ \bibnamefont {Conradie}}, \bibinfo {author} {\bibfnamefont
  {M.}~\bibnamefont {Couder}}, \bibinfo {author} {\bibfnamefont {R.~W.}\
  \bibnamefont {Fearick}}, \bibinfo {author} {\bibfnamefont {S.~V.}\
  \bibnamefont {F\"{o}rtsch}}, \bibinfo {author} {\bibfnamefont {D.~T.}\
  \bibnamefont {Fourie}}, \bibinfo {author} {\bibfnamefont {Y.}~\bibnamefont
  {Fujita}}, \bibinfo {author} {\bibfnamefont {J.}~\bibnamefont {G\"{o}rres}},
  \bibinfo {author} {\bibfnamefont {K.}~\bibnamefont {Hatanaka}}, \bibinfo
  {author} {\bibfnamefont {M.}~\bibnamefont {Jingo}}, \bibinfo {author}
  {\bibfnamefont {A.~M.}\ \bibnamefont {Krumbholz}}, \bibinfo {author}
  {\bibfnamefont {C.~O.}\ \bibnamefont {Kureba}}, \bibinfo {author}
  {\bibfnamefont {J.~P.}\ \bibnamefont {Mira}}, \bibinfo {author}
  {\bibfnamefont {S.~H.~T.}\ \bibnamefont {Murray}}, \bibinfo {author}
  {\bibfnamefont {P.}~\bibnamefont {von Neumann-Cosel}}, \bibinfo {author}
  {\bibfnamefont {S.}~\bibnamefont {O'Brien}}, \bibinfo {author} {\bibfnamefont
  {P.}~\bibnamefont {Papka}}, \bibinfo {author} {\bibfnamefont
  {I.}~\bibnamefont {Poltoratska}}, \bibinfo {author} {\bibfnamefont
  {A.}~\bibnamefont {Richter}}, \bibinfo {author} {\bibfnamefont
  {E.}~\bibnamefont {Sideras-Haddad}}, \bibinfo {author} {\bibfnamefont
  {J.~A.}\ \bibnamefont {Swartz}}, \bibinfo {author} {\bibfnamefont
  {A.}~\bibnamefont {Tamii}}, \bibinfo {author} {\bibfnamefont {I.~T.}\
  \bibnamefont {Usman}}, \ and\ \bibinfo {author} {\bibfnamefont {J.~J.}\
  \bibnamefont {van Zyl}},\ }\href@noop {} {\bibfield  {journal} {\bibinfo
  {journal} {Nucl. Instrum. Methods A}\ }\textbf {\bibinfo {volume} {654}},\
  \bibinfo {pages} {29} (\bibinfo {year} {2011}{\natexlab{a}})}\BibitemShut
  {NoStop}%
\bibitem [{\citenamefont {Tamii}\ \emph {et~al.}(2009)\citenamefont {Tamii},
  \citenamefont {Fujita}, \citenamefont {Matsubara}, \citenamefont {Adachi},
  \citenamefont {Carter}, \citenamefont {Dozono}, \citenamefont {Fujita},
  \citenamefont {Fujita}, \citenamefont {Hashimoto}, \citenamefont {Hatanaka},
  \citenamefont {Itahashi}, \citenamefont {Itoh}, \citenamefont {Kawabata},
  \citenamefont {Nakanishi}, \citenamefont {Ninomiya}, \citenamefont
  {Perez-Cerdan}, \citenamefont {Popescu}, \citenamefont {Rubio}, \citenamefont
  {Saito}, \citenamefont {Sakaguchi}, \citenamefont {Sakemi}, \citenamefont
  {Sasamoto}, \citenamefont {Shimbara}, \citenamefont {Shimizu}, \citenamefont
  {Smit}, \citenamefont {Tameshige}, \citenamefont {Yosoi},\ and\ \citenamefont
  {Zenhiro}}]{Tam09}%
  \BibitemOpen
  \bibfield  {author} {\bibinfo {author} {\bibfnamefont {A.}~\bibnamefont
  {Tamii}}, \bibinfo {author} {\bibfnamefont {Y.}~\bibnamefont {Fujita}},
  \bibinfo {author} {\bibfnamefont {H.}~\bibnamefont {Matsubara}}, \bibinfo
  {author} {\bibfnamefont {T.}~\bibnamefont {Adachi}}, \bibinfo {author}
  {\bibfnamefont {J.}~\bibnamefont {Carter}}, \bibinfo {author} {\bibfnamefont
  {M.}~\bibnamefont {Dozono}}, \bibinfo {author} {\bibfnamefont
  {H.}~\bibnamefont {Fujita}}, \bibinfo {author} {\bibfnamefont
  {K.}~\bibnamefont {Fujita}}, \bibinfo {author} {\bibfnamefont
  {H.}~\bibnamefont {Hashimoto}}, \bibinfo {author} {\bibfnamefont
  {K.}~\bibnamefont {Hatanaka}}, \bibinfo {author} {\bibfnamefont
  {T.}~\bibnamefont {Itahashi}}, \bibinfo {author} {\bibfnamefont
  {M.}~\bibnamefont {Itoh}}, \bibinfo {author} {\bibfnamefont {T.}~\bibnamefont
  {Kawabata}}, \bibinfo {author} {\bibfnamefont {K.}~\bibnamefont {Nakanishi}},
  \bibinfo {author} {\bibfnamefont {S.}~\bibnamefont {Ninomiya}}, \bibinfo
  {author} {\bibfnamefont {A.~B.}\ \bibnamefont {Perez-Cerdan}}, \bibinfo
  {author} {\bibfnamefont {L.}~\bibnamefont {Popescu}}, \bibinfo {author}
  {\bibfnamefont {B.}~\bibnamefont {Rubio}}, \bibinfo {author} {\bibfnamefont
  {T.}~\bibnamefont {Saito}}, \bibinfo {author} {\bibfnamefont
  {H.}~\bibnamefont {Sakaguchi}}, \bibinfo {author} {\bibfnamefont
  {Y.}~\bibnamefont {Sakemi}}, \bibinfo {author} {\bibfnamefont
  {Y.}~\bibnamefont {Sasamoto}}, \bibinfo {author} {\bibfnamefont
  {Y.}~\bibnamefont {Shimbara}}, \bibinfo {author} {\bibfnamefont
  {Y.}~\bibnamefont {Shimizu}}, \bibinfo {author} {\bibfnamefont {F.~D.}\
  \bibnamefont {Smit}}, \bibinfo {author} {\bibfnamefont {Y.}~\bibnamefont
  {Tameshige}}, \bibinfo {author} {\bibfnamefont {M.}~\bibnamefont {Yosoi}}, \
  and\ \bibinfo {author} {\bibfnamefont {J.}~\bibnamefont {Zenhiro}},\
  }\href@noop {} {\bibfield  {journal} {\bibinfo  {journal} {Nucl. Instrum.
  Methods A}\ }\textbf {\bibinfo {volume} {605}},\ \bibinfo {pages} {326}
  (\bibinfo {year} {2009})}\BibitemShut {NoStop}%
\bibitem [{\citenamefont {Tamii}\ \emph {et~al.}(2011)\citenamefont {Tamii},
  \citenamefont {Poltoratska}, \citenamefont {von Neumann-Cosel}, \citenamefont
  {Fujita}, \citenamefont {Adachi}, \citenamefont {Bertulani}, \citenamefont
  {Carter}, \citenamefont {Dozono}, \citenamefont {Fujita}, \citenamefont
  {Fujita}, \citenamefont {Hatanaka}, \citenamefont {Ishikawa}, \citenamefont
  {Itoh}, \citenamefont {Kawabata}, \citenamefont {Kalmykov}, \citenamefont
  {Krumbholz}, \citenamefont {Litvinova}, \citenamefont {Matsubara},
  \citenamefont {Nakanishi}, \citenamefont {Neveling}, \citenamefont {Okamura},
  \citenamefont {Ong}, \citenamefont {\"Ozel-Tashenov}, \citenamefont
  {Ponomarev}, \citenamefont {Richter}, \citenamefont {Rubio}, \citenamefont
  {Sakaguchi}, \citenamefont {Sakemi}, \citenamefont {Sasamoto}, \citenamefont
  {Shimbara}, \citenamefont {Shimizu}, \citenamefont {Smit}, \citenamefont
  {Suzuki}, \citenamefont {Tameshige}, \citenamefont {Wambach}, \citenamefont
  {Yamada}, \citenamefont {Yosoi},\ and\ \citenamefont {Zenihiro}}]{Tamii2011}%
  \BibitemOpen
  \bibfield  {author} {\bibinfo {author} {\bibfnamefont {A.}~\bibnamefont
  {Tamii}}, \bibinfo {author} {\bibfnamefont {I.}~\bibnamefont {Poltoratska}},
  \bibinfo {author} {\bibfnamefont {P.}~\bibnamefont {von Neumann-Cosel}},
  \bibinfo {author} {\bibfnamefont {Y.}~\bibnamefont {Fujita}}, \bibinfo
  {author} {\bibfnamefont {T.}~\bibnamefont {Adachi}}, \bibinfo {author}
  {\bibfnamefont {C.~A.}\ \bibnamefont {Bertulani}}, \bibinfo {author}
  {\bibfnamefont {J.}~\bibnamefont {Carter}}, \bibinfo {author} {\bibfnamefont
  {M.}~\bibnamefont {Dozono}}, \bibinfo {author} {\bibfnamefont
  {H.}~\bibnamefont {Fujita}}, \bibinfo {author} {\bibfnamefont
  {K.}~\bibnamefont {Fujita}}, \bibinfo {author} {\bibfnamefont
  {K.}~\bibnamefont {Hatanaka}}, \bibinfo {author} {\bibfnamefont
  {D.}~\bibnamefont {Ishikawa}}, \bibinfo {author} {\bibfnamefont
  {M.}~\bibnamefont {Itoh}}, \bibinfo {author} {\bibfnamefont {T.}~\bibnamefont
  {Kawabata}}, \bibinfo {author} {\bibfnamefont {Y.}~\bibnamefont {Kalmykov}},
  \bibinfo {author} {\bibfnamefont {A.~M.}\ \bibnamefont {Krumbholz}}, \bibinfo
  {author} {\bibfnamefont {E.}~\bibnamefont {Litvinova}}, \bibinfo {author}
  {\bibfnamefont {H.}~\bibnamefont {Matsubara}}, \bibinfo {author}
  {\bibfnamefont {K.}~\bibnamefont {Nakanishi}}, \bibinfo {author}
  {\bibfnamefont {R.}~\bibnamefont {Neveling}}, \bibinfo {author}
  {\bibfnamefont {H.}~\bibnamefont {Okamura}}, \bibinfo {author} {\bibfnamefont
  {H.~J.}\ \bibnamefont {Ong}}, \bibinfo {author} {\bibfnamefont
  {B.}~\bibnamefont {\"Ozel-Tashenov}}, \bibinfo {author} {\bibfnamefont
  {V.~Y.}\ \bibnamefont {Ponomarev}}, \bibinfo {author} {\bibfnamefont
  {A.}~\bibnamefont {Richter}}, \bibinfo {author} {\bibfnamefont
  {B.}~\bibnamefont {Rubio}}, \bibinfo {author} {\bibfnamefont
  {H.}~\bibnamefont {Sakaguchi}}, \bibinfo {author} {\bibfnamefont
  {Y.}~\bibnamefont {Sakemi}}, \bibinfo {author} {\bibfnamefont
  {Y.}~\bibnamefont {Sasamoto}}, \bibinfo {author} {\bibfnamefont
  {Y.}~\bibnamefont {Shimbara}}, \bibinfo {author} {\bibfnamefont
  {Y.}~\bibnamefont {Shimizu}}, \bibinfo {author} {\bibfnamefont {F.~D.}\
  \bibnamefont {Smit}}, \bibinfo {author} {\bibfnamefont {T.}~\bibnamefont
  {Suzuki}}, \bibinfo {author} {\bibfnamefont {Y.}~\bibnamefont {Tameshige}},
  \bibinfo {author} {\bibfnamefont {J.}~\bibnamefont {Wambach}}, \bibinfo
  {author} {\bibfnamefont {R.}~\bibnamefont {Yamada}}, \bibinfo {author}
  {\bibfnamefont {M.}~\bibnamefont {Yosoi}}, \ and\ \bibinfo {author}
  {\bibfnamefont {J.}~\bibnamefont {Zenihiro}},\ }\href {\doibase
  10.1103/PhysRevLett.107.062502} {\bibfield  {journal} {\bibinfo  {journal}
  {Phys. Rev. Lett.}\ }\textbf {\bibinfo {volume} {107}},\ \bibinfo {pages}
  {062502} (\bibinfo {year} {2011})}\BibitemShut {NoStop}%
\bibitem [{\citenamefont {Poltoratska}\ \emph {et~al.}(2012)\citenamefont
  {Poltoratska}, \citenamefont {von Neumann-Cosel}, \citenamefont {Tamii},
  \citenamefont {Adachi}, \citenamefont {Bertulani}, \citenamefont {Carter},
  \citenamefont {Dozono}, \citenamefont {Fujita}, \citenamefont {Fujita},
  \citenamefont {Fujita}, \citenamefont {Hatanaka}, \citenamefont {Itoh},
  \citenamefont {Kawabata}, \citenamefont {Kalmykov}, \citenamefont
  {Krumbholz}, \citenamefont {Litvinova}, \citenamefont {Matsubara},
  \citenamefont {Nakanishi}, \citenamefont {Neveling}, \citenamefont {Okamura},
  \citenamefont {Ong}, \citenamefont {\"Ozel-Tashenov}, \citenamefont
  {Ponomarev}, \citenamefont {Richter}, \citenamefont {Rubio}, \citenamefont
  {Sakaguchi}, \citenamefont {Sakemi}, \citenamefont {Sasamoto}, \citenamefont
  {Shimbara}, \citenamefont {Shimizu}, \citenamefont {Smit}, \citenamefont
  {Suzuki}, \citenamefont {Tameshige}, \citenamefont {Wambach}, \citenamefont
  {Yosoi},\ and\ \citenamefont {Zenihiro}}]{Pol2012}%
  \BibitemOpen
  \bibfield  {author} {\bibinfo {author} {\bibfnamefont {I.}~\bibnamefont
  {Poltoratska}}, \bibinfo {author} {\bibfnamefont {P.}~\bibnamefont {von
  Neumann-Cosel}}, \bibinfo {author} {\bibfnamefont {A.}~\bibnamefont {Tamii}},
  \bibinfo {author} {\bibfnamefont {T.}~\bibnamefont {Adachi}}, \bibinfo
  {author} {\bibfnamefont {C.~A.}\ \bibnamefont {Bertulani}}, \bibinfo {author}
  {\bibfnamefont {J.}~\bibnamefont {Carter}}, \bibinfo {author} {\bibfnamefont
  {M.}~\bibnamefont {Dozono}}, \bibinfo {author} {\bibfnamefont
  {H.}~\bibnamefont {Fujita}}, \bibinfo {author} {\bibfnamefont
  {K.}~\bibnamefont {Fujita}}, \bibinfo {author} {\bibfnamefont
  {Y.}~\bibnamefont {Fujita}}, \bibinfo {author} {\bibfnamefont
  {K.}~\bibnamefont {Hatanaka}}, \bibinfo {author} {\bibfnamefont
  {M.}~\bibnamefont {Itoh}}, \bibinfo {author} {\bibfnamefont {T.}~\bibnamefont
  {Kawabata}}, \bibinfo {author} {\bibfnamefont {Y.}~\bibnamefont {Kalmykov}},
  \bibinfo {author} {\bibfnamefont {A.~M.}\ \bibnamefont {Krumbholz}}, \bibinfo
  {author} {\bibfnamefont {E.}~\bibnamefont {Litvinova}}, \bibinfo {author}
  {\bibfnamefont {H.}~\bibnamefont {Matsubara}}, \bibinfo {author}
  {\bibfnamefont {K.}~\bibnamefont {Nakanishi}}, \bibinfo {author}
  {\bibfnamefont {R.}~\bibnamefont {Neveling}}, \bibinfo {author}
  {\bibfnamefont {H.}~\bibnamefont {Okamura}}, \bibinfo {author} {\bibfnamefont
  {H.~J.}\ \bibnamefont {Ong}}, \bibinfo {author} {\bibfnamefont
  {B.}~\bibnamefont {\"Ozel-Tashenov}}, \bibinfo {author} {\bibfnamefont
  {V.~Y.}\ \bibnamefont {Ponomarev}}, \bibinfo {author} {\bibfnamefont
  {A.}~\bibnamefont {Richter}}, \bibinfo {author} {\bibfnamefont
  {B.}~\bibnamefont {Rubio}}, \bibinfo {author} {\bibfnamefont
  {H.}~\bibnamefont {Sakaguchi}}, \bibinfo {author} {\bibfnamefont
  {Y.}~\bibnamefont {Sakemi}}, \bibinfo {author} {\bibfnamefont
  {Y.}~\bibnamefont {Sasamoto}}, \bibinfo {author} {\bibfnamefont
  {Y.}~\bibnamefont {Shimbara}}, \bibinfo {author} {\bibfnamefont
  {Y.}~\bibnamefont {Shimizu}}, \bibinfo {author} {\bibfnamefont {F.~D.}\
  \bibnamefont {Smit}}, \bibinfo {author} {\bibfnamefont {T.}~\bibnamefont
  {Suzuki}}, \bibinfo {author} {\bibfnamefont {Y.}~\bibnamefont {Tameshige}},
  \bibinfo {author} {\bibfnamefont {J.}~\bibnamefont {Wambach}}, \bibinfo
  {author} {\bibfnamefont {M.}~\bibnamefont {Yosoi}}, \ and\ \bibinfo {author}
  {\bibfnamefont {J.}~\bibnamefont {Zenihiro}},\ }\href {\doibase
  10.1103/PhysRevC.85.041304} {\bibfield  {journal} {\bibinfo  {journal} {Phys.
  Rev. C}\ }\textbf {\bibinfo {volume} {85}},\ \bibinfo {pages} {041304(R)}
  (\bibinfo {year} {2012})}\BibitemShut {NoStop}%
\bibitem [{\citenamefont {Krumbholz}\ \emph {et~al.}(2015)\citenamefont
  {Krumbholz}, \citenamefont {von Neumann-Cosel}, \citenamefont {Hashimoto},
  \citenamefont {Tamii}, \citenamefont {Adachi}, \citenamefont {Bertulani},
  \citenamefont {Fujita}, \citenamefont {Fujita}, \citenamefont
  {Ganio\ifmmode~\check{g}\else \v{g}\fi{}lu}, \citenamefont {Hatanaka},
  \citenamefont {Iwamoto}, \citenamefont {Kawabata}, \citenamefont {Khai},
  \citenamefont {Krugmann}, \citenamefont {Martin}, \citenamefont {Matsubara},
  \citenamefont {Neveling}, \citenamefont {Okamura}, \citenamefont {Ong},
  \citenamefont {Poltoratska}, \citenamefont {Ponomarev}, \citenamefont
  {Richter}, \citenamefont {Sakaguchi}, \citenamefont {Shimbara}, \citenamefont
  {Shimizu}, \citenamefont {Simonis}, \citenamefont {Smit}, \citenamefont
  {S\"usoy}, \citenamefont {Thies}, \citenamefont {Suzuki}, \citenamefont
  {Yosoi},\ and\ \citenamefont {Zenihiro}}]{Krumbholz2015}%
  \BibitemOpen
  \bibfield  {author} {\bibinfo {author} {\bibfnamefont {A.~M.}\ \bibnamefont
  {Krumbholz}}, \bibinfo {author} {\bibfnamefont {P.}~\bibnamefont {von
  Neumann-Cosel}}, \bibinfo {author} {\bibfnamefont {T.}~\bibnamefont
  {Hashimoto}}, \bibinfo {author} {\bibfnamefont {A.}~\bibnamefont {Tamii}},
  \bibinfo {author} {\bibfnamefont {T.}~\bibnamefont {Adachi}}, \bibinfo
  {author} {\bibfnamefont {C.~A.}\ \bibnamefont {Bertulani}}, \bibinfo {author}
  {\bibfnamefont {H.}~\bibnamefont {Fujita}}, \bibinfo {author} {\bibfnamefont
  {Y.}~\bibnamefont {Fujita}}, \bibinfo {author} {\bibfnamefont
  {E.}~\bibnamefont {Ganio\ifmmode~\check{g}\else \v{g}\fi{}lu}}, \bibinfo
  {author} {\bibfnamefont {K.}~\bibnamefont {Hatanaka}}, \bibinfo {author}
  {\bibfnamefont {C.}~\bibnamefont {Iwamoto}}, \bibinfo {author} {\bibfnamefont
  {T.}~\bibnamefont {Kawabata}}, \bibinfo {author} {\bibfnamefont {N.~T.}\
  \bibnamefont {Khai}}, \bibinfo {author} {\bibfnamefont {A.}~\bibnamefont
  {Krugmann}}, \bibinfo {author} {\bibfnamefont {D.}~\bibnamefont {Martin}},
  \bibinfo {author} {\bibfnamefont {H.}~\bibnamefont {Matsubara}}, \bibinfo
  {author} {\bibfnamefont {R.}~\bibnamefont {Neveling}}, \bibinfo {author}
  {\bibfnamefont {H.}~\bibnamefont {Okamura}}, \bibinfo {author} {\bibfnamefont
  {H.~J.}\ \bibnamefont {Ong}}, \bibinfo {author} {\bibfnamefont
  {I.}~\bibnamefont {Poltoratska}}, \bibinfo {author} {\bibfnamefont {V.~Y.}\
  \bibnamefont {Ponomarev}}, \bibinfo {author} {\bibfnamefont {A.}~\bibnamefont
  {Richter}}, \bibinfo {author} {\bibfnamefont {H.}~\bibnamefont {Sakaguchi}},
  \bibinfo {author} {\bibfnamefont {Y.}~\bibnamefont {Shimbara}}, \bibinfo
  {author} {\bibfnamefont {Y.}~\bibnamefont {Shimizu}}, \bibinfo {author}
  {\bibfnamefont {J.}~\bibnamefont {Simonis}}, \bibinfo {author} {\bibfnamefont
  {F.~D.}\ \bibnamefont {Smit}}, \bibinfo {author} {\bibfnamefont
  {G.}~\bibnamefont {S\"usoy}}, \bibinfo {author} {\bibfnamefont {J.~H.}\
  \bibnamefont {Thies}}, \bibinfo {author} {\bibfnamefont {T.}~\bibnamefont
  {Suzuki}}, \bibinfo {author} {\bibfnamefont {M.}~\bibnamefont {Yosoi}}, \
  and\ \bibinfo {author} {\bibfnamefont {J.}~\bibnamefont {Zenihiro}},\
  }\href@noop {} {\bibfield  {journal} {\bibinfo  {journal} {Phys. Lett. B}\
  }\textbf {\bibinfo {volume} {744}},\ \bibinfo {pages} {7 } (\bibinfo {year}
  {2015})}\BibitemShut {NoStop}%
\bibitem [{\citenamefont {Hashimoto}\ \emph {et~al.}(2015)\citenamefont
  {Hashimoto}, \citenamefont {Krumbholz}, \citenamefont {Reinhard},
  \citenamefont {Tamii}, \citenamefont {von Neumann-Cosel}, \citenamefont
  {Adachi}, \citenamefont {Aoi}, \citenamefont {Bertulani}, \citenamefont
  {Fujita}, \citenamefont {Fujita}, \citenamefont {Ganio\ifmmode~\check{g}\else
  \v{g}\fi{}lu}, \citenamefont {Hatanaka}, \citenamefont {Ideguchi},
  \citenamefont {Iwamoto}, \citenamefont {Kawabata}, \citenamefont {Khai},
  \citenamefont {Krugmann}, \citenamefont {Martin}, \citenamefont {Matsubara},
  \citenamefont {Miki}, \citenamefont {Neveling}, \citenamefont {Okamura},
  \citenamefont {Ong}, \citenamefont {Poltoratska}, \citenamefont {Ponomarev},
  \citenamefont {Richter}, \citenamefont {Sakaguchi}, \citenamefont {Shimbara},
  \citenamefont {Shimizu}, \citenamefont {Simonis}, \citenamefont {Smit},
  \citenamefont {S\"usoy}, \citenamefont {Suzuki}, \citenamefont {Thies},
  \citenamefont {Yosoi},\ and\ \citenamefont {Zenihiro}}]{Hashimoto2015}%
  \BibitemOpen
  \bibfield  {author} {\bibinfo {author} {\bibfnamefont {T.}~\bibnamefont
  {Hashimoto}}, \bibinfo {author} {\bibfnamefont {A.~M.}\ \bibnamefont
  {Krumbholz}}, \bibinfo {author} {\bibfnamefont {P.-G.}\ \bibnamefont
  {Reinhard}}, \bibinfo {author} {\bibfnamefont {A.}~\bibnamefont {Tamii}},
  \bibinfo {author} {\bibfnamefont {P.}~\bibnamefont {von Neumann-Cosel}},
  \bibinfo {author} {\bibfnamefont {T.}~\bibnamefont {Adachi}}, \bibinfo
  {author} {\bibfnamefont {N.}~\bibnamefont {Aoi}}, \bibinfo {author}
  {\bibfnamefont {C.~A.}\ \bibnamefont {Bertulani}}, \bibinfo {author}
  {\bibfnamefont {H.}~\bibnamefont {Fujita}}, \bibinfo {author} {\bibfnamefont
  {Y.}~\bibnamefont {Fujita}}, \bibinfo {author} {\bibfnamefont
  {E.}~\bibnamefont {Ganio\ifmmode~\check{g}\else \v{g}\fi{}lu}}, \bibinfo
  {author} {\bibfnamefont {K.}~\bibnamefont {Hatanaka}}, \bibinfo {author}
  {\bibfnamefont {E.}~\bibnamefont {Ideguchi}}, \bibinfo {author}
  {\bibfnamefont {C.}~\bibnamefont {Iwamoto}}, \bibinfo {author} {\bibfnamefont
  {T.}~\bibnamefont {Kawabata}}, \bibinfo {author} {\bibfnamefont {N.~T.}\
  \bibnamefont {Khai}}, \bibinfo {author} {\bibfnamefont {A.}~\bibnamefont
  {Krugmann}}, \bibinfo {author} {\bibfnamefont {D.}~\bibnamefont {Martin}},
  \bibinfo {author} {\bibfnamefont {H.}~\bibnamefont {Matsubara}}, \bibinfo
  {author} {\bibfnamefont {K.}~\bibnamefont {Miki}}, \bibinfo {author}
  {\bibfnamefont {R.}~\bibnamefont {Neveling}}, \bibinfo {author}
  {\bibfnamefont {H.}~\bibnamefont {Okamura}}, \bibinfo {author} {\bibfnamefont
  {H.~J.}\ \bibnamefont {Ong}}, \bibinfo {author} {\bibfnamefont
  {I.}~\bibnamefont {Poltoratska}}, \bibinfo {author} {\bibfnamefont {V.~Y.}\
  \bibnamefont {Ponomarev}}, \bibinfo {author} {\bibfnamefont {A.}~\bibnamefont
  {Richter}}, \bibinfo {author} {\bibfnamefont {H.}~\bibnamefont {Sakaguchi}},
  \bibinfo {author} {\bibfnamefont {Y.}~\bibnamefont {Shimbara}}, \bibinfo
  {author} {\bibfnamefont {Y.}~\bibnamefont {Shimizu}}, \bibinfo {author}
  {\bibfnamefont {J.}~\bibnamefont {Simonis}}, \bibinfo {author} {\bibfnamefont
  {F.~D.}\ \bibnamefont {Smit}}, \bibinfo {author} {\bibfnamefont
  {G.}~\bibnamefont {S\"usoy}}, \bibinfo {author} {\bibfnamefont
  {T.}~\bibnamefont {Suzuki}}, \bibinfo {author} {\bibfnamefont {J.~H.}\
  \bibnamefont {Thies}}, \bibinfo {author} {\bibfnamefont {M.}~\bibnamefont
  {Yosoi}}, \ and\ \bibinfo {author} {\bibfnamefont {J.}~\bibnamefont
  {Zenihiro}},\ }\href {\doibase 10.1103/PhysRevC.92.031305} {\bibfield
  {journal} {\bibinfo  {journal} {Phys. Rev. C}\ }\textbf {\bibinfo {volume}
  {92}},\ \bibinfo {pages} {031305(R)} (\bibinfo {year} {2015})}\BibitemShut
  {NoStop}%
\bibitem [{\citenamefont {Donaldson}\ \emph {et~al.}(2018)\citenamefont
  {Donaldson}, \citenamefont {Bertulani}, \citenamefont {Carter}, \citenamefont
  {Nesterenko}, \citenamefont {von Neumann-Cosel}, \citenamefont {Neveling},
  \citenamefont {Ponomarev}, \citenamefont {Reinhard}, \citenamefont {Usman},
  \citenamefont {Adsley}, \citenamefont {Br{\"{u}}mmer}, \citenamefont
  {Buthelezi}, \citenamefont {Cooper}, \citenamefont {Fearick}, \citenamefont
  {F{\"{o}}rtsch}, \citenamefont {Fujita}, \citenamefont {Fujita},
  \citenamefont {Jingo}, \citenamefont {Kleinig}, \citenamefont {Kureba},
  \citenamefont {Kvasil}, \citenamefont {Latif}, \citenamefont {Li},
  \citenamefont {Mira}, \citenamefont {Nemulodi}, \citenamefont {Papka},
  \citenamefont {Pellegri}, \citenamefont {Pietralla}, \citenamefont {Richter},
  \citenamefont {Sideras-Haddad}, \citenamefont {Smit}, \citenamefont {Steyn},
  \citenamefont {Swartz},\ and\ \citenamefont {Tamii}}]{Don18}%
  \BibitemOpen
  \bibfield  {author} {\bibinfo {author} {\bibfnamefont {L.~M.}\ \bibnamefont
  {Donaldson}}, \bibinfo {author} {\bibfnamefont {C.~A.}\ \bibnamefont
  {Bertulani}}, \bibinfo {author} {\bibfnamefont {J.}~\bibnamefont {Carter}},
  \bibinfo {author} {\bibfnamefont {V.~O.}\ \bibnamefont {Nesterenko}},
  \bibinfo {author} {\bibfnamefont {P.}~\bibnamefont {von Neumann-Cosel}},
  \bibinfo {author} {\bibfnamefont {R.}~\bibnamefont {Neveling}}, \bibinfo
  {author} {\bibfnamefont {V.~Y.}\ \bibnamefont {Ponomarev}}, \bibinfo {author}
  {\bibfnamefont {P.~G.}\ \bibnamefont {Reinhard}}, \bibinfo {author}
  {\bibfnamefont {I.~T.}\ \bibnamefont {Usman}}, \bibinfo {author}
  {\bibfnamefont {P.}~\bibnamefont {Adsley}}, \bibinfo {author} {\bibfnamefont
  {J.~W.}\ \bibnamefont {Br{\"{u}}mmer}}, \bibinfo {author} {\bibfnamefont
  {E.~Z.}\ \bibnamefont {Buthelezi}}, \bibinfo {author} {\bibfnamefont
  {G.~R.~J.}\ \bibnamefont {Cooper}}, \bibinfo {author} {\bibfnamefont {R.~W.}\
  \bibnamefont {Fearick}}, \bibinfo {author} {\bibfnamefont {S.~V.}\
  \bibnamefont {F{\"{o}}rtsch}}, \bibinfo {author} {\bibfnamefont
  {H.}~\bibnamefont {Fujita}}, \bibinfo {author} {\bibfnamefont
  {Y.}~\bibnamefont {Fujita}}, \bibinfo {author} {\bibfnamefont
  {M.}~\bibnamefont {Jingo}}, \bibinfo {author} {\bibfnamefont
  {W.}~\bibnamefont {Kleinig}}, \bibinfo {author} {\bibfnamefont {C.~O.}\
  \bibnamefont {Kureba}}, \bibinfo {author} {\bibfnamefont {J.}~\bibnamefont
  {Kvasil}}, \bibinfo {author} {\bibfnamefont {M.}~\bibnamefont {Latif}},
  \bibinfo {author} {\bibfnamefont {K.~C.~W.}\ \bibnamefont {Li}}, \bibinfo
  {author} {\bibfnamefont {J.~P.}\ \bibnamefont {Mira}}, \bibinfo {author}
  {\bibfnamefont {F.}~\bibnamefont {Nemulodi}}, \bibinfo {author}
  {\bibfnamefont {P.}~\bibnamefont {Papka}}, \bibinfo {author} {\bibfnamefont
  {L.}~\bibnamefont {Pellegri}}, \bibinfo {author} {\bibfnamefont
  {N.}~\bibnamefont {Pietralla}}, \bibinfo {author} {\bibfnamefont
  {A.}~\bibnamefont {Richter}}, \bibinfo {author} {\bibfnamefont
  {E.}~\bibnamefont {Sideras-Haddad}}, \bibinfo {author} {\bibfnamefont
  {F.~D.}\ \bibnamefont {Smit}}, \bibinfo {author} {\bibfnamefont {G.~F.}\
  \bibnamefont {Steyn}}, \bibinfo {author} {\bibfnamefont {J.~A.}\ \bibnamefont
  {Swartz}}, \ and\ \bibinfo {author} {\bibfnamefont {A.}~\bibnamefont
  {Tamii}},\ }\href {\doibase 10.1016/j.physletb.2017.11.025} {\bibfield
  {journal} {\bibinfo  {journal} {Phys. Lett. B}\ }\textbf {\bibinfo {volume}
  {776}},\ \bibinfo {pages} {133} (\bibinfo {year} {2018})}\BibitemShut
  {NoStop}%
\bibitem [{\citenamefont {Bassauer}\ \emph
  {et~al.}(2020{\natexlab{a}})\citenamefont {Bassauer}, \citenamefont {von
  Neumann-Cosel}, \citenamefont {Reinhard}, \citenamefont {Tamii},
  \citenamefont {Adachi}, \citenamefont {Bertulani}, \citenamefont {Chan},
  \citenamefont {Col\`o}, \citenamefont {D'Alessio}, \citenamefont {Fujioka},
  \citenamefont {Fujita}, \citenamefont {Fujita}, \citenamefont {Gey},
  \citenamefont {Hilcker}, \citenamefont {Hoang}, \citenamefont {Inoue},
  \citenamefont {Isaak}, \citenamefont {Iwamoto}, \citenamefont {Klaus},
  \citenamefont {Kobayashi}, \citenamefont {Maeda}, \citenamefont {Matsuda},
  \citenamefont {Nakatsuka}, \citenamefont {Noji}, \citenamefont {Ong},
  \citenamefont {Ou}, \citenamefont {Paar}, \citenamefont {Pietralla},
  \citenamefont {Ponomarev}, \citenamefont {Reen}, \citenamefont {Richter},
  \citenamefont {Roca-Maza}, \citenamefont {Singer}, \citenamefont
  {Steinhilber}, \citenamefont {Sudo}, \citenamefont {Togano}, \citenamefont
  {Tsumura}, \citenamefont {Watanabe},\ and\ \citenamefont
  {Werner}}]{Bassauer2020}%
  \BibitemOpen
  \bibfield  {author} {\bibinfo {author} {\bibfnamefont {S.}~\bibnamefont
  {Bassauer}}, \bibinfo {author} {\bibfnamefont {P.}~\bibnamefont {von
  Neumann-Cosel}}, \bibinfo {author} {\bibfnamefont {P.-G.}\ \bibnamefont
  {Reinhard}}, \bibinfo {author} {\bibfnamefont {A.}~\bibnamefont {Tamii}},
  \bibinfo {author} {\bibfnamefont {S.}~\bibnamefont {Adachi}}, \bibinfo
  {author} {\bibfnamefont {C.~A.}\ \bibnamefont {Bertulani}}, \bibinfo {author}
  {\bibfnamefont {P.~Y.}\ \bibnamefont {Chan}}, \bibinfo {author}
  {\bibfnamefont {G.}~\bibnamefont {Col\`o}}, \bibinfo {author} {\bibfnamefont
  {A.}~\bibnamefont {D'Alessio}}, \bibinfo {author} {\bibfnamefont
  {H.}~\bibnamefont {Fujioka}}, \bibinfo {author} {\bibfnamefont
  {H.}~\bibnamefont {Fujita}}, \bibinfo {author} {\bibfnamefont
  {Y.}~\bibnamefont {Fujita}}, \bibinfo {author} {\bibfnamefont
  {G.}~\bibnamefont {Gey}}, \bibinfo {author} {\bibfnamefont {M.}~\bibnamefont
  {Hilcker}}, \bibinfo {author} {\bibfnamefont {T.~H.}\ \bibnamefont {Hoang}},
  \bibinfo {author} {\bibfnamefont {A.}~\bibnamefont {Inoue}}, \bibinfo
  {author} {\bibfnamefont {J.}~\bibnamefont {Isaak}}, \bibinfo {author}
  {\bibfnamefont {C.}~\bibnamefont {Iwamoto}}, \bibinfo {author} {\bibfnamefont
  {T.}~\bibnamefont {Klaus}}, \bibinfo {author} {\bibfnamefont
  {N.}~\bibnamefont {Kobayashi}}, \bibinfo {author} {\bibfnamefont
  {Y.}~\bibnamefont {Maeda}}, \bibinfo {author} {\bibfnamefont
  {M.}~\bibnamefont {Matsuda}}, \bibinfo {author} {\bibfnamefont
  {N.}~\bibnamefont {Nakatsuka}}, \bibinfo {author} {\bibfnamefont
  {S.}~\bibnamefont {Noji}}, \bibinfo {author} {\bibfnamefont {H.~J.}\
  \bibnamefont {Ong}}, \bibinfo {author} {\bibfnamefont {I.}~\bibnamefont
  {Ou}}, \bibinfo {author} {\bibfnamefont {N.}~\bibnamefont {Paar}}, \bibinfo
  {author} {\bibfnamefont {N.}~\bibnamefont {Pietralla}}, \bibinfo {author}
  {\bibfnamefont {V.~Y.}\ \bibnamefont {Ponomarev}}, \bibinfo {author}
  {\bibfnamefont {M.~S.}\ \bibnamefont {Reen}}, \bibinfo {author}
  {\bibfnamefont {A.}~\bibnamefont {Richter}}, \bibinfo {author} {\bibfnamefont
  {X.}~\bibnamefont {Roca-Maza}}, \bibinfo {author} {\bibfnamefont
  {M.}~\bibnamefont {Singer}}, \bibinfo {author} {\bibfnamefont
  {G.}~\bibnamefont {Steinhilber}}, \bibinfo {author} {\bibfnamefont
  {T.}~\bibnamefont {Sudo}}, \bibinfo {author} {\bibfnamefont {Y.}~\bibnamefont
  {Togano}}, \bibinfo {author} {\bibfnamefont {M.}~\bibnamefont {Tsumura}},
  \bibinfo {author} {\bibfnamefont {Y.}~\bibnamefont {Watanabe}}, \ and\
  \bibinfo {author} {\bibfnamefont {V.}~\bibnamefont {Werner}},\ }\href@noop {}
  {\bibfield  {journal} {\bibinfo  {journal} {Phys. Lett. B}\ }\textbf
  {\bibinfo {volume} {812}},\ \bibinfo {pages} {135804} (\bibinfo {year}
  {2020}{\natexlab{a}})}\BibitemShut {NoStop}%
\bibitem [{\citenamefont {Bassauer}\ \emph
  {et~al.}(2020{\natexlab{b}})\citenamefont {Bassauer}, \citenamefont {von
  Neumann-Cosel}, \citenamefont {Reinhard}, \citenamefont {Tamii},
  \citenamefont {Adachi}, \citenamefont {Bertulani}, \citenamefont {Chan},
  \citenamefont {D'Alessio}, \citenamefont {Fujioka}, \citenamefont {Fujita},
  \citenamefont {Fujita}, \citenamefont {Gey}, \citenamefont {Hilcker},
  \citenamefont {Hoang}, \citenamefont {Inoue}, \citenamefont {Isaak},
  \citenamefont {Iwamoto}, \citenamefont {Klaus}, \citenamefont {Kobayashi},
  \citenamefont {Maeda}, \citenamefont {Matsuda}, \citenamefont {Nakatsuka},
  \citenamefont {Noji}, \citenamefont {Ong}, \citenamefont {Ou}, \citenamefont
  {Pietralla}, \citenamefont {Ponomarev}, \citenamefont {Reen}, \citenamefont
  {Richter}, \citenamefont {Singer}, \citenamefont {Steinhilber}, \citenamefont
  {Sudo}, \citenamefont {Togano}, \citenamefont {Tsumura}, \citenamefont
  {Watanabe},\ and\ \citenamefont {Werner}}]{Bassauer2020a}%
  \BibitemOpen
  \bibfield  {author} {\bibinfo {author} {\bibfnamefont {S.}~\bibnamefont
  {Bassauer}}, \bibinfo {author} {\bibfnamefont {P.}~\bibnamefont {von
  Neumann-Cosel}}, \bibinfo {author} {\bibfnamefont {P.-G.}\ \bibnamefont
  {Reinhard}}, \bibinfo {author} {\bibfnamefont {A.}~\bibnamefont {Tamii}},
  \bibinfo {author} {\bibfnamefont {S.}~\bibnamefont {Adachi}}, \bibinfo
  {author} {\bibfnamefont {C.~A.}\ \bibnamefont {Bertulani}}, \bibinfo {author}
  {\bibfnamefont {P.~Y.}\ \bibnamefont {Chan}}, \bibinfo {author}
  {\bibfnamefont {A.}~\bibnamefont {D'Alessio}}, \bibinfo {author}
  {\bibfnamefont {H.}~\bibnamefont {Fujioka}}, \bibinfo {author} {\bibfnamefont
  {H.}~\bibnamefont {Fujita}}, \bibinfo {author} {\bibfnamefont
  {Y.}~\bibnamefont {Fujita}}, \bibinfo {author} {\bibfnamefont
  {G.}~\bibnamefont {Gey}}, \bibinfo {author} {\bibfnamefont {M.}~\bibnamefont
  {Hilcker}}, \bibinfo {author} {\bibfnamefont {T.~H.}\ \bibnamefont {Hoang}},
  \bibinfo {author} {\bibfnamefont {A.}~\bibnamefont {Inoue}}, \bibinfo
  {author} {\bibfnamefont {J.}~\bibnamefont {Isaak}}, \bibinfo {author}
  {\bibfnamefont {C.}~\bibnamefont {Iwamoto}}, \bibinfo {author} {\bibfnamefont
  {T.}~\bibnamefont {Klaus}}, \bibinfo {author} {\bibfnamefont
  {N.}~\bibnamefont {Kobayashi}}, \bibinfo {author} {\bibfnamefont
  {Y.}~\bibnamefont {Maeda}}, \bibinfo {author} {\bibfnamefont
  {M.}~\bibnamefont {Matsuda}}, \bibinfo {author} {\bibfnamefont
  {N.}~\bibnamefont {Nakatsuka}}, \bibinfo {author} {\bibfnamefont
  {S.}~\bibnamefont {Noji}}, \bibinfo {author} {\bibfnamefont {H.~J.}\
  \bibnamefont {Ong}}, \bibinfo {author} {\bibfnamefont {I.}~\bibnamefont
  {Ou}}, \bibinfo {author} {\bibfnamefont {N.}~\bibnamefont {Pietralla}},
  \bibinfo {author} {\bibfnamefont {V.~Y.}\ \bibnamefont {Ponomarev}}, \bibinfo
  {author} {\bibfnamefont {M.~S.}\ \bibnamefont {Reen}}, \bibinfo {author}
  {\bibfnamefont {A.}~\bibnamefont {Richter}}, \bibinfo {author} {\bibfnamefont
  {M.}~\bibnamefont {Singer}}, \bibinfo {author} {\bibfnamefont
  {G.}~\bibnamefont {Steinhilber}}, \bibinfo {author} {\bibfnamefont
  {T.}~\bibnamefont {Sudo}}, \bibinfo {author} {\bibfnamefont {Y.}~\bibnamefont
  {Togano}}, \bibinfo {author} {\bibfnamefont {M.}~\bibnamefont {Tsumura}},
  \bibinfo {author} {\bibfnamefont {Y.}~\bibnamefont {Watanabe}}, \ and\
  \bibinfo {author} {\bibfnamefont {V.}~\bibnamefont {Werner}},\ }\href
  {\doibase 10.1103/PhysRevC.102.034327} {\bibfield  {journal} {\bibinfo
  {journal} {Phys. Rev. C}\ }\textbf {\bibinfo {volume} {102}},\ \bibinfo
  {pages} {034327} (\bibinfo {year} {2020}{\natexlab{b}})}\BibitemShut
  {NoStop}%
\bibitem [{\citenamefont {Lacroix}\ and\ \citenamefont
  {Chomaz}(1999)}]{Lacroix1999}%
  \BibitemOpen
  \bibfield  {author} {\bibinfo {author} {\bibfnamefont {D.}~\bibnamefont
  {Lacroix}}\ and\ \bibinfo {author} {\bibfnamefont {P.}~\bibnamefont
  {Chomaz}},\ }\href@noop {} {\bibfield  {journal} {\bibinfo  {journal} {Phys.
  Rev. C}\ }\textbf {\bibinfo {volume} {60}},\ \bibinfo {pages} {064307}
  (\bibinfo {year} {1999})}\BibitemShut {NoStop}%
\bibitem [{\citenamefont {von Neumann-Cosel}\ \emph {et~al.}(2019)\citenamefont
  {von Neumann-Cosel}, \citenamefont {Ponomarev}, \citenamefont {Richter},\
  and\ \citenamefont {Wambach}}]{vonNeumann-Cosel2019b}%
  \BibitemOpen
  \bibfield  {author} {\bibinfo {author} {\bibfnamefont {P.}~\bibnamefont {von
  Neumann-Cosel}}, \bibinfo {author} {\bibfnamefont {V.}~\bibnamefont
  {Ponomarev}}, \bibinfo {author} {\bibfnamefont {A.}~\bibnamefont {Richter}},
  \ and\ \bibinfo {author} {\bibfnamefont {J.}~\bibnamefont {Wambach}},\
  }\href@noop {} {\bibfield  {journal} {\bibinfo  {journal} {Eur. Phys. J. A}\
  }\textbf {\bibinfo {volume} {55}},\ \bibinfo {pages} {224} (\bibinfo {year}
  {2019})}\BibitemShut {NoStop}%
\bibitem [{\citenamefont {Ishkhanov}\ and\ \citenamefont
  {Troshchiev}(2011)}]{Ish11}%
  \BibitemOpen
  \bibfield  {author} {\bibinfo {author} {\bibfnamefont {B.~S.}\ \bibnamefont
  {Ishkhanov}}\ and\ \bibinfo {author} {\bibfnamefont {S.~Y.}\ \bibnamefont
  {Troshchiev}},\ }\href@noop {} {\bibfield  {journal} {\bibinfo  {journal}
  {Moscow University Physics Bulletin}\ }\textbf {\bibinfo {volume} {66}},\
  \bibinfo {pages} {325} (\bibinfo {year} {2011})}\BibitemShut {NoStop}%
\bibitem [{\citenamefont {Berman}\ and\ \citenamefont {Fultz}(1975)}]{Berm75}%
  \BibitemOpen
  \bibfield  {author} {\bibinfo {author} {\bibfnamefont {B.}~\bibnamefont
  {Berman}}\ and\ \bibinfo {author} {\bibfnamefont {S.}~\bibnamefont {Fultz}},\
  }\href@noop {} {\bibfield  {journal} {\bibinfo  {journal} {Rev. Mod. Phys.}\
  }\textbf {\bibinfo {volume} {47}},\ \bibinfo {pages} {713} (\bibinfo {year}
  {1975})}\BibitemShut {NoStop}%
\bibitem [{\citenamefont {Bengtsson}\ \emph {et~al.}(1984)\citenamefont
  {Bengtsson}, \citenamefont {M\"{o}ller}, \citenamefont {Nix},\ and\
  \citenamefont {Zhang}}]{Beng84}%
  \BibitemOpen
  \bibfield  {author} {\bibinfo {author} {\bibfnamefont {R.}~\bibnamefont
  {Bengtsson}}, \bibinfo {author} {\bibfnamefont {P.}~\bibnamefont
  {M\"{o}ller}}, \bibinfo {author} {\bibfnamefont {J.~R.}\ \bibnamefont {Nix}},
  \ and\ \bibinfo {author} {\bibfnamefont {J.~Y.}\ \bibnamefont {Zhang}},\
  }\href@noop {} {\bibfield  {journal} {\bibinfo  {journal} {Phys. Scr.}\
  }\textbf {\bibinfo {volume} {29}},\ \bibinfo {pages} {402} (\bibinfo {year}
  {1984})}\BibitemShut {NoStop}%
\bibitem [{\citenamefont {Casten}\ and\ \citenamefont {Zamfir}(2001)}]{Casten}%
  \BibitemOpen
  \bibfield  {author} {\bibinfo {author} {\bibfnamefont {R.~F.}\ \bibnamefont
  {Casten}}\ and\ \bibinfo {author} {\bibfnamefont {N.~V.}\ \bibnamefont
  {Zamfir}},\ }\href@noop {} {\bibfield  {journal} {\bibinfo  {journal} {Phys.
  Rev. Lett.}\ }\textbf {\bibinfo {volume} {87}},\ \bibinfo {pages} {052503}
  (\bibinfo {year} {2001})}\BibitemShut {NoStop}%
\bibitem [{\citenamefont {Carlos}\ \emph {et~al.}(1971)\citenamefont {Carlos},
  \citenamefont {Beil}, \citenamefont {Berg\'{e}re}, \citenamefont
  {Lepr\^{e}tre},\ and\ \citenamefont {Veyssi\'{e}re}}]{Carlos1971}%
  \BibitemOpen
  \bibfield  {author} {\bibinfo {author} {\bibfnamefont {P.}~\bibnamefont
  {Carlos}}, \bibinfo {author} {\bibfnamefont {H.}~\bibnamefont {Beil}},
  \bibinfo {author} {\bibfnamefont {R.}~\bibnamefont {Berg\'{e}re}}, \bibinfo
  {author} {\bibfnamefont {A.}~\bibnamefont {Lepr\^{e}tre}}, \ and\ \bibinfo
  {author} {\bibfnamefont {A.}~\bibnamefont {Veyssi\'{e}re}},\ }\href@noop {}
  {\bibfield  {journal} {\bibinfo  {journal} {Nucl. Phys. A}\ }\textbf
  {\bibinfo {volume} {172}},\ \bibinfo {pages} {437} (\bibinfo {year}
  {1971})}\BibitemShut {NoStop}%
\bibitem [{\citenamefont {Carlos}\ \emph {et~al.}(1974)\citenamefont {Carlos},
  \citenamefont {Beil}, \citenamefont {Berg\'{e}re}, \citenamefont
  {Lepr\^{e}tre}, \citenamefont {{De Miniac}},\ and\ \citenamefont
  {Veyssi\'{e}re}}]{Carlos1974}%
  \BibitemOpen
  \bibfield  {author} {\bibinfo {author} {\bibfnamefont {P.}~\bibnamefont
  {Carlos}}, \bibinfo {author} {\bibfnamefont {H.}~\bibnamefont {Beil}},
  \bibinfo {author} {\bibfnamefont {R.}~\bibnamefont {Berg\'{e}re}}, \bibinfo
  {author} {\bibfnamefont {A.}~\bibnamefont {Lepr\^{e}tre}}, \bibinfo {author}
  {\bibfnamefont {A.}~\bibnamefont {{De Miniac}}}, \ and\ \bibinfo {author}
  {\bibfnamefont {A.}~\bibnamefont {Veyssi\'{e}re}},\ }\href@noop {} {\bibfield
   {journal} {\bibinfo  {journal} {Nucl. Phys. A}\ }\textbf {\bibinfo {volume}
  {225}},\ \bibinfo {pages} {171} (\bibinfo {year} {1974})}\BibitemShut
  {NoStop}%
\bibitem [{\citenamefont {Donaldson}(2016)}]{Don16}%
  \BibitemOpen
  \bibfield  {author} {\bibinfo {author} {\bibfnamefont {L.~M.}\ \bibnamefont
  {Donaldson}},\ }\href@noop {} {\bibinfo {type} {{PhD} thesis}},\ \bibinfo
  {school} {University of the Witwatersrand} (\bibinfo {year}
  {2016})\BibitemShut {NoStop}%
\bibitem [{\citenamefont {Jingo}\ \emph {et~al.}(2018)\citenamefont {Jingo},
  \citenamefont {Buthelezi}, \citenamefont {Carter}, \citenamefont {Cooper},
  \citenamefont {Fearick}, \citenamefont {F{\"{o}}rtsch}, \citenamefont
  {Kureba}, \citenamefont {Krumbholz}, \citenamefont {von Neumann-Cosel},
  \citenamefont {Neveling}, \citenamefont {Papka}, \citenamefont {Poltoratska},
  \citenamefont {Ponomarev}, \citenamefont {Richter}, \citenamefont
  {Sideras-Haddad}, \citenamefont {Smit}, \citenamefont {Swartz}, \citenamefont
  {Tamii},\ and\ \citenamefont {Usman}}]{Jingo2018}%
  \BibitemOpen
  \bibfield  {author} {\bibinfo {author} {\bibfnamefont {M.}~\bibnamefont
  {Jingo}}, \bibinfo {author} {\bibfnamefont {E.~Z.}\ \bibnamefont
  {Buthelezi}}, \bibinfo {author} {\bibfnamefont {J.}~\bibnamefont {Carter}},
  \bibinfo {author} {\bibfnamefont {G.~R.~J.}\ \bibnamefont {Cooper}}, \bibinfo
  {author} {\bibfnamefont {R.~W.}\ \bibnamefont {Fearick}}, \bibinfo {author}
  {\bibfnamefont {S.~V.}\ \bibnamefont {F{\"{o}}rtsch}}, \bibinfo {author}
  {\bibfnamefont {C.~O.}\ \bibnamefont {Kureba}}, \bibinfo {author}
  {\bibfnamefont {A.~M.}\ \bibnamefont {Krumbholz}}, \bibinfo {author}
  {\bibfnamefont {P.}~\bibnamefont {von Neumann-Cosel}}, \bibinfo {author}
  {\bibfnamefont {R.}~\bibnamefont {Neveling}}, \bibinfo {author}
  {\bibfnamefont {P.}~\bibnamefont {Papka}}, \bibinfo {author} {\bibfnamefont
  {I.}~\bibnamefont {Poltoratska}}, \bibinfo {author} {\bibfnamefont {V.~Y.}\
  \bibnamefont {Ponomarev}}, \bibinfo {author} {\bibfnamefont {A.}~\bibnamefont
  {Richter}}, \bibinfo {author} {\bibfnamefont {E.}~\bibnamefont
  {Sideras-Haddad}}, \bibinfo {author} {\bibfnamefont {F.~D.}\ \bibnamefont
  {Smit}}, \bibinfo {author} {\bibfnamefont {J.~A.}\ \bibnamefont {Swartz}},
  \bibinfo {author} {\bibfnamefont {A.}~\bibnamefont {Tamii}}, \ and\ \bibinfo
  {author} {\bibfnamefont {I.~T.}\ \bibnamefont {Usman}},\ }\href {\doibase
  10.1140/epja/i2018-12664-5} {\bibfield  {journal} {\bibinfo  {journal}
  {European Physical Journal A}\ }\textbf {\bibinfo {volume} {54}},\ \bibinfo
  {pages} {234} (\bibinfo {year} {2018})}\BibitemShut {NoStop}%
\bibitem [{\citenamefont {Fujita}\ \emph {et~al.}(2002)\citenamefont {Fujita},
  \citenamefont {Fujita}, \citenamefont {Berg}, \citenamefont {Bacher},
  \citenamefont {Foster}, \citenamefont {Hara}, \citenamefont {Hatanaka},
  \citenamefont {Kawabata}, \citenamefont {Noro}, \citenamefont {Sakaguchi},
  \citenamefont {Shimbara}, \citenamefont {Shinada}, \citenamefont
  {Stephenson}, \citenamefont {Ueno},\ and\ \citenamefont
  {Yosoi}}]{Fujita2002}%
  \BibitemOpen
  \bibfield  {author} {\bibinfo {author} {\bibfnamefont {H.}~\bibnamefont
  {Fujita}}, \bibinfo {author} {\bibfnamefont {Y.}~\bibnamefont {Fujita}},
  \bibinfo {author} {\bibfnamefont {G.~P.~A.}\ \bibnamefont {Berg}}, \bibinfo
  {author} {\bibfnamefont {A.~D.}\ \bibnamefont {Bacher}}, \bibinfo {author}
  {\bibfnamefont {C.~C.}\ \bibnamefont {Foster}}, \bibinfo {author}
  {\bibfnamefont {K.}~\bibnamefont {Hara}}, \bibinfo {author} {\bibfnamefont
  {K.}~\bibnamefont {Hatanaka}}, \bibinfo {author} {\bibfnamefont
  {T.}~\bibnamefont {Kawabata}}, \bibinfo {author} {\bibfnamefont
  {T.}~\bibnamefont {Noro}}, \bibinfo {author} {\bibfnamefont {H.}~\bibnamefont
  {Sakaguchi}}, \bibinfo {author} {\bibfnamefont {Y.}~\bibnamefont {Shimbara}},
  \bibinfo {author} {\bibfnamefont {T.}~\bibnamefont {Shinada}}, \bibinfo
  {author} {\bibfnamefont {E.~J.}\ \bibnamefont {Stephenson}}, \bibinfo
  {author} {\bibfnamefont {H.}~\bibnamefont {Ueno}}, \ and\ \bibinfo {author}
  {\bibfnamefont {M.}~\bibnamefont {Yosoi}},\ }\href@noop {} {\bibfield
  {journal} {\bibinfo  {journal} {Nucl. Instrum. Meth. A}\ }\textbf {\bibinfo
  {volume} {484}},\ \bibinfo {pages} {17} (\bibinfo {year} {2002})}\BibitemShut
  {NoStop}%
\bibitem [{\citenamefont {Raynal}(2007)}]{Raynal07}%
  \BibitemOpen
  \bibfield  {author} {\bibinfo {author} {\bibfnamefont {J.}~\bibnamefont
  {Raynal}},\ }\href@noop {} {}\bibinfo {howpublished} {{Computer Code: DWBA07,
  NEA Data Service NEA1209/008}} (\bibinfo {year} {2007})\BibitemShut {NoStop}%
\bibitem [{\citenamefont {Love}\ and\ \citenamefont {Franey}(1981)}]{Love81}%
  \BibitemOpen
  \bibfield  {author} {\bibinfo {author} {\bibfnamefont {W.~G.}\ \bibnamefont
  {Love}}\ and\ \bibinfo {author} {\bibfnamefont {M.~A.}\ \bibnamefont
  {Franey}},\ }\href {\doibase 10.1103/PhysRevC.24.1073} {\bibfield  {journal}
  {\bibinfo  {journal} {Phys. Rev. C}\ }\textbf {\bibinfo {volume} {24}},\
  \bibinfo {pages} {1073} (\bibinfo {year} {1981})}\BibitemShut {NoStop}%
\bibitem [{\citenamefont {Franey}\ and\ \citenamefont {Love}(1985)}]{Franey85}%
  \BibitemOpen
  \bibfield  {author} {\bibinfo {author} {\bibfnamefont {M.~A.}\ \bibnamefont
  {Franey}}\ and\ \bibinfo {author} {\bibfnamefont {W.~G.}\ \bibnamefont
  {Love}},\ }\href {\doibase 10.1103/PhysRevC.31.488} {\bibfield  {journal}
  {\bibinfo  {journal} {Phys. Rev. C}\ }\textbf {\bibinfo {volume} {31}},\
  \bibinfo {pages} {488} (\bibinfo {year} {1985})}\BibitemShut {NoStop}%
\bibitem [{\citenamefont {Kureba}(2014)}]{Kureba2014}%
  \BibitemOpen
  \bibfield  {author} {\bibinfo {author} {\bibfnamefont {C.~O.}\ \bibnamefont
  {Kureba}},\ }\href@noop {} {\bibinfo {type} {{PhD} thesis}},\ \bibinfo
  {school} {University of the Witwatersrand} (\bibinfo {year}
  {2014})\BibitemShut {NoStop}%
\bibitem [{\citenamefont {Kureba}\ \emph {et~al.}(2018)\citenamefont {Kureba},
  \citenamefont {Buthelezi}, \citenamefont {Carter}, \citenamefont {Cooper},
  \citenamefont {Fearick}, \citenamefont {F\"{o}rtsch}, \citenamefont {Jingo},
  \citenamefont {Kleinig}, \citenamefont {Krugmann}, \citenamefont {Krumbolz},
  \citenamefont {Kvasil}, \citenamefont {Mabiala}, \citenamefont {Mira},
  \citenamefont {Nesterenko}, \citenamefont {von Neumann-Cosel}, \citenamefont
  {Neveling}, \citenamefont {Papka}, \citenamefont {Reinhard}, \citenamefont
  {Richter}, \citenamefont {Sideras-Haddad}, \citenamefont {Smit},
  \citenamefont {Steyn}, \citenamefont {Swartz}, \citenamefont {Tamii},\ and\
  \citenamefont {Usman}}]{Kureba2018}%
  \BibitemOpen
  \bibfield  {author} {\bibinfo {author} {\bibfnamefont {C.~O.}\ \bibnamefont
  {Kureba}}, \bibinfo {author} {\bibfnamefont {Z.}~\bibnamefont {Buthelezi}},
  \bibinfo {author} {\bibfnamefont {J.}~\bibnamefont {Carter}}, \bibinfo
  {author} {\bibfnamefont {G.~R.~J.}\ \bibnamefont {Cooper}}, \bibinfo {author}
  {\bibfnamefont {R.~W.}\ \bibnamefont {Fearick}}, \bibinfo {author}
  {\bibfnamefont {S.~V.}\ \bibnamefont {F\"{o}rtsch}}, \bibinfo {author}
  {\bibfnamefont {M.}~\bibnamefont {Jingo}}, \bibinfo {author} {\bibfnamefont
  {W.}~\bibnamefont {Kleinig}}, \bibinfo {author} {\bibfnamefont
  {A.}~\bibnamefont {Krugmann}}, \bibinfo {author} {\bibfnamefont {A.~M.}\
  \bibnamefont {Krumbolz}}, \bibinfo {author} {\bibfnamefont {J.}~\bibnamefont
  {Kvasil}}, \bibinfo {author} {\bibfnamefont {J.}~\bibnamefont {Mabiala}},
  \bibinfo {author} {\bibfnamefont {J.~P.}\ \bibnamefont {Mira}}, \bibinfo
  {author} {\bibfnamefont {V.~O.}\ \bibnamefont {Nesterenko}}, \bibinfo
  {author} {\bibfnamefont {P.}~\bibnamefont {von Neumann-Cosel}}, \bibinfo
  {author} {\bibfnamefont {R.}~\bibnamefont {Neveling}}, \bibinfo {author}
  {\bibfnamefont {P.}~\bibnamefont {Papka}}, \bibinfo {author} {\bibfnamefont
  {P.~G.}\ \bibnamefont {Reinhard}}, \bibinfo {author} {\bibfnamefont
  {A.}~\bibnamefont {Richter}}, \bibinfo {author} {\bibfnamefont
  {E.}~\bibnamefont {Sideras-Haddad}}, \bibinfo {author} {\bibfnamefont
  {F.~D.}\ \bibnamefont {Smit}}, \bibinfo {author} {\bibfnamefont {G.~F.}\
  \bibnamefont {Steyn}}, \bibinfo {author} {\bibfnamefont {J.~A.}\ \bibnamefont
  {Swartz}}, \bibinfo {author} {\bibfnamefont {A.}~\bibnamefont {Tamii}}, \
  and\ \bibinfo {author} {\bibfnamefont {I.~T.}\ \bibnamefont {Usman}},\ }\href
  {\doibase https://doi.org/10.1016/j.physletb.2018.02.013} {\bibfield
  {journal} {\bibinfo  {journal} {Phys. Lett. B}\ }\textbf {\bibinfo {volume}
  {779}},\ \bibinfo {pages} {269 } (\bibinfo {year} {2018})}\BibitemShut
  {NoStop}%
\bibitem [{\citenamefont {Shevchenko}\ \emph {et~al.}(2004)\citenamefont
  {Shevchenko}, \citenamefont {Carter}, \citenamefont {Fearick}, \citenamefont
  {F\"ortsch}, \citenamefont {Fujita}, \citenamefont {Fujita}, \citenamefont
  {Kalmykov}, \citenamefont {Lacroix}, \citenamefont {Lawrie}, \citenamefont
  {von Neumann-Cosel}, \citenamefont {Neveling}, \citenamefont {Ponomarev},
  \citenamefont {Richter}, \citenamefont {Sideras-Haddad}, \citenamefont
  {Smit},\ and\ \citenamefont {Wambach}}]{Shevchenko2004}%
  \BibitemOpen
  \bibfield  {author} {\bibinfo {author} {\bibfnamefont {A.}~\bibnamefont
  {Shevchenko}}, \bibinfo {author} {\bibfnamefont {J.}~\bibnamefont {Carter}},
  \bibinfo {author} {\bibfnamefont {R.~W.}\ \bibnamefont {Fearick}}, \bibinfo
  {author} {\bibfnamefont {S.~V.}\ \bibnamefont {F\"ortsch}}, \bibinfo {author}
  {\bibfnamefont {H.}~\bibnamefont {Fujita}}, \bibinfo {author} {\bibfnamefont
  {Y.}~\bibnamefont {Fujita}}, \bibinfo {author} {\bibfnamefont
  {Y.}~\bibnamefont {Kalmykov}}, \bibinfo {author} {\bibfnamefont
  {D.}~\bibnamefont {Lacroix}}, \bibinfo {author} {\bibfnamefont {J.~J.}\
  \bibnamefont {Lawrie}}, \bibinfo {author} {\bibfnamefont {P.}~\bibnamefont
  {von Neumann-Cosel}}, \bibinfo {author} {\bibfnamefont {R.}~\bibnamefont
  {Neveling}}, \bibinfo {author} {\bibfnamefont {V.~Y.}\ \bibnamefont
  {Ponomarev}}, \bibinfo {author} {\bibfnamefont {A.}~\bibnamefont {Richter}},
  \bibinfo {author} {\bibfnamefont {E.}~\bibnamefont {Sideras-Haddad}},
  \bibinfo {author} {\bibfnamefont {F.~D.}\ \bibnamefont {Smit}}, \ and\
  \bibinfo {author} {\bibfnamefont {J.}~\bibnamefont {Wambach}},\ }\href@noop
  {} {\bibfield  {journal} {\bibinfo  {journal} {Phys. Rev. Lett.}\ }\textbf
  {\bibinfo {volume} {93}},\ \bibinfo {pages} {122501} (\bibinfo {year}
  {2004})}\BibitemShut {NoStop}%
\bibitem [{\citenamefont {Shevchenko}\ \emph {et~al.}(2009)\citenamefont
  {Shevchenko}, \citenamefont {Burda}, \citenamefont {Carter}, \citenamefont
  {Cooper}, \citenamefont {Fearick}, \citenamefont {F\"{o}rtsch}, \citenamefont
  {Fujita}, \citenamefont {Fujita}, \citenamefont {Kalmykov}, \citenamefont
  {Lacroix}, \citenamefont {Lawrie}, \citenamefont {von Neumann-Cosel},
  \citenamefont {Neveling}, \citenamefont {Ponomarev}, \citenamefont {Richter},
  \citenamefont {Sideras-Haddad}, \citenamefont {Smit},\ and\ \citenamefont
  {Wambach}}]{Shev2009}%
  \BibitemOpen
  \bibfield  {author} {\bibinfo {author} {\bibfnamefont {A.}~\bibnamefont
  {Shevchenko}}, \bibinfo {author} {\bibfnamefont {O.}~\bibnamefont {Burda}},
  \bibinfo {author} {\bibfnamefont {J.}~\bibnamefont {Carter}}, \bibinfo
  {author} {\bibfnamefont {G.~R.~J.}\ \bibnamefont {Cooper}}, \bibinfo {author}
  {\bibfnamefont {R.~W.}\ \bibnamefont {Fearick}}, \bibinfo {author}
  {\bibfnamefont {S.~V.}\ \bibnamefont {F\"{o}rtsch}}, \bibinfo {author}
  {\bibfnamefont {H.}~\bibnamefont {Fujita}}, \bibinfo {author} {\bibfnamefont
  {Y.}~\bibnamefont {Fujita}}, \bibinfo {author} {\bibfnamefont
  {Y.}~\bibnamefont {Kalmykov}}, \bibinfo {author} {\bibfnamefont
  {D.}~\bibnamefont {Lacroix}}, \bibinfo {author} {\bibfnamefont {J.~J.}\
  \bibnamefont {Lawrie}}, \bibinfo {author} {\bibfnamefont {P.}~\bibnamefont
  {von Neumann-Cosel}}, \bibinfo {author} {\bibfnamefont {R.}~\bibnamefont
  {Neveling}}, \bibinfo {author} {\bibfnamefont {V.~Y.}\ \bibnamefont
  {Ponomarev}}, \bibinfo {author} {\bibfnamefont {A.}~\bibnamefont {Richter}},
  \bibinfo {author} {\bibfnamefont {E.}~\bibnamefont {Sideras-Haddad}},
  \bibinfo {author} {\bibfnamefont {F.~D.}\ \bibnamefont {Smit}}, \ and\
  \bibinfo {author} {\bibfnamefont {J.}~\bibnamefont {Wambach}},\ }\href@noop
  {} {\bibfield  {journal} {\bibinfo  {journal} {Phys. Rev. C}\ }\textbf
  {\bibinfo {volume} {79}},\ \bibinfo {pages} {044305} (\bibinfo {year}
  {2009})}\BibitemShut {NoStop}%
\bibitem [{\citenamefont {Usman}\ \emph {et~al.}(2011)\citenamefont {Usman},
  \citenamefont {Buthelezi}, \citenamefont {Carter}, \citenamefont {Cooper},
  \citenamefont {Fearick}, \citenamefont {F\"ortsch}, \citenamefont {Fujita},
  \citenamefont {Fujita}, \citenamefont {Kalmykov}, \citenamefont {von
  Neumann-Cosel}, \citenamefont {Neveling}, \citenamefont {Papakonstantinou},
  \citenamefont {Richter}, \citenamefont {Roth}, \citenamefont {Shevchenko},
  \citenamefont {Sideras-Haddad},\ and\ \citenamefont {Smit}}]{Usman2011191}%
  \BibitemOpen
  \bibfield  {author} {\bibinfo {author} {\bibfnamefont {I.~T.}\ \bibnamefont
  {Usman}}, \bibinfo {author} {\bibfnamefont {Z.}~\bibnamefont {Buthelezi}},
  \bibinfo {author} {\bibfnamefont {J.}~\bibnamefont {Carter}}, \bibinfo
  {author} {\bibfnamefont {G.~R.~J.}\ \bibnamefont {Cooper}}, \bibinfo {author}
  {\bibfnamefont {R.~W.}\ \bibnamefont {Fearick}}, \bibinfo {author}
  {\bibfnamefont {S.~V.}\ \bibnamefont {F\"ortsch}}, \bibinfo {author}
  {\bibfnamefont {H.}~\bibnamefont {Fujita}}, \bibinfo {author} {\bibfnamefont
  {Y.}~\bibnamefont {Fujita}}, \bibinfo {author} {\bibfnamefont
  {Y.}~\bibnamefont {Kalmykov}}, \bibinfo {author} {\bibfnamefont
  {P.}~\bibnamefont {von Neumann-Cosel}}, \bibinfo {author} {\bibfnamefont
  {R.}~\bibnamefont {Neveling}}, \bibinfo {author} {\bibfnamefont
  {P.}~\bibnamefont {Papakonstantinou}}, \bibinfo {author} {\bibfnamefont
  {A.}~\bibnamefont {Richter}}, \bibinfo {author} {\bibfnamefont
  {R.}~\bibnamefont {Roth}}, \bibinfo {author} {\bibfnamefont {A.}~\bibnamefont
  {Shevchenko}}, \bibinfo {author} {\bibfnamefont {E.}~\bibnamefont
  {Sideras-Haddad}}, \ and\ \bibinfo {author} {\bibfnamefont {F.~D.}\
  \bibnamefont {Smit}},\ }\href@noop {} {\bibfield  {journal} {\bibinfo
  {journal} {Phys. Lett. B}\ }\textbf {\bibinfo {volume} {698}},\ \bibinfo
  {pages} {191} (\bibinfo {year} {2011})}\BibitemShut {NoStop}%
\bibitem [{\citenamefont {Itoh}\ \emph {et~al.}(2003)\citenamefont {Itoh},
  \citenamefont {Sakaguchi}, \citenamefont {Uchida}, \citenamefont {Ishikawa},
  \citenamefont {Kawabata}, \citenamefont {Murakami}, \citenamefont {Takeda},
  \citenamefont {Taki}, \citenamefont {Terashima}, \citenamefont {Tsukahara},
  \citenamefont {Yasuda}, \citenamefont {Yosoi}, \citenamefont {Garg},
  \citenamefont {Hedden}, \citenamefont {Kharraja}, \citenamefont {Koss},
  \citenamefont {Nayak}, \citenamefont {Zhu}, \citenamefont {Fujimura},
  \citenamefont {Fujiwara}, \citenamefont {Hara}, \citenamefont {Yoshida},
  \citenamefont {Akimune}, \citenamefont {Harakeh},\ and\ \citenamefont
  {Volkerts}}]{Itoh2003}%
  \BibitemOpen
  \bibfield  {author} {\bibinfo {author} {\bibfnamefont {M.}~\bibnamefont
  {Itoh}}, \bibinfo {author} {\bibfnamefont {H.}~\bibnamefont {Sakaguchi}},
  \bibinfo {author} {\bibfnamefont {M.}~\bibnamefont {Uchida}}, \bibinfo
  {author} {\bibfnamefont {T.}~\bibnamefont {Ishikawa}}, \bibinfo {author}
  {\bibfnamefont {T.}~\bibnamefont {Kawabata}}, \bibinfo {author}
  {\bibfnamefont {T.}~\bibnamefont {Murakami}}, \bibinfo {author}
  {\bibfnamefont {H.}~\bibnamefont {Takeda}}, \bibinfo {author} {\bibfnamefont
  {T.}~\bibnamefont {Taki}}, \bibinfo {author} {\bibfnamefont {S.}~\bibnamefont
  {Terashima}}, \bibinfo {author} {\bibfnamefont {N.}~\bibnamefont
  {Tsukahara}}, \bibinfo {author} {\bibfnamefont {Y.}~\bibnamefont {Yasuda}},
  \bibinfo {author} {\bibfnamefont {M.}~\bibnamefont {Yosoi}}, \bibinfo
  {author} {\bibfnamefont {U.}~\bibnamefont {Garg}}, \bibinfo {author}
  {\bibfnamefont {M.}~\bibnamefont {Hedden}}, \bibinfo {author} {\bibfnamefont
  {B.}~\bibnamefont {Kharraja}}, \bibinfo {author} {\bibfnamefont
  {M.}~\bibnamefont {Koss}}, \bibinfo {author} {\bibfnamefont {B.~K.}\
  \bibnamefont {Nayak}}, \bibinfo {author} {\bibfnamefont {S.}~\bibnamefont
  {Zhu}}, \bibinfo {author} {\bibfnamefont {H.}~\bibnamefont {Fujimura}},
  \bibinfo {author} {\bibfnamefont {M.}~\bibnamefont {Fujiwara}}, \bibinfo
  {author} {\bibfnamefont {K.}~\bibnamefont {Hara}}, \bibinfo {author}
  {\bibfnamefont {H.~P.}\ \bibnamefont {Yoshida}}, \bibinfo {author}
  {\bibfnamefont {H.}~\bibnamefont {Akimune}}, \bibinfo {author} {\bibfnamefont
  {M.~N.}\ \bibnamefont {Harakeh}}, \ and\ \bibinfo {author} {\bibfnamefont
  {M.}~\bibnamefont {Volkerts}},\ }\href {\doibase 10.1103/PhysRevC.68.064602}
  {\bibfield  {journal} {\bibinfo  {journal} {Phys. Rev. C}\ }\textbf {\bibinfo
  {volume} {68}},\ \bibinfo {pages} {064602} (\bibinfo {year}
  {2003})}\BibitemShut {NoStop}%
\bibitem [{\citenamefont {Poltoratska}\ \emph {et~al.}(2014)\citenamefont
  {Poltoratska}, \citenamefont {Fearick}, \citenamefont {Krumbholz},
  \citenamefont {Litvinova}, \citenamefont {Matsubara}, \citenamefont {von
  Neumann-Cosel}, \citenamefont {Ponomarev}, \citenamefont {Richter},\ and\
  \citenamefont {Tamii}}]{Pol14}%
  \BibitemOpen
  \bibfield  {author} {\bibinfo {author} {\bibfnamefont {I.}~\bibnamefont
  {Poltoratska}}, \bibinfo {author} {\bibfnamefont {R.~W.}\ \bibnamefont
  {Fearick}}, \bibinfo {author} {\bibfnamefont {A.~M.}\ \bibnamefont
  {Krumbholz}}, \bibinfo {author} {\bibfnamefont {E.}~\bibnamefont
  {Litvinova}}, \bibinfo {author} {\bibfnamefont {H.}~\bibnamefont
  {Matsubara}}, \bibinfo {author} {\bibfnamefont {P.}~\bibnamefont {von
  Neumann-Cosel}}, \bibinfo {author} {\bibfnamefont {V.~Y.}\ \bibnamefont
  {Ponomarev}}, \bibinfo {author} {\bibfnamefont {A.}~\bibnamefont {Richter}},
  \ and\ \bibinfo {author} {\bibfnamefont {A.}~\bibnamefont {Tamii}},\ }\href
  {\doibase 10.1103/PhysRevC.89.054322} {\bibfield  {journal} {\bibinfo
  {journal} {Phys. Rev. C}\ }\textbf {\bibinfo {volume} {89}},\ \bibinfo
  {pages} {054322} (\bibinfo {year} {2014})}\BibitemShut {NoStop}%
\bibitem [{\citenamefont {Bertulani}\ and\ \citenamefont
  {Baur}(1988)}]{Bert88}%
  \BibitemOpen
  \bibfield  {author} {\bibinfo {author} {\bibfnamefont {C.}~\bibnamefont
  {Bertulani}}\ and\ \bibinfo {author} {\bibfnamefont {G.}~\bibnamefont
  {Baur}},\ }\href@noop {} {\bibfield  {journal} {\bibinfo  {journal} {Phys.
  Rep.}\ }\textbf {\bibinfo {volume} {163}},\ \bibinfo {pages} {299} (\bibinfo
  {year} {1988})}\BibitemShut {NoStop}%
\bibitem [{\citenamefont {Bertulani}(2009)}]{Bertulani2009}%
  \BibitemOpen
  \bibfield  {author} {\bibinfo {author} {\bibfnamefont {C.}~\bibnamefont
  {Bertulani}},\ }\href@noop {} {} (\bibinfo {year} {2009}),\ \Eprint
  {http://arxiv.org/abs/0908.4307} {arXiv:0908.4307 [nucl-th]} \BibitemShut
  {NoStop}%
\bibitem [{\citenamefont {Bertulani}\ and\ \citenamefont
  {Nathan}(1993)}]{Bert93}%
  \BibitemOpen
  \bibfield  {author} {\bibinfo {author} {\bibfnamefont {C.}~\bibnamefont
  {Bertulani}}\ and\ \bibinfo {author} {\bibfnamefont {A.}~\bibnamefont
  {Nathan}},\ }\href@noop {} {\bibfield  {journal} {\bibinfo  {journal} {Nucl.
  Phys. A}\ }\textbf {\bibinfo {volume} {554}},\ \bibinfo {pages} {158}
  (\bibinfo {year} {1993})}\BibitemShut {NoStop}%
\bibitem [{\citenamefont {Bertulani}(1993)}]{Bertulani1993}%
  \BibitemOpen
  \bibfield  {author} {\bibinfo {author} {\bibfnamefont {C.~A.}\ \bibnamefont
  {Bertulani}},\ }\href {\doibase https://doi.org/10.1016/0370-2693(93)91745-9}
  {\bibfield  {journal} {\bibinfo  {journal} {Phys. Lett. B}\ }\textbf
  {\bibinfo {volume} {319}},\ \bibinfo {pages} {421 } (\bibinfo {year}
  {1993})}\BibitemShut {NoStop}%
\bibitem [{\citenamefont {Neveling}\ \emph
  {et~al.}(2011{\natexlab{b}})\citenamefont {Neveling}, \citenamefont {Fujita},
  \citenamefont {Smit}, \citenamefont {Adachi}, \citenamefont {Berg},
  \citenamefont {Buthelezi}, \citenamefont {Carter}, \citenamefont {Conradie},
  \citenamefont {Couder}, \citenamefont {Fearick}, \citenamefont {F\"{o}rtsch},
  \citenamefont {Fourie}, \citenamefont {Fujita}, \citenamefont {G\"{o}rres},
  \citenamefont {Hatanaka}, \citenamefont {Heilmann}, \citenamefont {Mira},
  \citenamefont {Murray}, \citenamefont {von Neumann-Cosel}, \citenamefont
  {O'Brien}, \citenamefont {Papka}, \citenamefont {Poltoratska}, \citenamefont
  {Richter}, \citenamefont {Sideras-Haddad}, \citenamefont {Swartz},
  \citenamefont {Tamii}, \citenamefont {Usman},\ and\ \citenamefont {van
  Zyl}}]{Neveling2011}%
  \BibitemOpen
  \bibfield  {author} {\bibinfo {author} {\bibfnamefont {R.}~\bibnamefont
  {Neveling}}, \bibinfo {author} {\bibfnamefont {H.}~\bibnamefont {Fujita}},
  \bibinfo {author} {\bibfnamefont {F.~D.}\ \bibnamefont {Smit}}, \bibinfo
  {author} {\bibfnamefont {T.}~\bibnamefont {Adachi}}, \bibinfo {author}
  {\bibfnamefont {G.~P.~A.}\ \bibnamefont {Berg}}, \bibinfo {author}
  {\bibfnamefont {E.~Z.}\ \bibnamefont {Buthelezi}}, \bibinfo {author}
  {\bibfnamefont {J.}~\bibnamefont {Carter}}, \bibinfo {author} {\bibfnamefont
  {J.~L.}\ \bibnamefont {Conradie}}, \bibinfo {author} {\bibfnamefont
  {M.}~\bibnamefont {Couder}}, \bibinfo {author} {\bibfnamefont {R.~W.}\
  \bibnamefont {Fearick}}, \bibinfo {author} {\bibfnamefont {S.~V.}\
  \bibnamefont {F\"{o}rtsch}}, \bibinfo {author} {\bibfnamefont
  {D.}~\bibnamefont {Fourie}}, \bibinfo {author} {\bibfnamefont
  {Y.}~\bibnamefont {Fujita}}, \bibinfo {author} {\bibfnamefont
  {J.}~\bibnamefont {G\"{o}rres}}, \bibinfo {author} {\bibfnamefont
  {K.}~\bibnamefont {Hatanaka}}, \bibinfo {author} {\bibfnamefont {A.~M.}\
  \bibnamefont {Heilmann}}, \bibinfo {author} {\bibfnamefont {J.~P.}\
  \bibnamefont {Mira}}, \bibinfo {author} {\bibfnamefont {S.~H.~T.}\
  \bibnamefont {Murray}}, \bibinfo {author} {\bibfnamefont {P.}~\bibnamefont
  {von Neumann-Cosel}}, \bibinfo {author} {\bibfnamefont {S.}~\bibnamefont
  {O'Brien}}, \bibinfo {author} {\bibfnamefont {P.}~\bibnamefont {Papka}},
  \bibinfo {author} {\bibfnamefont {I.}~\bibnamefont {Poltoratska}}, \bibinfo
  {author} {\bibfnamefont {A.}~\bibnamefont {Richter}}, \bibinfo {author}
  {\bibfnamefont {E.}~\bibnamefont {Sideras-Haddad}}, \bibinfo {author}
  {\bibfnamefont {J.~A.}\ \bibnamefont {Swartz}}, \bibinfo {author}
  {\bibfnamefont {A.}~\bibnamefont {Tamii}}, \bibinfo {author} {\bibfnamefont
  {I.~T.}\ \bibnamefont {Usman}}, \ and\ \bibinfo {author} {\bibfnamefont
  {J.~J.}\ \bibnamefont {van Zyl}},\ }\href@noop {} {\bibfield  {journal}
  {\bibinfo  {journal} {J. Phys. Conf. Series}\ }\textbf {\bibinfo {volume}
  {312}},\ \bibinfo {pages} {052016} (\bibinfo {year}
  {2011}{\natexlab{b}})}\BibitemShut {NoStop}%
\bibitem [{\citenamefont {Graps}(1995)}]{Gra95}%
  \BibitemOpen
  \bibfield  {author} {\bibinfo {author} {\bibfnamefont {A.}~\bibnamefont
  {Graps}},\ }\href@noop {} {\bibfield  {journal} {\bibinfo  {journal} {IEEE
  Comput. Sci. Eng.}\ }\textbf {\bibinfo {volume} {2}},\ \bibinfo {pages} {50}
  (\bibinfo {year} {1995})}\BibitemShut {NoStop}%
\bibitem [{\citenamefont {Shevchenko}\ \emph {et~al.}(2008)\citenamefont
  {Shevchenko}, \citenamefont {Carter}, \citenamefont {Cooper}, \citenamefont
  {Fearick}, \citenamefont {Kalmykov}, \citenamefont {von Neumann-Cosel},
  \citenamefont {Ponomarev}, \citenamefont {Richter}, \citenamefont {Usman},\
  and\ \citenamefont {Wambach}}]{Shevchenko2008}%
  \BibitemOpen
  \bibfield  {author} {\bibinfo {author} {\bibfnamefont {A.}~\bibnamefont
  {Shevchenko}}, \bibinfo {author} {\bibfnamefont {J.}~\bibnamefont {Carter}},
  \bibinfo {author} {\bibfnamefont {G.~R.~J.}\ \bibnamefont {Cooper}}, \bibinfo
  {author} {\bibfnamefont {R.~W.}\ \bibnamefont {Fearick}}, \bibinfo {author}
  {\bibfnamefont {Y.}~\bibnamefont {Kalmykov}}, \bibinfo {author}
  {\bibfnamefont {P.}~\bibnamefont {von Neumann-Cosel}}, \bibinfo {author}
  {\bibfnamefont {V.~Y.}\ \bibnamefont {Ponomarev}}, \bibinfo {author}
  {\bibfnamefont {A.}~\bibnamefont {Richter}}, \bibinfo {author} {\bibfnamefont
  {I.}~\bibnamefont {Usman}}, \ and\ \bibinfo {author} {\bibfnamefont
  {J.}~\bibnamefont {Wambach}},\ }\href@noop {} {\bibfield  {journal} {\bibinfo
   {journal} {Phys. Rev. C}\ }\textbf {\bibinfo {volume} {77}},\ \bibinfo
  {pages} {024302} (\bibinfo {year} {2008})}\BibitemShut {NoStop}%
\bibitem [{\citenamefont {Cooper}\ and\ \citenamefont {Cowan}(2008)}]{Coo08}%
  \BibitemOpen
  \bibfield  {author} {\bibinfo {author} {\bibfnamefont {G.~R.~J.}\
  \bibnamefont {Cooper}}\ and\ \bibinfo {author} {\bibfnamefont {D.~R.}\
  \bibnamefont {Cowan}},\ }\href@noop {} {\bibfield  {journal} {\bibinfo
  {journal} {Comput. Geosci.}\ }\textbf {\bibinfo {volume} {34}},\ \bibinfo
  {pages} {95} (\bibinfo {year} {2008})}\BibitemShut {NoStop}%
\bibitem [{\citenamefont {Teolis}(1998)}]{Teolis98}%
  \BibitemOpen
  \bibfield  {author} {\bibinfo {author} {\bibfnamefont {A.}~\bibnamefont
  {Teolis}},\ }\enquote {\bibinfo {title} {{Continuous Wavelet and Gabor
  Transforms}},}\ in\ \href@noop {} {\emph {\bibinfo {booktitle} {Computational
  Signal Processing with Wavelets}}}\ (\bibinfo  {publisher} {Birkhauser
  Boston},\ \bibinfo {year} {1998})\ pp.\ \bibinfo {pages} {59--88}\BibitemShut
  {NoStop}%
\bibitem [{\citenamefont {Petermann}\ \emph {et~al.}(2010)\citenamefont
  {Petermann}, \citenamefont {Langanke}, \citenamefont {Mart\'{\i}nez-Pinedo},
  \citenamefont {von Neumann-Cosel}, \citenamefont {Nowacki},\ and\
  \citenamefont {Richter}}]{Petermann2010}%
  \BibitemOpen
  \bibfield  {author} {\bibinfo {author} {\bibfnamefont {I.}~\bibnamefont
  {Petermann}}, \bibinfo {author} {\bibfnamefont {K.}~\bibnamefont {Langanke}},
  \bibinfo {author} {\bibfnamefont {G.}~\bibnamefont {Mart\'{\i}nez-Pinedo}},
  \bibinfo {author} {\bibfnamefont {P.}~\bibnamefont {von Neumann-Cosel}},
  \bibinfo {author} {\bibfnamefont {F.}~\bibnamefont {Nowacki}}, \ and\
  \bibinfo {author} {\bibfnamefont {A.}~\bibnamefont {Richter}},\ }\href
  {\doibase 10.1103/PhysRevC.81.014308} {\bibfield  {journal} {\bibinfo
  {journal} {Phys. Rev. C}\ }\textbf {\bibinfo {volume} {81}},\ \bibinfo
  {pages} {014308} (\bibinfo {year} {2010})}\BibitemShut {NoStop}%
\bibitem [{\citenamefont {Usman}\ \emph {et~al.}(2016)\citenamefont {Usman},
  \citenamefont {Buthelezi}, \citenamefont {Carter}, \citenamefont {Cooper},
  \citenamefont {Fearick}, \citenamefont {F\"ortsch}, \citenamefont {Fujita},
  \citenamefont {Fujita}, \citenamefont {von Neumann-Cosel}, \citenamefont
  {Neveling}, \citenamefont {Papakonstantinou}, \citenamefont {Pysmenetska},
  \citenamefont {Richter}, \citenamefont {Roth}, \citenamefont
  {Sideras-Haddad},\ and\ \citenamefont {Smit}}]{Usman2016}%
  \BibitemOpen
  \bibfield  {author} {\bibinfo {author} {\bibfnamefont {I.~T.}\ \bibnamefont
  {Usman}}, \bibinfo {author} {\bibfnamefont {Z.}~\bibnamefont {Buthelezi}},
  \bibinfo {author} {\bibfnamefont {J.}~\bibnamefont {Carter}}, \bibinfo
  {author} {\bibfnamefont {G.~R.~J.}\ \bibnamefont {Cooper}}, \bibinfo {author}
  {\bibfnamefont {R.~W.}\ \bibnamefont {Fearick}}, \bibinfo {author}
  {\bibfnamefont {S.~V.}\ \bibnamefont {F\"ortsch}}, \bibinfo {author}
  {\bibfnamefont {H.}~\bibnamefont {Fujita}}, \bibinfo {author} {\bibfnamefont
  {Y.}~\bibnamefont {Fujita}}, \bibinfo {author} {\bibfnamefont
  {P.}~\bibnamefont {von Neumann-Cosel}}, \bibinfo {author} {\bibfnamefont
  {R.}~\bibnamefont {Neveling}}, \bibinfo {author} {\bibfnamefont
  {P.}~\bibnamefont {Papakonstantinou}}, \bibinfo {author} {\bibfnamefont
  {I.}~\bibnamefont {Pysmenetska}}, \bibinfo {author} {\bibfnamefont
  {A.}~\bibnamefont {Richter}}, \bibinfo {author} {\bibfnamefont
  {R.}~\bibnamefont {Roth}}, \bibinfo {author} {\bibfnamefont {E.}~\bibnamefont
  {Sideras-Haddad}}, \ and\ \bibinfo {author} {\bibfnamefont {F.~D.}\
  \bibnamefont {Smit}},\ }\href {\doibase 10.1103/PhysRevC.94.024308}
  {\bibfield  {journal} {\bibinfo  {journal} {Phys. Rev. C}\ }\textbf {\bibinfo
  {volume} {94}},\ \bibinfo {pages} {024308} (\bibinfo {year}
  {2016})}\BibitemShut {NoStop}%
\bibitem [{\citenamefont {Fearick}\ \emph {et~al.}(2018)\citenamefont
  {Fearick}, \citenamefont {Erler}, \citenamefont {Matsubara}, \citenamefont
  {von Neumann-Cosel}, \citenamefont {Richter}, \citenamefont {Roth},\ and\
  \citenamefont {Tamii}}]{Fearick2018}%
  \BibitemOpen
  \bibfield  {author} {\bibinfo {author} {\bibfnamefont {R.~W.}\ \bibnamefont
  {Fearick}}, \bibinfo {author} {\bibfnamefont {B.}~\bibnamefont {Erler}},
  \bibinfo {author} {\bibfnamefont {H.}~\bibnamefont {Matsubara}}, \bibinfo
  {author} {\bibfnamefont {P.}~\bibnamefont {von Neumann-Cosel}}, \bibinfo
  {author} {\bibfnamefont {A.}~\bibnamefont {Richter}}, \bibinfo {author}
  {\bibfnamefont {R.}~\bibnamefont {Roth}}, \ and\ \bibinfo {author}
  {\bibfnamefont {A.}~\bibnamefont {Tamii}},\ }\href {\doibase
  10.1103/PhysRevC.97.044325} {\bibfield  {journal} {\bibinfo  {journal} {Phys.
  Rev. C}\ }\textbf {\bibinfo {volume} {97}},\ \bibinfo {pages} {044325}
  (\bibinfo {year} {2018})}\BibitemShut {NoStop}%
\bibitem [{\citenamefont {Kalmykov}\ \emph {et~al.}(2006)\citenamefont
  {Kalmykov}, \citenamefont {Adachi}, \citenamefont {Berg}, \citenamefont
  {Fujita}, \citenamefont {Fujita}, \citenamefont {Fujita}, \citenamefont
  {Hatanaka}, \citenamefont {Kamiya}, \citenamefont {Nakanishi}, \citenamefont
  {von Neumann-Cosel}, \citenamefont {Ponomarev}, \citenamefont {Richter},
  \citenamefont {Sakamoto}, \citenamefont {Sakemi}, \citenamefont {Shevchenko},
  \citenamefont {Shimbara}, \citenamefont {Shimizu}, \citenamefont {Smit},
  \citenamefont {Wakasa}, \citenamefont {Wambach},\ and\ \citenamefont
  {Yosoi}}]{Kalmykov2006}%
  \BibitemOpen
  \bibfield  {author} {\bibinfo {author} {\bibfnamefont {Y.}~\bibnamefont
  {Kalmykov}}, \bibinfo {author} {\bibfnamefont {T.}~\bibnamefont {Adachi}},
  \bibinfo {author} {\bibfnamefont {G.~P.~A.}\ \bibnamefont {Berg}}, \bibinfo
  {author} {\bibfnamefont {H.}~\bibnamefont {Fujita}}, \bibinfo {author}
  {\bibfnamefont {K.}~\bibnamefont {Fujita}}, \bibinfo {author} {\bibfnamefont
  {Y.}~\bibnamefont {Fujita}}, \bibinfo {author} {\bibfnamefont
  {K.}~\bibnamefont {Hatanaka}}, \bibinfo {author} {\bibfnamefont
  {J.}~\bibnamefont {Kamiya}}, \bibinfo {author} {\bibfnamefont
  {K.}~\bibnamefont {Nakanishi}}, \bibinfo {author} {\bibfnamefont
  {P.}~\bibnamefont {von Neumann-Cosel}}, \bibinfo {author} {\bibfnamefont
  {V.~Y.}\ \bibnamefont {Ponomarev}}, \bibinfo {author} {\bibfnamefont
  {A.}~\bibnamefont {Richter}}, \bibinfo {author} {\bibfnamefont
  {N.}~\bibnamefont {Sakamoto}}, \bibinfo {author} {\bibfnamefont
  {Y.}~\bibnamefont {Sakemi}}, \bibinfo {author} {\bibfnamefont
  {A.}~\bibnamefont {Shevchenko}}, \bibinfo {author} {\bibfnamefont
  {Y.}~\bibnamefont {Shimbara}}, \bibinfo {author} {\bibfnamefont
  {Y.}~\bibnamefont {Shimizu}}, \bibinfo {author} {\bibfnamefont {F.~D.}\
  \bibnamefont {Smit}}, \bibinfo {author} {\bibfnamefont {T.}~\bibnamefont
  {Wakasa}}, \bibinfo {author} {\bibfnamefont {J.}~\bibnamefont {Wambach}}, \
  and\ \bibinfo {author} {\bibfnamefont {M.}~\bibnamefont {Yosoi}},\
  }\href@noop {} {\bibfield  {journal} {\bibinfo  {journal} {Phys. Rev. Lett.}\
  }\textbf {\bibinfo {volume} {96}},\ \bibinfo {pages} {012502} (\bibinfo
  {year} {2006})}\BibitemShut {NoStop}%
\bibitem [{\citenamefont {Wakasa}\ \emph {et~al.}(2007)\citenamefont {Wakasa},
  \citenamefont {Ihara}, \citenamefont {Fujita}, \citenamefont {Funaki},
  \citenamefont {Hatanaka}, \citenamefont {Horiuchi}, \citenamefont {Itoh},
  \citenamefont {Kamiya}, \citenamefont {R\"{o}pke}, \citenamefont {Sakaguchi},
  \citenamefont {Sakamoto}, \citenamefont {Sakemi}, \citenamefont {Schuck},
  \citenamefont {Shimizu}, \citenamefont {Takashina}, \citenamefont
  {Terashima}, \citenamefont {Tohsaki}, \citenamefont {Uchida}, \citenamefont
  {Yoshida},\ and\ \citenamefont {Yosoi}}]{Wakasa2007}%
  \BibitemOpen
  \bibfield  {author} {\bibinfo {author} {\bibfnamefont {T.}~\bibnamefont
  {Wakasa}}, \bibinfo {author} {\bibfnamefont {E.}~\bibnamefont {Ihara}},
  \bibinfo {author} {\bibfnamefont {K.}~\bibnamefont {Fujita}}, \bibinfo
  {author} {\bibfnamefont {Y.}~\bibnamefont {Funaki}}, \bibinfo {author}
  {\bibfnamefont {K.}~\bibnamefont {Hatanaka}}, \bibinfo {author}
  {\bibfnamefont {H.}~\bibnamefont {Horiuchi}}, \bibinfo {author}
  {\bibfnamefont {M.}~\bibnamefont {Itoh}}, \bibinfo {author} {\bibfnamefont
  {J.}~\bibnamefont {Kamiya}}, \bibinfo {author} {\bibfnamefont
  {G.}~\bibnamefont {R\"{o}pke}}, \bibinfo {author} {\bibfnamefont
  {H.}~\bibnamefont {Sakaguchi}}, \bibinfo {author} {\bibfnamefont
  {N.}~\bibnamefont {Sakamoto}}, \bibinfo {author} {\bibfnamefont
  {Y.}~\bibnamefont {Sakemi}}, \bibinfo {author} {\bibfnamefont
  {P.}~\bibnamefont {Schuck}}, \bibinfo {author} {\bibfnamefont
  {Y.}~\bibnamefont {Shimizu}}, \bibinfo {author} {\bibfnamefont
  {M.}~\bibnamefont {Takashina}}, \bibinfo {author} {\bibfnamefont
  {S.}~\bibnamefont {Terashima}}, \bibinfo {author} {\bibfnamefont
  {A.}~\bibnamefont {Tohsaki}}, \bibinfo {author} {\bibfnamefont
  {M.}~\bibnamefont {Uchida}}, \bibinfo {author} {\bibfnamefont {H.~P.}\
  \bibnamefont {Yoshida}}, \ and\ \bibinfo {author} {\bibfnamefont
  {M.}~\bibnamefont {Yosoi}},\ }\href {\doibase
  https://doi.org/10.1016/j.physletb.2007.08.016} {\bibfield  {journal}
  {\bibinfo  {journal} {Phys. Lett. B}\ }\textbf {\bibinfo {volume} {653}},\
  \bibinfo {pages} {173} (\bibinfo {year} {2007})}\BibitemShut {NoStop}%
\bibitem [{\citenamefont {Bailey}\ \emph {et~al.}(2019)\citenamefont {Bailey},
  \citenamefont {Kokalova}, \citenamefont {Freer}, \citenamefont {Wheldon},
  \citenamefont {Smith}, \citenamefont {Walshe}, \citenamefont {Curtis},
  \citenamefont {Soi\ifmmode~\acute{c}\else \'{c}\fi{}}, \citenamefont
  {Prepolec}, \citenamefont {Toki\ifmmode~\acute{c}\else \'{c}\fi{}},
  \citenamefont {Marqu\'es}, \citenamefont {Achouri}, \citenamefont {Delaunay},
  \citenamefont {Deshayes}, \citenamefont {Parlog}, \citenamefont
  {Fern\'andez-Dominguez}, \citenamefont {Jacquot},\ and\ \citenamefont
  {Soylu}}]{Bailey2019}%
  \BibitemOpen
  \bibfield  {author} {\bibinfo {author} {\bibfnamefont {S.}~\bibnamefont
  {Bailey}}, \bibinfo {author} {\bibfnamefont {T.}~\bibnamefont {Kokalova}},
  \bibinfo {author} {\bibfnamefont {M.}~\bibnamefont {Freer}}, \bibinfo
  {author} {\bibfnamefont {C.}~\bibnamefont {Wheldon}}, \bibinfo {author}
  {\bibfnamefont {R.}~\bibnamefont {Smith}}, \bibinfo {author} {\bibfnamefont
  {J.}~\bibnamefont {Walshe}}, \bibinfo {author} {\bibfnamefont
  {N.}~\bibnamefont {Curtis}}, \bibinfo {author} {\bibfnamefont
  {N.}~\bibnamefont {Soi\ifmmode~\acute{c}\else \'{c}\fi{}}}, \bibinfo {author}
  {\bibfnamefont {L.}~\bibnamefont {Prepolec}}, \bibinfo {author}
  {\bibfnamefont {V.}~\bibnamefont {Toki\ifmmode~\acute{c}\else \'{c}\fi{}}},
  \bibinfo {author} {\bibfnamefont {F.~M.}\ \bibnamefont {Marqu\'es}}, \bibinfo
  {author} {\bibfnamefont {L.}~\bibnamefont {Achouri}}, \bibinfo {author}
  {\bibfnamefont {F.}~\bibnamefont {Delaunay}}, \bibinfo {author}
  {\bibfnamefont {Q.}~\bibnamefont {Deshayes}}, \bibinfo {author}
  {\bibfnamefont {M.}~\bibnamefont {Parlog}}, \bibinfo {author} {\bibfnamefont
  {B.}~\bibnamefont {Fern\'andez-Dominguez}}, \bibinfo {author} {\bibfnamefont
  {B.}~\bibnamefont {Jacquot}}, \ and\ \bibinfo {author} {\bibfnamefont
  {A.}~\bibnamefont {Soylu}},\ }\href {\doibase 10.1103/PhysRevC.100.051302}
  {\bibfield  {journal} {\bibinfo  {journal} {Phys. Rev. C}\ }\textbf {\bibinfo
  {volume} {100}},\ \bibinfo {pages} {051302(R)} (\bibinfo {year}
  {2019})}\BibitemShut {NoStop}%
\bibitem [{\citenamefont {Nesterenko}\ \emph {et~al.}(2006)\citenamefont
  {Nesterenko}, \citenamefont {Kleinig}, \citenamefont {Kvasil}, \citenamefont
  {Vesely}, \citenamefont {Reinhard},\ and\ \citenamefont {Dolci}}]{Nes06}%
  \BibitemOpen
  \bibfield  {author} {\bibinfo {author} {\bibfnamefont {V.~O.}\ \bibnamefont
  {Nesterenko}}, \bibinfo {author} {\bibfnamefont {W.}~\bibnamefont {Kleinig}},
  \bibinfo {author} {\bibfnamefont {J.}~\bibnamefont {Kvasil}}, \bibinfo
  {author} {\bibfnamefont {P.}~\bibnamefont {Vesely}}, \bibinfo {author}
  {\bibfnamefont {P.~G.}\ \bibnamefont {Reinhard}}, \ and\ \bibinfo {author}
  {\bibfnamefont {D.~S.}\ \bibnamefont {Dolci}},\ }\href@noop {} {\bibfield
  {journal} {\bibinfo  {journal} {Phys. Rev. C}\ }\textbf {\bibinfo {volume}
  {74}},\ \bibinfo {pages} {064306} (\bibinfo {year} {2006})}\BibitemShut
  {NoStop}%
\bibitem [{\citenamefont {Nesterenko}\ \emph {et~al.}(2008)\citenamefont
  {Nesterenko}, \citenamefont {Kleinig}, \citenamefont {Kvasil}, \citenamefont
  {Vesely},\ and\ \citenamefont {Reinhard}}]{Nes08}%
  \BibitemOpen
  \bibfield  {author} {\bibinfo {author} {\bibfnamefont {V.~O.}\ \bibnamefont
  {Nesterenko}}, \bibinfo {author} {\bibfnamefont {W.}~\bibnamefont {Kleinig}},
  \bibinfo {author} {\bibfnamefont {J.}~\bibnamefont {Kvasil}}, \bibinfo
  {author} {\bibfnamefont {P.}~\bibnamefont {Vesely}}, \ and\ \bibinfo {author}
  {\bibfnamefont {P.-G.}\ \bibnamefont {Reinhard}},\ }\href {\doibase
  10.1142/S0218301308009586} {\bibfield  {journal} {\bibinfo  {journal} {Int.
  J. Mod. Phys. E}\ }\textbf {\bibinfo {volume} {17}},\ \bibinfo {pages} {89}
  (\bibinfo {year} {2008})}\BibitemShut {NoStop}%
\bibitem [{\citenamefont {Kleinig}\ \emph {et~al.}(2008)\citenamefont
  {Kleinig}, \citenamefont {Nesterenko}, \citenamefont {Kvasil}, \citenamefont
  {Reinhard},\ and\ \citenamefont {Vesely}}]{SSRPA08}%
  \BibitemOpen
  \bibfield  {author} {\bibinfo {author} {\bibfnamefont {W.}~\bibnamefont
  {Kleinig}}, \bibinfo {author} {\bibfnamefont {V.~O.}\ \bibnamefont
  {Nesterenko}}, \bibinfo {author} {\bibfnamefont {J.}~\bibnamefont {Kvasil}},
  \bibinfo {author} {\bibfnamefont {P.-G.}\ \bibnamefont {Reinhard}}, \ and\
  \bibinfo {author} {\bibfnamefont {P.}~\bibnamefont {Vesely}},\ }\href@noop {}
  {\bibfield  {journal} {\bibinfo  {journal} {Phys. Rev. C}\ }\textbf {\bibinfo
  {volume} {78}},\ \bibinfo {pages} {044313} (\bibinfo {year}
  {2008})}\BibitemShut {NoStop}%
\bibitem [{\citenamefont {Chabanat}\ \emph {et~al.}(1998)\citenamefont
  {Chabanat}, \citenamefont {Bonche}, \citenamefont {Haensel}, \citenamefont
  {Meyer},\ and\ \citenamefont {Schaeffer}}]{Chabanat98}%
  \BibitemOpen
  \bibfield  {author} {\bibinfo {author} {\bibfnamefont {E.}~\bibnamefont
  {Chabanat}}, \bibinfo {author} {\bibfnamefont {P.}~\bibnamefont {Bonche}},
  \bibinfo {author} {\bibfnamefont {P.}~\bibnamefont {Haensel}}, \bibinfo
  {author} {\bibfnamefont {J.}~\bibnamefont {Meyer}}, \ and\ \bibinfo {author}
  {\bibfnamefont {R.}~\bibnamefont {Schaeffer}},\ }\href {\doibase
  https://doi.org/10.1016/S0375-9474(98)00180-8} {\bibfield  {journal}
  {\bibinfo  {journal} {Nucl. Phys. A}\ }\textbf {\bibinfo {volume} {635}},\
  \bibinfo {pages} {231 } (\bibinfo {year} {1998})}\BibitemShut {NoStop}%
\bibitem [{\citenamefont {Bender}\ \emph {et~al.}(2000)\citenamefont {Bender},
  \citenamefont {Rutz}, \citenamefont {Reinhard},\ and\ \citenamefont
  {Maruhn}}]{Bender00}%
  \BibitemOpen
  \bibfield  {author} {\bibinfo {author} {\bibfnamefont {M.}~\bibnamefont
  {Bender}}, \bibinfo {author} {\bibfnamefont {K.}~\bibnamefont {Rutz}},
  \bibinfo {author} {\bibfnamefont {P.-G.}\ \bibnamefont {Reinhard}}, \ and\
  \bibinfo {author} {\bibfnamefont {J.}~\bibnamefont {Maruhn}},\ }\href@noop {}
  {\bibfield  {journal} {\bibinfo  {journal} {Eur. Phys. J. A}\ }\textbf
  {\bibinfo {volume} {8}},\ \bibinfo {pages} {59} (\bibinfo {year}
  {2000})}\BibitemShut {NoStop}%
\bibitem [{\citenamefont {Ring}\ and\ \citenamefont {Schuck}(1980)}]{Ring1980}%
  \BibitemOpen
  \bibfield  {author} {\bibinfo {author} {\bibfnamefont {P.}~\bibnamefont
  {Ring}}\ and\ \bibinfo {author} {\bibfnamefont {P.}~\bibnamefont {Schuck}},\
  }\href@noop {} {\emph {\bibinfo {title} {The Nuclear Many-Body Problem}}}\
  (\bibinfo  {publisher} {Springer},\ \bibinfo {year} {1980})\BibitemShut
  {NoStop}%
\bibitem [{\citenamefont {Soloviev}(1992)}]{Sol}%
  \BibitemOpen
  \bibfield  {author} {\bibinfo {author} {\bibfnamefont {V.}~\bibnamefont
  {Soloviev}},\ }\href@noop {} {\emph {\bibinfo {title} {Theory of Atomic
  Nuclei: Quasiparticles and Phonons}}}\ (\bibinfo  {publisher} {Institute of
  Physics},\ \bibinfo {year} {1992})\BibitemShut {NoStop}%
\bibitem [{\citenamefont {Ponomarev}\ \emph {et~al.}(1979)\citenamefont
  {Ponomarev}, \citenamefont {Soloviev}, \citenamefont {Stoyanov},\ and\
  \citenamefont {Vdovin}}]{WSs}%
  \BibitemOpen
  \bibfield  {author} {\bibinfo {author} {\bibfnamefont {V.~Y.}\ \bibnamefont
  {Ponomarev}}, \bibinfo {author} {\bibfnamefont {V.~G.}\ \bibnamefont
  {Soloviev}}, \bibinfo {author} {\bibfnamefont {C.}~\bibnamefont {Stoyanov}},
  \ and\ \bibinfo {author} {\bibfnamefont {A.~I.}\ \bibnamefont {Vdovin}},\
  }\href@noop {} {\bibfield  {journal} {\bibinfo  {journal} {Nucl. Phys. A}\
  }\textbf {\bibinfo {volume} {323}},\ \bibinfo {pages} {446} (\bibinfo {year}
  {1979})}\BibitemShut {NoStop}%
\bibitem [{\citenamefont {Gareev}\ \emph {et~al.}(1973)\citenamefont {Gareev},
  \citenamefont {Ivanova}, \citenamefont {Soloviev},\ and\ \citenamefont
  {Fedotov}}]{WSd}%
  \BibitemOpen
  \bibfield  {author} {\bibinfo {author} {\bibfnamefont {F.~A.}\ \bibnamefont
  {Gareev}}, \bibinfo {author} {\bibfnamefont {S.~P.}\ \bibnamefont {Ivanova}},
  \bibinfo {author} {\bibfnamefont {V.~G.}\ \bibnamefont {Soloviev}}, \ and\
  \bibinfo {author} {\bibfnamefont {S.~I.}\ \bibnamefont {Fedotov}},\
  }\href@noop {} {\bibfield  {journal} {\bibinfo  {journal} {Sov. J. Part.
  Nucl.}\ }\textbf {\bibinfo {volume} {4}},\ \bibinfo {pages} {357} (\bibinfo
  {year} {1973})}\BibitemShut {NoStop}%
\bibitem [{\citenamefont {Tselyaev}\ \emph {et~al.}(2016)\citenamefont
  {Tselyaev}, \citenamefont {Lyutorovich}, \citenamefont {Speth}, \citenamefont
  {Krewald},\ and\ \citenamefont {Reinhard}}]{Tselyaev2016}%
  \BibitemOpen
  \bibfield  {author} {\bibinfo {author} {\bibfnamefont {V.}~\bibnamefont
  {Tselyaev}}, \bibinfo {author} {\bibfnamefont {N.}~\bibnamefont
  {Lyutorovich}}, \bibinfo {author} {\bibfnamefont {J.}~\bibnamefont {Speth}},
  \bibinfo {author} {\bibfnamefont {S.}~\bibnamefont {Krewald}}, \ and\
  \bibinfo {author} {\bibfnamefont {P.-G.}\ \bibnamefont {Reinhard}},\ }\href
  {http://dx.doi.org/10.1103/PhysRevC.94.034306} {\bibfield  {journal}
  {\bibinfo  {journal} {Phys. Rev. C}\ }\textbf {\bibinfo {volume} {94}},\
  \bibinfo {pages} {034306} (\bibinfo {year} {2016})}\BibitemShut {NoStop}%
\end{thebibliography}%

\end{document}